\documentclass[prd,superscriptaddress,showpacs,showkeys]{revtex4}

\usepackage{amsmath}
\usepackage{graphicx}
\usepackage{amssymb}
\usepackage{wasysym}
\usepackage{hyperref}

\newcommand{\half}{{\textstyle \frac{1}{2}}}
\newcommand{\wt}[1]{\widetilde{#1}}

\newlength{\defaultfig}
\newlength{\widefig}
\newlength{\narrowfig}

\DeclareMathOperator{\tr}{Tr}

\graphicspath{{newfigs/}}

\begin{document}

\title{Two- and three-body color flux tubes in the \\
  Chromo Dielectric Model}

\date{\today}

\author{Gunnar Martens}
\email[e-mail: ]{Gunnar.Martens@theo.physik.uni-giessen.de}
\affiliation{Institut f\"ur Theoretische Physik, 
  Universit\"at Giessen, Germany  
}
\author{Carsten Greiner}
\affiliation{Institut f\"ur Theoretische Physik, 
  Universit\"at Frankfurt, Germany  }
\author{Stefan Leupold}
\author{Ulrich Mosel}
\affiliation{Institut f\"ur Theoretische Physik, 
  Universit\"at Giessen, Germany
}

\begin{abstract}
  Using the framework of the Chromo Dielectric Model we perform an
  analysis of color electric flux tubes in meson-like $q\bar{q}$ and
  baryon-like $qqq$ quark configurations. We discuss the Abelian color
  structure of the model and point out a symmetry in color space as
  a remnant of the SU(3) symmetry of QCD. 
  The generic features of the model are discussed by varying the model
  parameters. We fix these parameters by 
  reproducing the string tension  $\tau=980\,$MeV/fm and the
  transverse width $\rho=0.35\,$fm of the $q\bar{q}$ flux tube
  obtained in  lattice calculations. We use a bag
  constant $B^{1/4}=(240-260)\,$MeV, a glueball mass $m_g =
  (1000-1700)\,$MeV and a strong coupling constant $C_F \alpha_s =
  0.2-0.3$. We show that the asymptotic string profile of an infinitely
  long flux tube is already reached for $q\bar{q}$ separations
  $R\ge1.0\,$fm. A connection to the Dual Color Superconductor is made
  by extracting a magnetic current from the model equations and a
  qualitative agreement between the two descriptions of confinement
  is shown. In the study of
  the $qqq$ system we observe a $\mathsf{\Delta}$-like geometry for the color
  electric fields and a \textsf{Y}-like geometry in the scalar fields both in
  the energy density distribution and in the corresponding
  potentials. The resulting total $qqq$ potential is described neither
  by the $\mathsf{\Delta}$-picture nor by the \textsf{Y}-picture alone.
\end{abstract}



\maketitle

\section{Introduction}
\label{sec:intro}

\setlength{\defaultfig}{0.7\textwidth}
\setlength{\widefig}{\textwidth}
\setlength{\narrowfig}{0.45\columnwidth}


The structure of hadrons is still a subject of discussion. It is
widely accepted that Quantum Chromodynamics (QCD) is the right theory
for strong interactions, and that the properties of hadrons can be
described 
within this framework. QCD has been tested successfully in the region
of large momentum transfer, where perturbative methods work due to
asymptotic freedom. However, in the region of small relative momentum,
where the formation of hadrons sets in and confinement plays a
dominant role, a calculation from first principles is still limited
to lattice techniques.  

In order to understand the formation of hadrons out of quarks and gluons
dynamically, and in turn the interactions of hadrons with each other,
one still has to rely on models that include the phenomenon of
confinement. Such phenomenological models are for example the bag model
\cite{Chodos:1974je,Chodos:1974pn}, where quarks are treated as free
particles restricted to predefined bags, the quark molecular dynamics
model \cite{Hofmann:1999jx,Scherer:2001ap}, where colored quarks
obey the classical Hamilton dynamics bound by a linear confining
potential, or the model of the dual superconductor
\cite{Baker:1991bc,Ripka04:dual_super}, where confinement is achieved
by monopole condensation \cite{'tHooft:1974qc,Polyakov:1974ek} and
an accompanying dual supercurrent. The model of the stochastic vacuum
relies on the calculation of Wilson-loops in a Gaussian approximation
\cite{Dosch:1987sk,Shoshi:2002rd} which leads to a linear potential
between quarks and anti-quarks. 

In this work we adopt the Chromo Dielectric Model (CDM)
\cite{Friedberg:1977eg,Friedberg:1977xf,wilets:1989}, which is an
extension of the bag model in the sense that bags are formed
dynamically in the presence of quarks. Mack \cite{Mack:1983yi} and
Pirner et al.\ \cite{Pirner:1984hd,Mathiot:1989tu,Pirner:1992im} have
given a renormalization group  derivation of the lattice
colordielectric model, which also has a scalar field modelling
confinement, but keeps strongly coupled non-Abelian fields in the
large distance action. A Monte-Carlo calculation within the lattice
colordielectric model was done in
\cite{Pirner:1987qj,Grossmann:1990ws}. The model has already been 
used to calculate hadron properties like low lying baryon masses, the nucleon
magnetic moments and the (axial-vector)/vector coupling constant ratio
\cite{Goldflam:1982tg} and  nucleon--nucleon
interactions in vacuum and in nuclear matter
\cite{Pirner:1984hd,Achtzehnter:1985ur}. 
\cite{Schuh:1986mi,Koepf:1994uq,Pepin:1996mp}. In another 
approach the description has been used within a transport
theoretical scheme to describe the dynamics
\cite{Vetter:1995gp,Kalmbach:1993sp,Loh:1997sk} of quarks bound in
nucleons and strings. In \cite{Traxler:1998bk} a full molecular
dynamics simulation for colored quarks was performed, showing the
ability of the model to produce color neutral hadrons out of a gas of
colored quarks, thus giving for the first time a microscopic description of
hadronization from a quark gluon plasma.

The parameters used in
\cite{Vetter:1995gp,Kalmbach:1993sp,Loh:1997sk,Traxler:1998bk}, which 
define the model, were
mostly motivated by phenomenological arguments and not subject to a
further investigation. The resulting color flux tubes are rather
large objects with a string radius up to 1 fm. In addition the
linear rising $q\bar{q}$ potential was only seen for quark
separations $R > 1\,$fm. On the other hand, lattice calculations
\cite{Bali:1995de,Trottier:1993jv}  indicate, that the radius
of a colored flux tube is much smaller than 1 fm, and that the
$q\bar{q}$ potential already develops a linear rising term for
separations larger than $R=0.2\,$fm. In addition, on the lattice a
clear Coulomb-like potential was observed for quark separations
$R<0.2\,$fm, which has not been resolved in \cite{Traxler:1998bk}. In
the present work we model the results from lattice calculations as well as
possible within the framework of the CDM.
On the lattice the most accurate
results were obtained in SU(2) \cite{Bali:1995de}. We assume that
the shape of the color flux tube does not depend significantly on the underlying
Lie algebra. Therefore we compare the results of our
calculations to those obtained in lattice SU(2) theory. The quantities
that we want to reproduce are the transverse shape of a flux tube of
given length, and  the linear coefficient of the $q\bar{q}$ 
potential obtained in meson spectroscopy
\cite{Eichten:1975af,Quigg:1979vr,Eichten:1980ms}. 

Having established a set of parameters matching the criteria above, we 
then analyze the structure of the flux tubes. The emphasis will be on 
the question how
the string is build up when the constituents of the string are
separated from each other. In varying the $q\bar{q}$ distance $R$ we
probe both the perturbative (small $R$) and  the
non-perturbative region (large $R$). We will also see how fast the
transition from one to the other sets in and from which distances on
the string picture holds. 

The formation of strings is not restricted to $q\bar{q}$ objects. In
QCD one has the possibility to build up color neutral objects from
three quarks. When the pairwise quark separations are large compared
to the characteristic width of the string, color flux tubes will
stretch between the quarks. In general two geometrically different
pictures are possible. The first is the so called \textsf{Y}-geometry,
where three flux tubes meet at a central point
\cite{Artru:1975zn,Brambilla:1994zw}. The second one is the
$\mathsf{\Delta}$-geometry, where the three quarks are connected
pairwise 
\cite{Cornwall:1977xd,Cornwall:1996xr}. The three-quark potential emerging
from these two pictures has been compared to lattice results in
\cite{Alexandrou:2001ip} and \cite{Takahashi:2002bw}. Due to the lack of
numerical precision the two groups obtained different results.
Therefore lattice calculations cannot clearly rule out one of the
pictures so far. Within our model we
are able to describe those baryonic quark configurations not only on
the level of the $qqq$ potential but also on the level of the energy
distributions. We can study the structure of the formed flux tubes and
discriminate between the two geometries. 

The main goal of this work is to fix the model parameters on lattice
data of $q\bar{q}$ flux tubes. The parameters obtained will be used to
describe both the shape and the potential of the three quark system
within CDM.
The structure of this work is as follows. In section \ref{sec:cdm} we present
the model and its specific mechanism of confinement. In section
\ref{sec:analysis} we analyze the dependence of the shape and the potential of 
a $q\bar{q}$ string on the variation of the parameters introduced in the 
model. We compare our numerical $q\bar{q}$ results 
for three specific sets of parameters to those obtained within lattice
gauge calculations in section~\ref{sec:strings} and extend the analysis to
baryon-like three-quark systems in section~\ref{sec:baryons}. 
In the appendix \ref{sec:numerics} we give a description of the
algorithm used in our numerical calculations.

\section{The Chromo dielectric Model}
\label{sec:cdm}

Presumably the non-Abelian gluon interactions of QCD are responsible for a
highly structured non-perturbative vacuum. Although it is difficult to
disentangle these interactions from first principles, the large scale
behavior of strong interactions might be simple. 
In the Chromo Dielectric Model (CDM) one assumes, that the QCD vacuum 
behaves as a perfect dielectric medium, i.e. as a medium with
vanishing dielectric constant. Colored quarks embedded in this
vacuum produce electric color fields. In the 
presence of the dielectric medium these fiels are compressed in well
defined flux tubes connecting quarks with opposite charge. 

To be more specific, the medium is described by a colorless scalar
field $\sigma$ which mediates the vacuum properties via the
dielectric function $\kappa(\sigma)$. The confinement field itself is
evolving in the presence of a scalar self interaction $U(\sigma)$.

As the dynamics and the non-Abelian interactions of the gluon sector 
are merged in the confinement field $\sigma$ and its dielectric
coupling $\kappa(\sigma)$ one is left with a set of two Abelian gluon
fields $A^{\mu,a}$ that interacts with the dielectric medium. In principle it
is possible to formulate the model with dynamical quarks, described by
a Dirac-like Lagrangian
\cite{Goldflam:1982tg,Schuh:1986mi,Fai:1988ak}, but in this work we
concentrate on the 
structure of color flux tubes. Therefore we treat the quarks as
external sources of color fields. 
The CDM can now be defined by the following Lagrange density:
\begin{subequations}
  \label{eq:lagrangian}
  \begin{eqnarray}
    \label{subeq:lagrangean}
    {\cal L} &=& {\cal L}_g + {\cal L}_{\sigma} \quad ,\\
    \label{subeq:gluon}
    {\cal L}_g &=& - {\textstyle \frac{1}{4}} {\kappa(\sigma)}
    F_{\mu\nu}^{a} F^{\mu\nu, {a}}-g_s \;j_\mu^a \,A^{\mu,a} \quad , \\
    \label{subeq:sigma}
    {\cal L}_{\sigma} &=& {{\textstyle \frac{1}{2}}
      \partial_\mu\sigma\partial^\mu\sigma  - U(\sigma) } \quad , \\
    \label{subeq:gluontensor}
    F^{\mu\nu, { a}} &=& \partial^\mu A^{\nu,a} 
    - \partial^\nu A^{\mu,a}, \;\;\; a \in\{3,8\}  \quad .
  \end{eqnarray}  
\end{subequations}
The color fields $A^{\mu,a}$ couple to the color charge current
$j^{\mu,a} = (\rho^a,\vec{\jmath}^a)$ classically with a strong
coupling constant $g_s$ which we have not included in the definition
of the current. The charge density is defined as a sum over
all charged sources, i.e. the quarks, $\rho^a(\vec{x}) = \sum_k q_k^a
w(\vec{x}-\vec{x}_k)$, where the $q^a$ are the color charges of the
quarks. In the static case we are interested in, the spatial part of
the color current vanishes  $\vec{\jmath}=0$. 
The quarks are in principle point-like objects but in our numerical
analysis described in sec.~\ref{sec:analysis} we assign a finite
Gaussian width $w(\vec{x}) = (2\pi r_0^2)^{-3/2} \exp(-\vec{x}^2/2
r_0^2)$ to each quark. The width is introduced for numerical reasons
(see appendix \ref{sec:fas}). Throughout this work we adopt a
value of $r_0 = 0.02\,$fm, which is large enough to resolve the
Gaussian distribution $w(\vec{x})$ and small compared to the
dimensions of the flux tubes.

\begin{figure}[htbp]
  \begin{center}
    \includegraphics[width=0.6\defaultfig,keepaspectratio,clip]{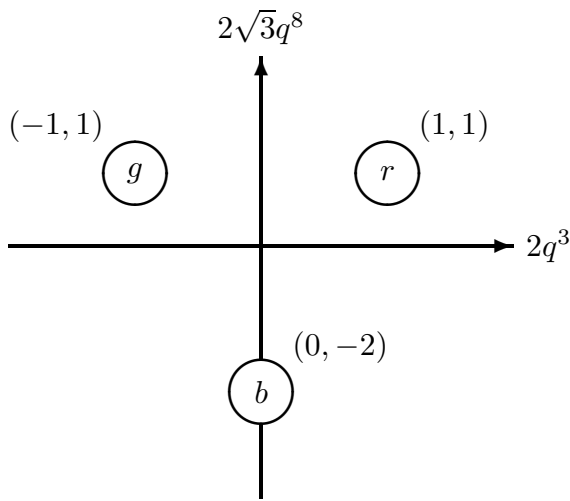}
    \caption{The color charge $q^{3,8}$ with respect to the color
      fields $A^{3,8}$.} 
    \label{fig:triplett}
  \end{center}
\end{figure}

As we are working in an Abelian model inspired by QCD, we have three
different colors interacting with only two  
Abelian color fields. In the color space we choose an arbitrary but
fixed base for the quarks in the fundamental representation,
$|\mbox{red}\rangle = (1,0,0)^T$, $|\mbox{green}\rangle = (0,1,0)^T$
and $|\mbox{blue}\rangle = (0,0,1)^T$. The color charges $q^a$ are then
defined as the diagonal entries of the corresponding generators $t^a$
of the color group in the same representation $q_c^a = \langle
c|t^a|c\rangle$, where $c \in \{\mbox{red (r), green (g),
  blue (b)}\}$ and $a \in\{3,8\}$. The numerical values of the color
charges can be read off from table \ref{tab:charges} and are depicted
in figure \ref{fig:triplett}. The generators are normalized according
to $\tr\, t^at^b = \delta^{ab}/2$. Note that by this
definition the $q_a$ are reduced by a factor of two as compared to
\cite{Traxler:1998bk} and that the strong coupling parameter $g_s$
is enhanced by the same factor leaving $g_s q_a$ fixed.

\begin{table}[htbp]
  \begin{ruledtabular}
    \begin{center}
      \begin{tabular}{lrr}
        color    &   $q^3$   &   $q^8$ \\\hline
        red      &   $1/2$     &   $1/(2\sqrt{3})$ \\
        green    &  $-1/2$     &   $1/(2\sqrt{3})$ \\
        blue     &   $0$     &   $-1/\sqrt{3}$
      \end{tabular}
      \caption{The color charges $q^a$ of the three colors with respect
        to the two Abelian color fields.}
      \label{tab:charges}
    \end{center}
  \end{ruledtabular}
\end{table}

In the Abelian projected theory one is left with two independent color
fields and a remaining U(1)$\times$U(1) gauge symmetry. The
corresponding gauge transformations are
\begin{eqnarray}
  \label{eq:u1trafo}
  |c\rangle &\rightarrow& U(x) |c\rangle = 
  \text{diag}\left(e^{ig_s\chi_1},e^{ig_s\chi_2},
    e^{-ig_s(\chi_1+\chi_2)}\right) |c\rangle \\
  A^\mu &\rightarrow& \left(A_a^\mu + \partial^\mu \theta_a\right) t_a, 
\end{eqnarray}
where $\chi_1 = \tfrac{\theta_3}{2} + \tfrac{\theta_8}{2\sqrt{3}}$ and
$\chi_2 = -\tfrac{\theta_3}{2} + \tfrac{\theta_8}{2\sqrt{3}}$.
This means that
the color charges $q^{3/8}$ are conserved independently and in
turn the color fields $F^{\mu\nu, a}$ would be observable
fields. However in the Abelian approximation there is a
further symmetry, namely a symmetry under special discrete rotations
in color space
\begin{equation}
  \label{eq:discrete_color}
  |c\rangle \rightarrow |c'\rangle = V |c\rangle \quad .
\end{equation}
In matrix form the color rotations are given explicitly as:

\begin{subequations}
  \label{eq:eq:colorrotation2}
  \begin{eqnarray}
    \label{subeq:id}
    V_1:\quad V_{11}^1 = \epsilon_1, \quad V_{22}^1 = \epsilon_2, \quad V_{33}^1 &=&
    \epsilon_1\epsilon_2\\ 
    \label{subeq:cyc1}
    V_2:\quad V_{12}^2 = \epsilon_1, \quad V_{23}^2 = \epsilon_2, \quad V_{31}^2 &=&
    \epsilon_1\epsilon_2\\
    \label{subeq:cyc2}
    V_3:\quad V_{13}^3 = \epsilon_1, \quad V_{21}^3 = \epsilon_2, \quad V_{32}^3 &=&
    \epsilon_1\epsilon_2\\
    \label{subeq:perm1}
    V_4:\quad V_{11}^4 = \epsilon_1, \quad V_{23}^4 = \epsilon_2, \quad V_{32}^4 &=&
    -\epsilon_1\epsilon_2\\ 
    \label{subeq:perm2}
    V_5:\quad V_{13}^5 = \epsilon_1, \quad V_{22}^5 = \epsilon_2, \quad V_{31}^5 &=&
    -\epsilon_1\epsilon_2\\ 
    \label{subeq:perm3}
    V_6:\quad V_{12}^6 = \epsilon_1, \quad V_{21}^6 = \epsilon_2, \quad V_{33}^6 &=&
    -\epsilon_1\epsilon_2,
  \end{eqnarray}  
\end{subequations}
where $\epsilon_{1/2} = \pm 1$ and all other matrix elements being
zero. There are therefore 4 different copies of each of the $V_i$
differing from each other in the signs $\epsilon_{1/2}$. 
These rotations $V$ act either as a cyclic exchange of all colors
$r\rightarrow g\rightarrow b$ \eqref{subeq:cyc1} and $b\rightarrow
g\rightarrow r$ \eqref{subeq:cyc2} or as a pairwise exchange of two 
colors $g\leftrightarrow b$ \eqref{subeq:perm1}, $r\leftrightarrow b$
\eqref{subeq:perm2} and  $r\leftrightarrow g$ \eqref{subeq:perm3} with
additional phases $\epsilon_{1/2}$. The transformations themselves
form a global subgroup $D$ of SU(3), but are not independent from the
former U(1)$\times$U(1) gauge group in the sense that out of the set
of four copies of $V_i$ three of them can be constructed by a
combination $V_i U(x)$ with $U(x) \in U(1)\times U(1)$. The discrete
color rotations $V_i$ transform the gauge fields $A^{\mu,3}$ and
$A^{\mu, 8}$ only into each other without mixing to the non-Abelian
gauge fields: 
\begin{equation}
  \label{eq:gauge_rot}
  A^\mu = A^{\mu,3} t_3 + A^{\mu,8} t_8 \rightarrow V A^\mu V^\dagger =
  A'^{\mu,3} t_3 + A'^{\mu,8} t_8,
\end{equation}
with $(A^{\mu,3})^2 + (A^{\mu,8})^2 = (A'^{\mu,3})^2 +
(A'^{\mu,8})^2$ and $A'^{\mu,a} \neq A^{\mu,a}$. The same is true for
the color fields $F^{\mu\nu,3}$ and $F^{\mu\nu,8}$. They are
therefore not invariant under the rotations $V \in D$. This is a
relict of the full SU(3) gauge symmetry. Of course,
the action density \eqref{eq:lagrangian} and the corresponding energy
density, which are the only physical meaningful quantities in the
model, are invariant under $V\in D$.

\begin{figure}[htbp]
  \centering
  \includegraphics[width=\defaultfig,keepaspectratio,clip]{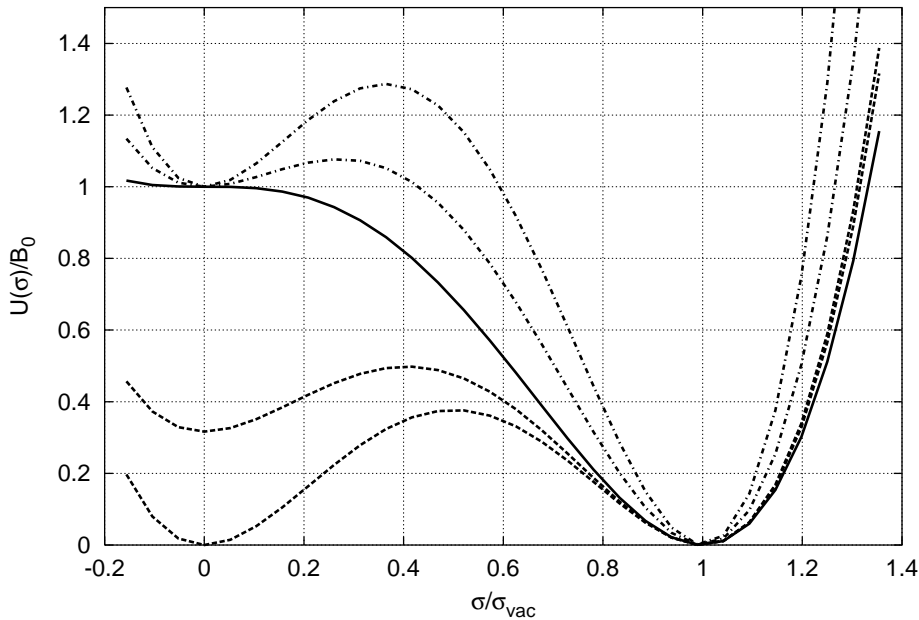}
  \caption{The scalar potential $U(\sigma)$ with $m_g$ fulfilling
  the equality in eq.~\eqref{eq:inflection} (solid curve),
  fixed $m_g$ (dashed curves) and fixed $B$ (dash-dotted curves).} 
  \label{fig:U_pot}
\end{figure}

In the confinement part of the Lagrangian \eqref{subeq:sigma} the
scalar self interaction $U(\sigma)$ is of a quartic form, i.e. 
\begin{equation}
  \label{eq:uscalar}
  U(\sigma) = B + a\sigma^2 + b\sigma^3 + c\sigma^4.
\end{equation}
The form is chosen to develop two stable points. A metastable one at
$\sigma = 0$ and a stable one at the vacuum expectation value $\sigma =
\sigma_\text{vac}$ (see fig.~\ref{fig:U_pot}). 
The requirement that $U$ has an absolute minimum at $\sigma =
\sigma_\text{vac}$, together with $U(\sigma_\text{vac})=0$, leaves only
two additional free
parameters which we choose to be the bag constant $B$ and the
curvature $m_g^2 = U''(\sigma_\text{vac})$ of the potential
$U(\sigma)$ at the absolute minimum. Since the confinement field
$\sigma$ absorbs non-Abelian gluon properties of QCD, we can interpret
$m_g$ as the mass of the lowest collective gluon excitation, i.e. the glueball
mass. 
The parameters $a$, $b$ and $c$ can therefore be
expressed by the quantities $B$, $m_g$ and $\sigma_\text{vac}$:
\begin{subequations}
  \label{eq:Uparameter}
  \begin{eqnarray}
    \label{subeq:Uapar}
    a &=& \frac{1}{2} \frac{m_g^2\sigma_\text{vac}^2 -
      12B}{\sigma_\text{vac}^2} \quad , \\
    \label{subeq:Ubpar}
    b &=& - \;\frac{m_g^2\sigma_\text{vac}^2 -
      8B}{\sigma_\text{vac}^3} \quad , \\    
    \label{subeq:Ucpar} 
    c &=& \frac{1}{2} \frac{m_g^2\sigma_\text{vac}^2 -
      6B}{\sigma_\text{vac}^4} \quad .
  \end{eqnarray}
\end{subequations}
In order to have a local minimum at $\sigma = 0$ we must fulfill $a\ge 
0$ or   
\begin{equation}
  \label{eq:inflection}
  m_g^2\sigma_\text{vac}^2 \ge 12B  
\end{equation}
The generic form of $U(\sigma)$ is shown in fig.~\ref{fig:U_pot}.
For the equality in eq.~\eqref{eq:inflection} the potential $U$ has
only an inflection point at $\sigma = 0$ (solid curve). For fixed
$m_g$ and $\sigma_\text{vac}$ we can vary $B$ up to an upper limit
given by eq.~\eqref{eq:inflection} (dashed curves). Alternatively we
can fix $B$ and $\sigma_\text{vac}$ and vary $m_g$ starting from a
lower bound given by eq.~\eqref{eq:inflection} (dash-dotted curves).

\begin{figure}[htbp]
  \begin{center}
    \includegraphics[width=\defaultfig]{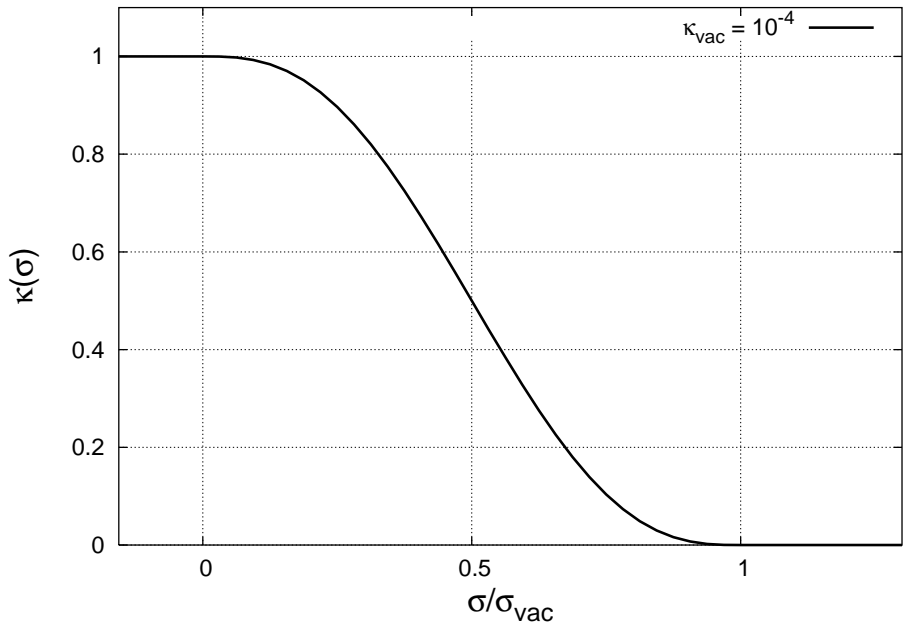}
    \caption{The dielectric function drops from its perturbative value
      $\kappa = 1$ at $\sigma=0$ to its non-perturbative value
      $\kappa_\text{vac}\ll 1$ at $\sigma=\sigma_\text{vac}$. Here we have
      used $\kappa_\text{vac}= 10^{-4}$.
    } 
    \label{fig:kappa}
  \end{center}
\end{figure}

The two (meta-)stable points separate two different phases of 
the vacuum. In the perturbative vacuum at $\sigma = 0$ where the dynamics
is driven by short range interactions electric fields can propagate
freely, i.e.\/ the vacuum is described by a dielectric constant $\kappa =
1$. In the non-perturbative phase $\sigma=\sigma_\text{vac}$
the dynamics is dominated by long range interactions and the
dielectric constant nearly vanishes ($\kappa = \kappa_\text{vac} \ll 1$).
In the limit $\kappa_\text{vac}\rightarrow 0$ the non-perturbative
vacuum behaves as a perfect dielectric. Between the two vacua
the dielectric function $\kappa(\sigma)$ drops continuously from 1 to
$\kappa_\text{vac}$ (fig.~\ref{fig:kappa}). We choose for its
parameterization a 5th order polynomial
\begin{equation}
  \label{eq:kappa}
  \kappa(s) = \left\{
    \begin{array}{r@{\quad,\quad}l}
      1 + k_3 s^3 + k_4  s^4 + k_5 s^5 
      & 0 \le s \le 1\\
      1 & s < 0\\
      \kappa_\text{vac} & s > 1
    \end{array}\right. \quad ,
\end{equation}
with $s = \sigma/\sigma_\text{vac}$ and with coefficients
\begin{eqnarray}
  \label{eq:kappa_koeff}
  \begin{array}{r@{\quad =\quad}l}
    k_3 & {\displaystyle \frac{1}{2} \left(29\kappa_\text{vac}- 20\right)}\\
    k_4 & {\displaystyle \phantom{\frac{1}{2}} \left(15 - 23\kappa_\text{vac}\right)}\\
    k_5 & {\displaystyle \frac{1}{2} \left(19\kappa_\text{vac}- 12\right)}
  \end{array}
  \stackrel{\kappa_\text{vac}\rightarrow\, 0}{\longrightarrow}
  \begin{array}{r@{\quad =\quad}r}
    k_3 & {\displaystyle \vphantom{\frac{1}{2}} -10}\\
    k_4 & {\displaystyle \vphantom{\frac{1}{2}} 15}\\
    k_5 & {\displaystyle \vphantom{\frac{1}{2}} -6}\\
  \end{array}
  \quad .
\end{eqnarray}
The coefficients are chosen such that the first two derivatives of 
$\kappa$ at $\sigma=0$ vanish and  both $\kappa$ and its first two
derivatives at $\sigma=\sigma_\text{vac}$ are proportional to
$\kappa_\text{vac}$ and thus vanish in the limit
$\kappa_\text{vac}\rightarrow 0$. The form of this parameterization is
only weakly sensitive to the value of $\kappa_\text{vac}$. We will see
that physical quantities do not depend on the actual value of
$\kappa_\text{vac}$ once it is chosen small enough. In
fig.~\ref{fig:kappa} we have chosen a value $\kappa_\text{vac} =
10^{-4}$ and the difference to $\kappa_\text{vac} = 10^{-3}$ is hidden
within the linewidth. Of course other parameterizations are
possible. In \cite{Martens:2003yy} we used
$\kappa(\sigma)=\kappa_\text{vac}^{-x^3}$ but this functional form
depends more strongly on $\kappa_\text{vac}$ and thus has more
influence on physical quantities. The dielectric function approaches
$\kappa_\text{vac}$ faster for decreasing $\kappa_\text{vac}$. As a
consequence the transverse shape of the color
flux tube becomes steeper with vanishing $\kappa_\text{vac}$ and
the energy of the system still increases even for very small values of
$\kappa_\text{vac}$. The actual polynomial choice is very similar to
that in \cite{Goldflam:1982tg} and \cite{Loh:1996nh}. 
All parameters defined in eqs.~\eqref{eq:lagrangian},
\eqref{eq:uscalar}, \eqref{eq:Uparameter} are subject to an
investigation described in sec.~\ref{sec:parameter}. The choice of
$\kappa_\text{vac}$ is discussed in appendix \ref{sec:kappa-dep}. 

From a variational principle we can derive the equations of motion for
the gluon fields $A^{\mu,a} = (\phi^a, \vec{A}^a)$ and the scalar
confinement field $\sigma$:
\begin{subequations}
  \label{eq:eom}
  \begin{eqnarray}
    \label{subeq:eom_color}
    \partial_\mu \left(\kappa F^{\mu\nu,a}\right) &=& g_s j^{\nu,a}
    \quad , \\
    \label{subeq:eom_scalar}
    \partial_\mu \partial^\mu \sigma &=& -U'(\sigma) - {\textstyle
    \frac{1}{4}} \kappa'(\sigma) F_{\mu\nu}^a F^{\mu\nu,a} \quad ,
  \end{eqnarray}
\end{subequations}
where the prime denotes a differentiation with respect to $\sigma$.
The color field tensor $F^{\mu\nu,a}$ in
eq.~\eqref{subeq:gluontensor} determines the color electric and
magnetic fields $\vec{E}_i^a = - F^{0i,a} = \left(-\nabla \phi^a-\partial_t
\vec{A}^a\right)_i$ and $\vec{B}_i^a = -\half \varepsilon_{ijk} F^{jk,a} = (\nabla
\times \vec{A}^a)_i$ respectively. With the help of the electric and
magnetic fields we can recast eq.~\eqref{subeq:eom_color} into the
two sets of inhomogeneous Maxwell equations, which we supplement with
the two homogenous ones, that are fulfilled automatically by the
definition of the field tensor in \eqref{subeq:gluontensor}
\begin{subequations}
  \label{eq:inh_maxwell}
  \begin{eqnarray}
    \label{subeq:gauss}
    \nabla \cdot \vec{D}^a &=& g_s \rho^a  \\
    \label{subeq:ampere}
    \nabla \times \vec{H}^a - \partial_t \vec{D}^a &=& g_s
    \vec{\jmath}^{\;a} \\ 
    \label{subeq:faraday}
    \nabla \times \vec{E}^a + \partial_t \vec{B}^a &=& 0 \\
    \label{subeq:nomonopoles}
    \nabla \cdot \vec{B} &=& 0 \quad ,
  \end{eqnarray}
\end{subequations}
where we have introduced the electric displacement $\vec{D}^a =
\kappa(\sigma) \vec{E}^a$ and the magnetic field $\vec{H}^a =
\kappa(\sigma) \vec{B}^a$. 
The energy of the system is given by 
\begin{equation}
  \label{eq:totenergy}
  E_\text{tot} = \int d^3x \half (\vec{E}^a\cdot\vec{D}^a + \vec{B}^a\cdot\vec{H}^a)
  + \half (\partial_t \sigma)^2 + \half (\nabla \sigma)^2 + U(\sigma)
  \quad .
\end{equation}
In this work we are interested in static
solutions of given quark configurations, i.e.\/ $\vec{\jmath}=0$, and one
can assume that $A^{\mu,a}$ and $\sigma$ are
also time independent. In this case the energy is minimized
with $\vec{B}^a = 0$ and eq.~\eqref{eq:totenergy} reduces to
\begin{subequations}
  \label{eq:energy_decomp}
  \begin{eqnarray}
    \label{eq:energy_1}
    E_\text{tot} &=& E_\text{el} + E_\text{vol} + E_\text{sur} \\
    \label{eq:energy_el}
    E_\text{el} &=& \half \int d^3r \, \vec{E}^a\cdot\vec{D}^a \\
    \label{eq:energy_vol}
    E_\text{vol} &=& \int d^3r\, U(\sigma) \\
    \label{eq:energy_sur}
    E_\text{sur} &=& \half \int d^3r \, \left(\nabla \sigma\right)^2
  \end{eqnarray}  
\end{subequations}
subject to the constraint in form of Gauss's law
\eqref{subeq:gauss}. The equations of motion now read
\begin{subequations}
  \label{eq:static_eoms}
  \begin{eqnarray}
    \label{eq:poisson}
    \nabla\cdot(\kappa(\sigma) \nabla\phi^a) &=& -g_s \rho^a \\
    \label{eq:static_sigma}
    \nabla^2 \sigma - U'(\sigma) &=& - \half \kappa'(\sigma)
    \vec{E}^a\cdot\vec{E}^a \quad ,
  \end{eqnarray}
\end{subequations}
Note that Gauss's law has to be fulfilled always
and that it is conserved by the dynamics of the system. Note also that
in the electric part of the  energy we have included the self energy
of a charge distribution which diverges for point-like particles. The
absence of the magnetic field might be due to the neglect of quantum
effects. On the lattice \cite{Bali:1995de,Pennanen:1997qm} it is seen, that energy
density and action density differ to some extent, i.e.\
$(\vec{B}^a)^2\neq 0$. As pointed out in \cite{Shoshi:2002rd} in the
framework of the stochastic vacuum model, the squared magnetic field is
dependent on the renormalization scale. In the cited work the
calculation of a $q\bar{q}$ flux tube was performed at a scale, where
the magnetic field vanishes as well. However, in the work of Pirner in
a non-Abelian version of the dielectric model \cite{Pirner:1992im}
a non-vanishing magnetic field arises naturally. 

The coupled system of equations \eqref{eq:static_eoms} for the two
color electric potentials $\phi^a$ and the confinement field $\sigma$
are the fundamental equations to be solved in this work. This is achieved
numerically by the \emph{Full Approximation Storage} 
algorithm described in appendix \ref{sec:fas}.

There is one comment about the validity of the classical treatment of
the model. By describing the quark fields by classical
charge distributions one might expect some modification of the results
due to the neglect of quantum mechanical aspects. One famous
modification of string-like objects is for example the L\"uscher-term
showing up in the $q\bar{q}$-potential \cite{Luscher:1980ac} and
corrections to it \cite{Arvis:1983fp}, which is due to quantum
fluctuations around the classical solution. We will discuss the
$q\bar{q}$-potential within our model later in
sec.~\ref{sec:generic}. Another modification might be expected due to
the neglect of possible superpositions of states in color space. 
We will show now, that in the Abelian approximation the  
energy density is unaffected by changing from a quantum mechanical 
superposition to the classical analog used in our calculations.

To be explicit regard a meson type state 
$|q\bar{q}\rangle$ and a baryon type state $|qqq\rangle$
\begin{subequations}
  \label{eq:qmstates}
  \begin{eqnarray}
    \label{subeq:qm_meson}
    |q\bar{q}\rangle &=& |M\rangle = \frac{1}{\sqrt{3}} \sum_c |(c\psi)\,
    (\bar{c}\bar{\psi})\rangle \\
    \label{subeq:qm_baryon}
    |qqq\rangle &=& |B\rangle = \frac{1}{\sqrt{3!}} \sum_{ijk}
    \epsilon_{ijk} |(c_i \psi_1)\,(c_j \psi_2)\,(c_k \psi_3)\rangle
    \quad , 
  \end{eqnarray}
\end{subequations}
where $c$($\bar{c}$) denotes the color of the (anti) quarks and
$\psi$($\bar{\psi}$) the spatial part of the (anti) quark
wavefunction. Here we have neglected all other quark quantum numbers
like spin and flavor as they do not enter the Lagrangian. 
The contraction with the total anti-symmetric tensor
$\epsilon_{ijk}$ in eq.~\eqref{subeq:qm_baryon} ensures the
anti-symmetry in the color part of the baryonic wavefunction.

The classical analogs of these states in our model are
\begin{subequations}
  \label{eq:cdm_states}
  \begin{eqnarray}
    \label{subeq:cdm_meson}
    |q\bar{q}\rangle &=& |M_\text{cl}\rangle =
    |(r\psi)\,(\bar{r}\bar{\psi})\rangle \\ 
    \label{subeq:cdm_baryon}
    |qqq\rangle &=& |B_\text{cl}\rangle =  |(r \psi_1)\,(g \psi_2)\,(b
    \psi_3)\rangle 
  \end{eqnarray}
\end{subequations}
where $r, (\bar{r}), g, b$ are the (anti-)colors of the particles. 
The explicit choice of colors indeed is irrelevant due to the global
color symmetry discussed before. The particles are supposed to be
located at different positions, i.e.\/ $|M(M_\text{cl})\rangle$ and
$|B(B_\text{cl})\rangle$ describe extended objects. 

The charge density $\rho^a(x)$ is the expectation value of
$\widehat{\rho}^a(x) = \sum_n \widehat{\rho}_n(x) t^a_n$ in a given
hadronic state, where $\widehat{\rho}_n(x) =
\delta(\vec{x}-\vec{x}_n)$ is the one-particle density operator.
The sum runs over all quarks and anti-quarks and the index $n$
indicates that the operator acts on the $n$th particle in the state.

It turns out that the charge densities of the quantum-mechanical states
\eqref{eq:qmstates} vanish everywhere, whereas they are finite for
the classical analogs. Note that the total color charge vanishes in
both cases.
However, the charge density is not an observable quantity due
to the global color symmetry and one better should look at the energy
density. If one rewrites the electric part of the  energy
\eqref{eq:energy_el} in terms of the charge density one gets: 
\begin{equation}
  \label{eq:el_density_rho}
  E_\text{el} = \tfrac{1}{2} \int d^3x d^3y
  \frac{\rho^a(x)\rho^a(y)}{|x-y|}.
\end{equation}
Here we have used the perturbative expression ($\kappa = 1$) and
$\phi^a(\vec{x}) = \int d^3y \rho^a(\vec{y})/|\vec{x} - \vec{y}|$. The
modification of the electric energy due to the scalar function
$\kappa(\sigma)$ in the non-perturbative case does not change our
argument. 
Thus the interesting part of the electric energy is given by
the expectation value of the squared charge operator $\widehat{\rho}^a(x)
\widehat{\rho}^a(y)$. In the Abelian approximation ($a \in \{3,8\}$) it
turns out that
\begin{subequations}
  \label{eq:hadron-energy}
  \begin{eqnarray}
    \label{subeq:mes_energy}
    \langle M | \widehat{\rho}^a(x) \widehat{\rho}^a(y) | M\rangle 
    &=& \hphantom{-} C_F^\text{ab} \left[\langle \psi |
      \widehat{\rho}(x)\widehat{\rho}(y) | 
      \psi\rangle \right. \nonumber\\
    && \hphantom{-C_F^\text{ab} \left[\right.}+\left. \langle \bar{\psi} |
      \widehat{\rho}(x)\widehat{\rho}(y) | \bar{\psi}\rangle\right] 
    \nonumber\\
    && - C_F^\text{ab} \left[\langle\psi | \widehat{\rho}(x) | \psi\rangle
      \langle\bar{\psi} | \widehat{\rho}(y) | \bar{\psi} \rangle\right. \nonumber\\
    && \hphantom {- C_F^\text{ab} \left[\right.}\left.+ \langle\psi |
      \widehat{\rho}(y) | \psi\rangle 
      \langle\bar{\psi} | \widehat{\rho}(x) | \bar{\psi} \rangle
    \right] \nonumber\\
    &=& \langle M_\text{cl} | \widehat{\rho}^a(x) \widehat{\rho}^a(y) |
    M_\text{cl}\rangle \\[1ex]
    \label{subeq:bar_energy}
    \langle B | \widehat{\rho}^a(x)\widehat{\rho}^a(y) | B\rangle 
    &=& \hphantom{- \frac{1}{2}} C_F^\text{ab} \sum_{n=1}^3 \langle \psi_n |
    \widehat{\rho}(x)\widehat{\rho}(y) | \psi_n\rangle \nonumber\\
    && - \frac{1}{2} C_F^\text{ab} 
    \sum_{\genfrac{}{}{0pt}{}{\{m,n\} = 1}{m\neq n}}^3   
    \langle \psi_m | \widehat{\rho}(x) | \psi_m\rangle
    \langle\psi_n | \widehat{\rho}(y) | \psi_n \rangle
    \nonumber \\
        &=& \langle B_\text{cl} | \widehat{\rho}^a(x)\widehat{\rho}^a(y) |
    B_\text{cl}\rangle \quad . 
  \end{eqnarray}
\end{subequations}
Here $C_F^\text{ab} = 1/3$ is the eigenvalue of the quadratic Casimir
operator in the fundamental (3-dimensional) representation \emph{in the
Abelian approximation} defined by $\sum_a t^at^a = C_F^\text{ab}
\mathbf{1}_3$, $a\in\{3,8\}$. 
The first term in both equations denotes the self energy of the
particles and the second one the two particle interaction. In the baryon
case, the interaction is accompanied by an additional color factor
1/2 due to the interaction between two quarks whereas in the meson
case it is a quark-antiquark interaction.
Note that the equalities in eqs.~\eqref{eq:hadron-energy} between the
quantum-mechanical and the classical expressions are
only valid in the Abelian approximation where $a\in\{3,8\}$. The
interaction in the states $|M\rangle$ and $|B\rangle$ is accompanied
by a color factor $- \tr\,t^at^a/N$, where $N$ is the normalization
constant appearing in eq.~\eqref{eq:qmstates}.
In the classical analogs the same factor amounts to $- \langle
c|t^a|c\rangle \langle c'|t^a|c'\rangle$. By explicit calculations
one sees, that these expressions are the same when summing over
$a\in\{3,8\}$ but differ by a factor of four when summing over
$a\in\{1\ldots 8\}$.
We conclude that the classical treatment of $q\bar{q}$ and $qqq$
states is reasonable in the Abelian approximation as
it does not influence the observable energy density.

\section{Analysis of the model}
\label{sec:analysis}

The confining properties of the model are ruled by the interaction
of the color fields with the dielectric medium
\cite{lee:1981,Traxler:1998bk}. In the absence of any 
colored quarks, all color fields vanish and the confinement field will
take on its vacuum expectation value $\sigma_\text{vac}$. If quarks
are added to the system color electric fields are created due to
eq.~\eqref{subeq:gauss} and one can distinguish two different
situations: those with non-vanishing and those with vanishing total
color charge. The prototypes of these two configurations are an isolated
quark and a $q\bar{q}$ configuration, respectively. In both systems a
bag with $\kappa\approx 1$ in its interior develops which is
stabilized by the vacuum pressure $B$. 
In the former case there exists only a monopole term and Gauss's law can
be solved for the electric displacement $\vec{D}^a$ exactly. Due to
the radial symmetry the fields are perpendicular to the bag surface
and $\vec{D}^a$ is a smooth function of the radial distance $r$. Using
radial symmetry the electric energy of the quark can be calculated
\begin{equation}
  \label{eq:quark-energy}
  \vec{D}^a = \frac{g_s q^a}{4\pi\,r^2}\,\vec{e}_r,
  \quad E_\text{el}^q 
  = \frac{1}{2} \int d^3r \,
  \frac{\vec{D}^a\cdot\vec{D}^a}{\kappa(r)}
  = \frac{1}{8\pi} \int dr \, \frac{g_s^2 C_F^\text{ab}}{r^2\kappa(r)} \quad . 
\end{equation}
In our numerical realization the diverging self energy 
is regulated due to the
finite quark width $w(\vec{x})$. But the energy diverges also in the
long range limit as soon as $\kappa(r)$ vanishes more rapidly than
$r^{-1}$ \cite{Goldflam:1982tg}. In this case the vacuum pressure
cannot balance the energy and the bag radius diverges as well.  
In the $q\bar{q}$ case the field lines start and end at the
quark and the anti-quark. Thus they can arrange to be completely
parallel to the bag surface and the electric field $\vec{E}^a$ is a
smooth function in space. The electric energy can be expressed as 
\begin{equation}
  \label{eq:qqbar-energy}
  E_\text{el}^{q\bar{q}} = \frac{1}{2} \int d^3r \,
  \kappa(\vec{r}) \, \vec{E}^a\cdot\vec{E}^a \quad .
\end{equation}
As $\kappa(\vec{r})$ vanishes rapidly outside the bag the energy
density is localized within the bag and the electric energy stays
finite. This geometry already can only be solved 
numerically. Solutions for the transverse profile of $q\bar{q}$ fields
for large quark separations $R$ with axial symmetry are
given in \cite{Bickeboeller:1985xa}. To illustrate the second scenario
we will first present a qualitative picture based on a simple bag-like
model, where the bag is a cylindrical tube with axial radius $\rho$
and with a sharp boundary, i.e.\/ $\kappa = 1$ inside and
$\kappa=\kappa_\text{vac}=0$ outside of the bag, respectively. 
Afterwards the quantitative analysis will be based on the CDM where
$\kappa$ becomes a smooth function of $\vec{r}$. In the bag model the 
electric displacement $\vec{D}^a$  vanishes exactly outside of the tube.
Further we will assume that the electric field is
homogeneous and constant inside the bag. The strength of the electric
field inside is given by eq.~\eqref{subeq:gauss} as $D^a = E^a = g_s q^a/(\pi
\rho^2)$ and the total energy in a central slice of the tube with
thickness $\Delta\ell$ is $E_\text{tot} =
E_\text{el}+E_\text{vol}$, with  $E_\text{el} = \half g_s^2
C_F^\text{ab}/(\pi \rho^2)\, \Delta\ell$ and $E_\text{vol} = B\pi \rho^2
\Delta\ell$. Minimizing the energy with respect to the radius $\rho$ yields  
\begin{subequations}
  \label{eq:bag_min}
  \begin{eqnarray}
    \label{subeq:R_min}
    \rho_0^4 &=& g_s^2 C_F^\text{ab}/(2\pi^2 B) \quad ,\\
    \label{subeq:W_min}
    E_0 &=& E_\text{el}^0 + E_\text{vol}^0 = \sqrt{2 g_s^2
      C_F^\text{ab} B} \, \Delta\ell =: \tau \Delta\ell \quad , \\
    \label{subeq:balance}
    E_\text{el}^0 &=& E_\text{vol}^0 = \sqrt{g_s^2
      C_F^\text{ab} B/2} \, \Delta\ell
  \end{eqnarray}
\end{subequations}
for the minimizing radius $\rho_0$ and the
corresponding energy $E_0$. Here we have introduced the
concept of the string tension $\tau$. 
The identification of the string
tension with the integrated energy density in the central slice of the
string is only valid for strings with constant width, i.e.\/ for
infinite large quark separations $R$. As we will see, the radius
increases for finite values of $R$. Therefore in the central slice the
volume energy increases while the electric energy decreases. A more
suited definition for the string tension in this case is $\tau =
\frac{dE_\text{tot}}{dR}$, where $E_\text{tot}$ is the total energy of
the string including the end caps.
At the stable point $\rho_0$ the electric and the
volume energy balance each other exactly
(cf. eq.~\eqref{subeq:balance}. For increasing values of $B$ the 
dielectric vacuum compresses the 
electric flux into thinner flux tubes while for increasing electric
flux, i.e.\/ for increasing $g_s$, the string radius grows
(cf. eq.~\eqref{subeq:R_min}). 
Note that the string tension scales linearly
with the coupling $g_s$ which is typical for all bag models
\cite{Johnson:1976sg,Lucini:2001nv,Hansson:1986um}. 

With a typical string tension $\tau = 980\,$MeV/fm taken from meson
spectroscopy \cite{Eichten:1975af,Quigg:1979vr,Eichten:1980ms} and a
tranverse radius of $\rho_0 = 0.35\,$fm taken from lattice
calculations \cite{Bali:1995de} we get a bag constant $B^{1/4} \approx
315\,$ MeV and a strong coupling constant $C_F^\text{ab}\alpha_s =
C_F^\text{ab}\,g_s^2/(4\pi) \approx 0.15$. Though the value of the bag
constant varies over a wide range in the 
literature from $B^{1/4} = 145\,$MeV in the MIT bag model
\cite{Chodos:1974je} to $B^{1/4} = 241\,$MeV taken from QCD sum rule
analysis \cite{Shifman:1979bx,Shifman:1979by}, the value found here is
rather high. On the other hand,  
the value of the coupling constant $\alpha_s$ is rather small 
as compared to that
obtained on the lattice $C_F\alpha_s = 0.3$ \cite{Bali:2000gf} or in
meson spectroscopy ranging from $C_F\alpha_s = 0.3-0.5$
\cite{Eichten:1977jk,Quigg:1979vr}. 
Note however, that we have presented a qualitative discussion here
based on a simple bag model with fixed and sharp boundary.
We will study in section \ref{sec:parameter} how the string tension
$\tau$ and the string radius $\rho$ depend on the model parameters
when the confinement field $\sigma$ and the electric fields are
calculated according to the equations of motions
\eqref{eq:static_eoms}.  
For a long flux tube, when the
quarks are located at the $x$-axis at $x= \pm x_0$ and separated by a
large distance, one may assume axial symmetry for the geometry of the string. 
In the central plane between the quarks at $x = 0$ the fields can be
described by $\sigma(\vec{r}) = \sigma(\rho)$ and $\phi^a = c_a
x$. The constant $c_a$ is determined by Gauss's law. In this case
the dielectric displacement has the simple form
\begin{equation}
  \label{eq:abrikosov}
  \vec{D}^a(\rho) = - c_a \kappa(\rho) \vec{e}_x.
\end{equation}
The electric field points along the flux tube axis and its profile is
proportional to the profile of $\kappa(\rho)=\kappa(\sigma(\rho))$. We
shall see in sec.~\ref{sec:strings} if and for which quark separations
$R$ this asymptotic behavior is reached.
The simple picture of sharp boundaries discussed above is reproduced if the
dielectric function $\kappa(\rho)$ is a step function. In this limit the
penetration depth of the electric field $\vec{D}^a$ into the
non-perturbative vacuum is zero. For any smooth profile of the
dielectric function the penetration depth stays finite and non-zero but the
electric fields are still screened.

\subsection{Generic features of CDM}
\label{sec:generic}

We proceed in studying in detail both the geometrical structure and the
energetic content of a $q\bar{q}$ flux tube of finite length. Our
major interest is concentrated on the transverse profile of the energy
density of this object as well as on the scaling of the total energy
with the $q\bar{q}$ separation. These quantities will be compared
later on to lattice results and experimental data. 
In this section we use a parameter set which we will call later PS-I
and which is given in tab.~\ref{tab:params} below. The corresponding
potential $U(\sigma)$ is shown in fig.~\ref{fig:U_ps123} (solid line).


It should be noted that the numerical realization is not 
restricted to the $q\bar{q}$ geometry. 
The $q\bar{q}$ configuration possesses an axial symmetry, and the
equations of motion \eqref{eq:static_eoms} can be reduced to two
dimensions. In the limit of an infinite long $q\bar{q}$ string, the
problem even can be reduced to one radial dimension.
But already the three-quark system does not have this symmetry
and therefore we need a three-dimensional algorithm, as presented
in appendix \ref{sec:fas}. 

\begin{figure}[htbp]
  \centering
  \includegraphics[width=\defaultfig,keepaspectratio,clip]{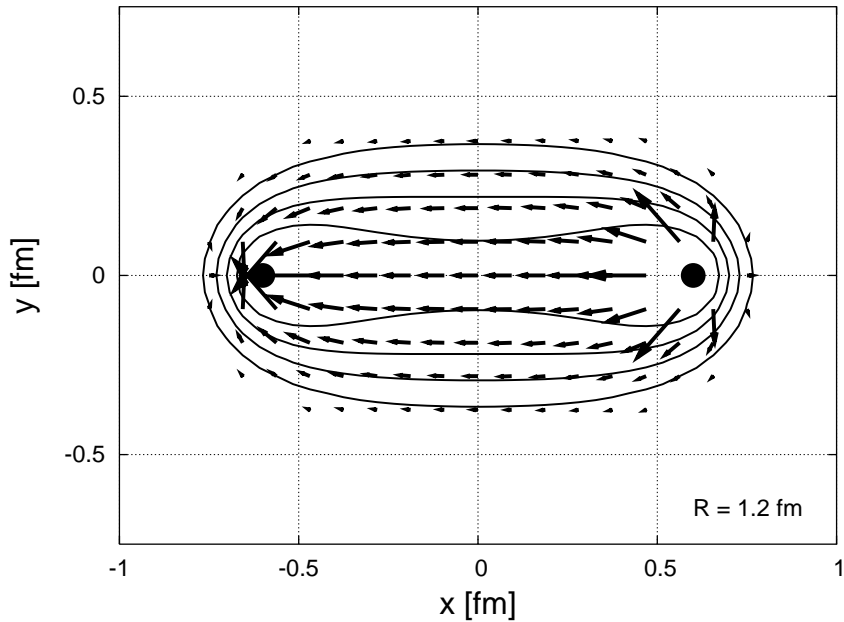}
  \caption{The arrows represent the electric displacement $\vec{D}^3$
  of a $r\bar{r}$ flux 
  tube. The dots at $x=\pm0.6\!$ fm represent the (anti-) quark
  position. The solid lines are the equidistant contour lines of the 
  dielectric function at $\kappa = (0.2,\ldots\/,0.8)$ starting from
  outside. The string has a radius of $\rho \approx 0.3\!$ fm. The  electric
  8-field of this configuration is $\vec{D}^8 =  (q^8/q^3) \vec{D}^3$.}
  \label{fig:flux-tube}
\end{figure}

To start with, we show the electric displacement $\vec{D}^3$ of such a
flux tube in fig.~\ref{fig:flux-tube}. The $q\bar{q}$ pair is located
on the $x$-axis at $x= \pm 0.6\,$fm. It is nicely seen that the field
lines emerge from the  
quark as a source for the 3-component of the field and end on the
anti-quark as a sink. Between the particles the field is nearly
homogeneous and drops off at an axial distance of about 0.3 fm. At
the positions of the particles one sees the expected Coulomb-like
field for point-like particle. In
the same figure we show also the contour lines of the dielectric
function $\kappa(\sigma)$ for $\kappa = \{0.2,\ldots\/,0.8\}$ starting
from the outside of the bag. Outside the bag where $\kappa < 0.2$ the
electric displacement nearly vanishes. 

\begin{figure}[htbp]
  \centering
  \includegraphics[width=\defaultfig,keepaspectratio,clip]{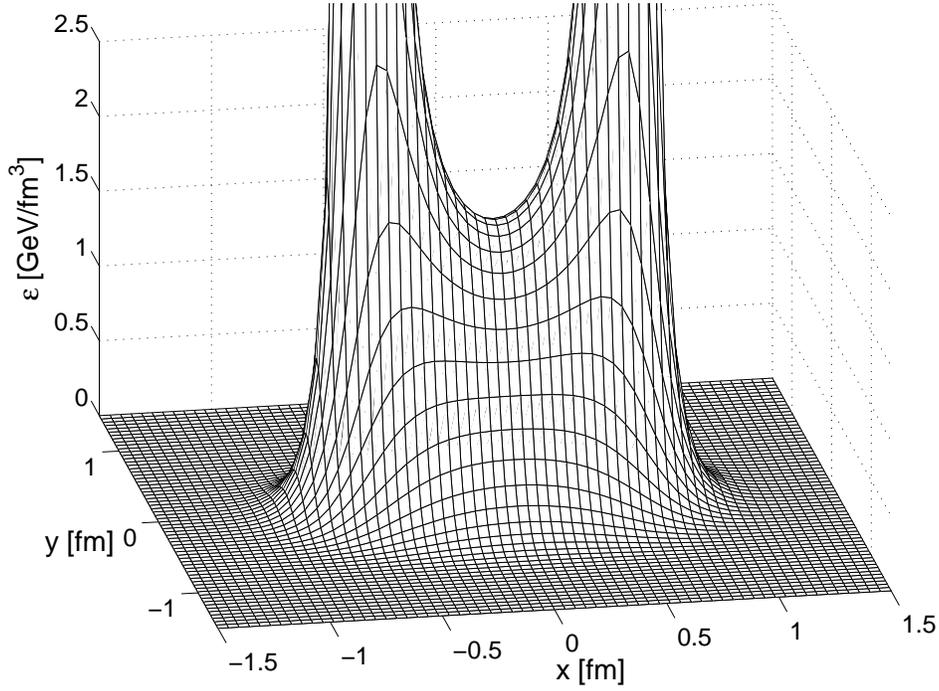}
  \caption{Energy density (integrand of eq.~\eqref{eq:energy_decomp}) of the
  same $q\bar{q}$ string as in fig.~\ref{fig:flux-tube}. A well
  defined flux tube is stretched between the particles. The scale is
  chosen to show the flux tube and therefore the strong Coulomb peaks
  at the quark positions are not seen.}
  \label{fig:energy_density}
\end{figure}

It should be noted that the
8-component of this very special configuration (namely a $r\bar{r}$
pair) has exactly the same geometrical shape but is reduced by
$q_r^8/q_r^3 = 1/\sqrt{3}$. 
Of course the electric fields for a different
color/anti-color (e.g. $b\bar{b}$) pair are different as the blue
quark has no 3-component of the charge (see
fig.~\ref{fig:triplett}). The two color configurations are connected
by a color rotation $V$ defined in eqs.~\eqref{eq:discrete_color}
and\eqref{eq:eq:colorrotation2} which leaves the energy (or action)
density invariant. The energy density is shown in
fig.~\ref{fig:energy_density}. At the quark positions the Coulomb
peaks develop, but we do not show them here to emphasize the structure
of the flux tube between the particles. The energy is distributed
over a well defined region outside of which the energy vanishes.

\begin{figure}[htbp]
  \centering
  \includegraphics[width=\narrowfig,keepaspectratio,clip]{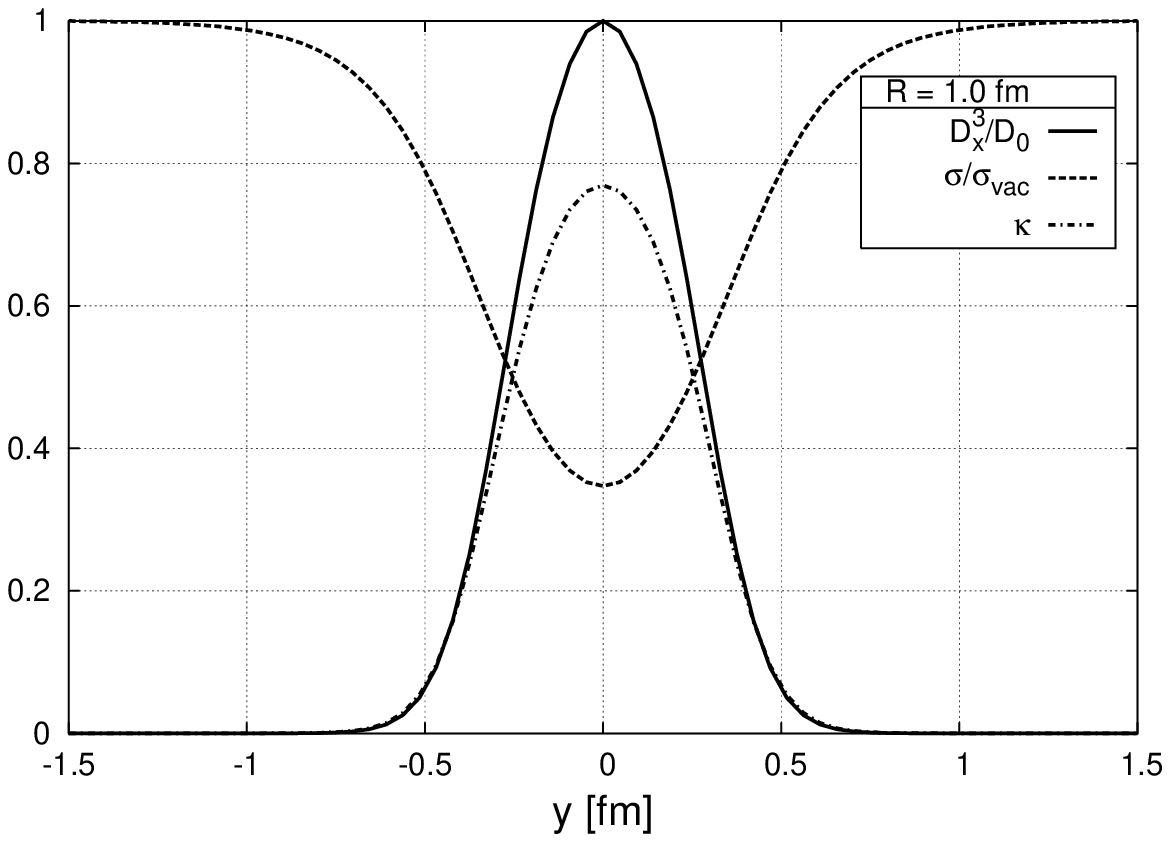}
  \includegraphics[width=\narrowfig,keepaspectratio,clip]{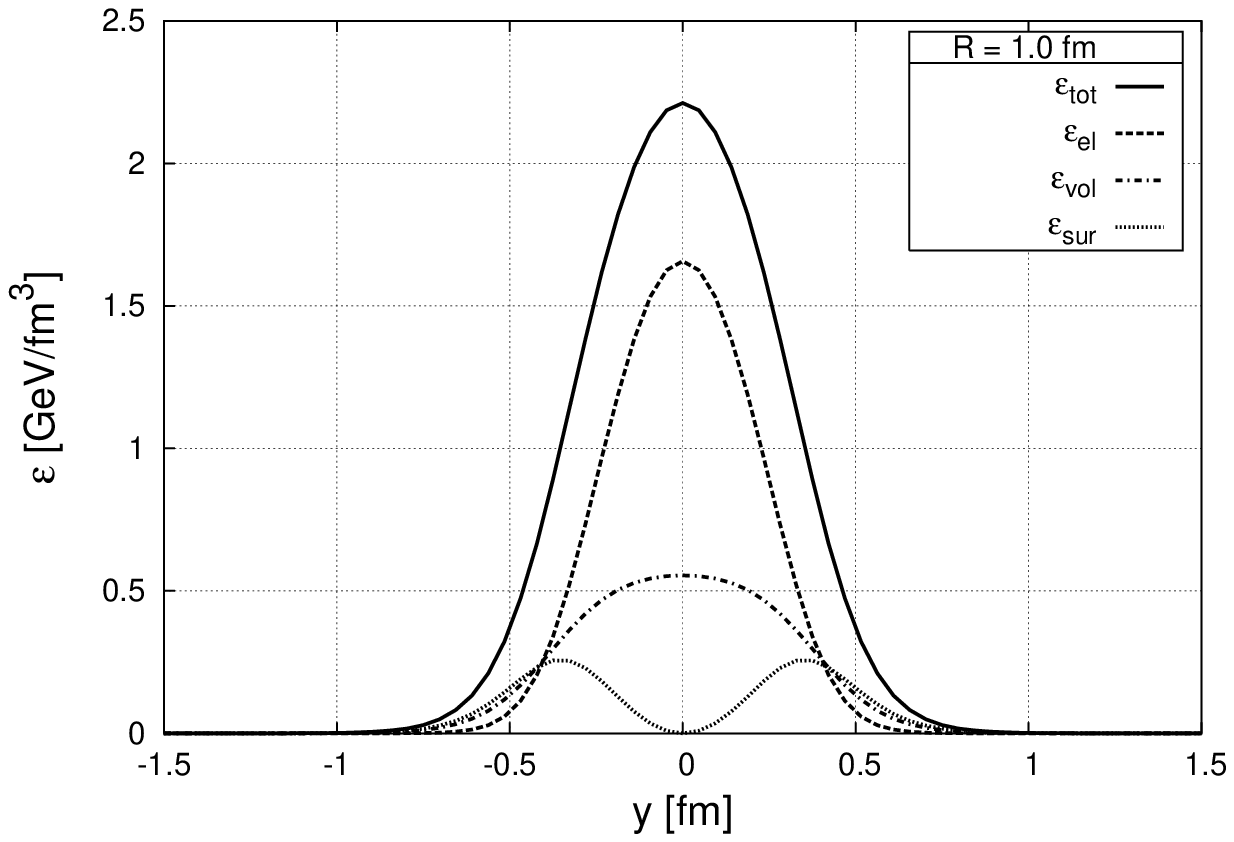}
  \caption{The profile of the fields (left) and the corresponding
  energies (right) of a 1 fm string. The electric field $D^3$ is
  scaled to its central value on the axis and the scalar field
  $\sigma$ to its vacuum value.} 
  \label{fig:flux_profile}
\end{figure}

The profiles of the underlying fields $\vec{D}^3$, the confinement
field $\sigma$ and the dielectric function $\kappa$ at $x = 0$ of a 1
fm long $q\bar{q}$ string are shown in fig.~\ref{fig:flux_profile}. On the
string axis $\kappa$ and $\sigma$ 
reach their maximal and minimal value, respectively, and both functions
approach to their vacuum  values $\kappa_\text{vac}$ and $\sigma_\text{vac}$,
respectively, for $\rho = \sqrt{y^2+z^2} > 0.5\,$fm. 
The confinement field does not drop
down to $\sigma = 0$ and therefore the fields are not completely in
the perturbative phase. 
The region where $\kappa$ deviates substantially from 
its vacuum expectation value ($\rho \le 0.3\,$fm) defines the bag and
consequently the electric flux falls off very fast at the bag
boundary.  

In the same figure on the right we show the corresponding
profile of the energy density. Here we have decomposed the
total energy \eqref{eq:totenergy} into an electric, a volume and a
surface term according to eq.~\eqref{eq:energy_decomp}.
The surface energy (dash-dotted line) is maximal on the bag boundary
at $\rho \approx 0.3\,$fm, justifying the expression \emph{surface}
energy. The volume energy (dotted line) is distributed more
homogeneously within the bag whereas the electric energy (dashed line)
falls off quite rapidly.

\begin{figure}[htbp]
  \begin{center}
    \includegraphics[width=\narrowfig, keepaspectratio,
    clip]{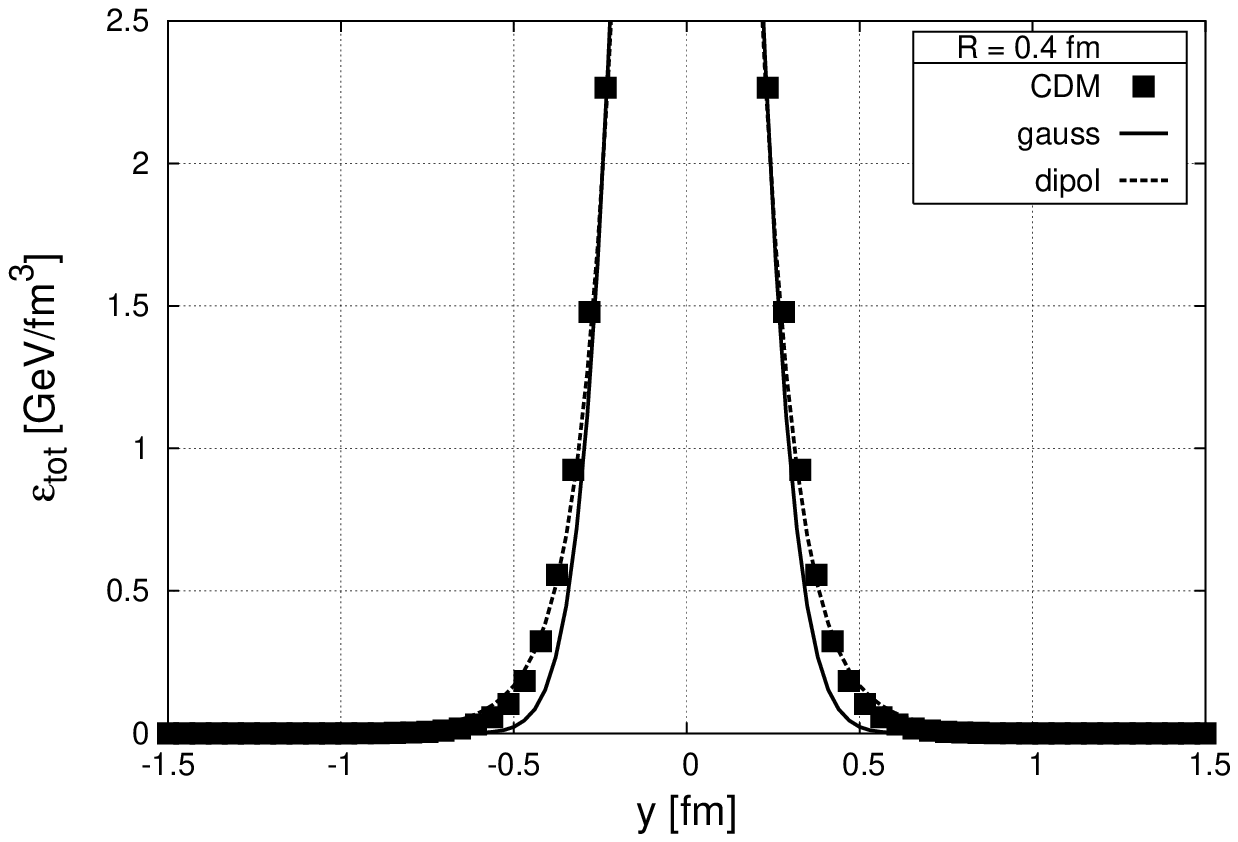} 
    \includegraphics[width=\narrowfig, keepaspectratio,
    clip]{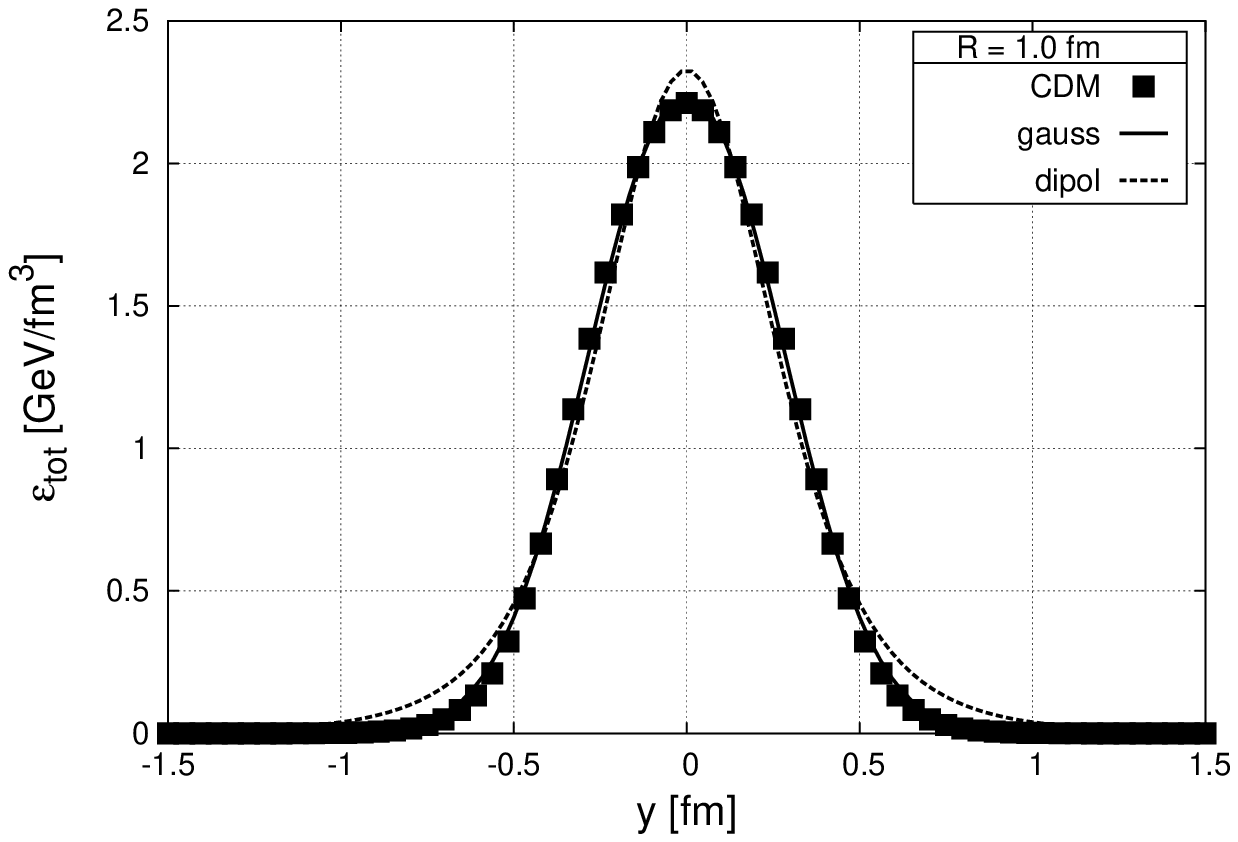} 
    \caption{The profile of the energy density together with the best
    fits of the parameterizations \eqref{eq:profile_fit} fitted over
    the whole range shown. The profile
    is better described by a dipole form for small $q\bar{q}$
    separations $R$ (left panel) and by a Gaussian form for large $R$
    (right panel). In the left fig. we have omitted the dipole peak
    at the axis to better show the characteristic dipole tail.}
    \label{fig:profile}
  \end{center}
\end{figure}

The energy density in the central plane between the charges at $x = 0$
is of special interest as it has been analyzed on the lattice in SU(2)
\cite{Bali:1995de} 
and  also in the framework of the dual color
superconductor \cite{Maedan:1990ju,Baker:1991bc} and in the Gaussian Stochastic
Model \cite{Kuzmenko:2000bq,Kuzmenko:2000rt,Shoshi:2002rd}. Because of
the symmetry all 
quantities depend only on 
the distance $\rho$ from the string axis and we may reduce our
analysis of the geometry of 
the flux tube to the shape of this profile function. We follow the
reasoning of \cite{Bali:1995de} and compare the energy profile to both a
dipole and Gaussian-like parameterization
\begin{subequations}
  \label{eq:profile_fit}
  \begin{eqnarray}
    \label{subeq:dipole_fit}
    f_d(\rho) &=& N_d (\rho^2 + \rho_d^2)^{-3} \\
    \label{subeq:gauss_fit}
    f_g(\rho) &=& N_g \exp[-\ln 2\,(\rho/\rho_g)^n] \quad ,
  \end{eqnarray}  
\end{subequations}
where $n$ in (\ref{subeq:gauss_fit}) is a parameter giving the steepness of
the profile. For $n=2$ $f_g(\rho)$ is a Gaussian.
For small quark separations $R$ perturbative QCD predicts Coulomb-like
fields with a characteristic dipole behavior~\eqref{subeq:dipole_fit}
whereas at large quark separations the field should fall off much
faster and the profile should be described better by a generalized Gaussian
\eqref{subeq:gauss_fit} with a width at half maximum $\rho_g$. 
We show the profile of the energy density with the parameter set PS-I
in fig.~\ref{fig:profile} together with
the best fit of both parameterizations. One can see
that the CDM results can be nicely described with a dipole shape for
small separations ($R=0.4\,$fm) and with a Gaussian shape ($n=2$) for large
separations ($R=1.0\,$fm). It should be noted that the Gaussian shape
is only a qualitative guess for the profile. The profile might fall
off even faster than described by a Gaussian. To quantify this we will
extract below the
parameter $n$ from the fit of (\ref{subeq:gauss_fit}) to the profile. A value
of $n=2$ indicates a pure Gaussian shape and a value $n>2$ is connected to a
sharper bounded bag.

In addition to
the shape of the flux tube, we will study below the dependence of the total
energy (see eq.~\eqref{eq:energy_decomp}) as a function of the quark
separation $R$, i.e.~the $q\bar{q}$ potential $V_{q\bar{q}}(R)$. We
compare the calculated CDM $q\bar{q}$ potential to the Cornell
potential 
\begin{equation}
  \label{eq:cornell}
  V_c(R) = 2 C_F^\text{ab} E_0 - C_F^\text{ab} \frac{\alpha}{R} + \tau R.
\end{equation}
This potential has been observed on the lattice in SU(2)
\cite{Bali:1995de} and SU(3) \cite{Bali:1992ab} and
has been used successfully in meson spectroscopy for heavy quarkonia
\cite{Eichten:1975af,Quigg:1979vr,Eichten:1980ms}. It contains the
three parameters $E_0$, $\alpha$ and $\tau$ which will be
determined by a fit to the CDM results. For short distances
$R$ one might expect that perturbative 
one-gluon exchange results of QCD are dominant and thus $V_c(R)$ shows
the characteristic Coulomb-like $1/R$ potential. To take into account
non-perturbative effects $\alpha$ is an effective coupling constant. 
It has been shown \cite{Luscher:1980ac,Arvis:1983fp} that there is a
$1/R$ correction in the $q\bar{q}$-potential due to
quantum-mechanical vibrations of strings around the classical
solution. It is beyond the scope of the present paper to predict the
corrections to the $R$-dependence within our model. A possible
correction of the Coulomb-type is included in the effective coupling
constant $\alpha$.
For large $R$ the
linear term in eq.~\eqref{eq:cornell} takes over. It is a consequence
of the formation of long linear flux tubes. The string tension $\tau$
expresses the strength of confinement. The constant term is due to the
self energy of the quarks. For point-like particles this should
diverge but in our numerical realization we have regulated it by the
non-zero quark width $r_0$. The potential will vanish as $R \rightarrow 0$,
because the equal and opposite charge  
distributions of the quark and the anti-quark start to overlap and to
cancel each other. Therefore a Cornell fit is only meaningful for
quark separations $R$ substantially larger than $r_0$.
For convenience we have separated from the Coulomb and the constant
term the factor $C_F^\text{ab} = 1/3$. The agreement between our
calculations and the Cornell parameterization in eq.~\eqref{eq:cornell}
will be demonstrated in the next section.

\subsection{Variation of the CDM parameters}
\label{sec:parameter}

In the remainder of this section we will discuss the influence of the
parameters $B$, $m_g$, $\sigma_\text{vac}$ and $g_s$ on both the energy
profile and the total energy. For this purpose we vary all
parameters one by one in a certain range while keeping the others at
a fixed value given in tab.~\ref{tab:vary-params}. We calculate the
profile for a $R=1.0\,$fm long string as well as the potential
$V_{q\bar{q}}(R)$ for varying $R$. 
From the generalized Gaussian fit to the energy profile we
extract the width $\rho_g$ and the steepness parameter $n$. 
From the Cornell fit to the CDM
$q\bar{q}$ potential we extract the string tension $\tau$ and the
effective coupling constant $\alpha$. The results are collected in
tab.~\ref{tab:vary-params}. 

We also study, if the equality between the electric part and the volume
part of the string tension is still valid in our
model. Note that this equality holds in the simple flux tube
model expressed in eq.~\eqref{subeq:balance}.
For that purpose we also perform a  Cornell fit to the
different energy fractions $E_\text{el}$, $E_\text{vol}$ and
$E_\text{sur}$ separately. From these fits we extract the different
parts of the string tension $\tau_\text{el}$, $\tau_\text{vol}$ and
$\tau_\text{sur}$. We present the ratios $\beta =
\tau_\text{el}/\tau_\text{vol}$ and $\gamma =
\tau_\text{sur}/\tau_\text{tot}$ also in
tab.~\ref{tab:vary-params}. We find for all parameter combinations a 
remarkable equality between $\tau_\text{el}$ and $\tau_\text{vol}$
and the ratio $\beta$ is 1 within a few percent. The surface part of
the string tension amounts to roughly $\gamma = 0.2$ except in the case when
$B\rightarrow 0$ and all other model parameters kept fixed (see first panel in
tab.~\ref{tab:vary-params}). 
In all simulations shown in this paper we have used $\kappa_\text{vac}
= 10^{-4}$. All string quantities approach its limiting values very
rapidly and do not change any more for value $\kappa_\text{vac}<
10^{-3}$ (see also appendix \ref{sec:kappa-dep}).

\begin{table}[htbp]
  \centering
  \begin{ruledtabular}
    \begin{tabular}{cccccc}

      $B^{1/4}\,$[MeV]&
      $\tau\,\left[\frac{\text{MeV}}{\text{fm}}\right]$ &  $\beta$ & 
       $\gamma$ & $\rho_g$ [fm] & $n$ \\\hline
      0            &   632      & 1.07 & 0.28 &  0.44   & 2.8 \\
      60           &   633      & 1.07 & 0.28 &  0.44   & 2.8 \\
      120          &   642      & 1.07 & 0.27 &  0.43   & 2.7\\
      180          &   675      & 1.06 & 0.24 &  0.40   & 2.5\\
      240          &   755      & 1.06 & 0.19 &  0.35   & 2.3
      \\\hline\hline      
      
      $m_g\,$[MeV] & $\tau\,\left[\frac{\text{MeV}}{\text{fm}}\right]$  
      &  $\beta$   & $\gamma$   & $\rho_g$ [fm] & $n$\\\hline
      1000         &   755      &  1.06  & 0.19 &  0.35   & 2.3\\
      1200         &   831      &  1.03  & 0.19 &  0.36   & 2.6\\
      1400         &   905      &  1.00  & 0.20 &  0.36   & 3.0\\
      1600         &   982      &  1.00  & 0.21 &  0.36   & 3.4\\
      1800         &   1053     &  1.00  & 0.22 &  0.36   & 4.0\\\hline\hline
      
      $\sigma_\text{vac}\,$[fm$^{-1}$] &   
      $\tau\,\left[\frac{\text{MeV}}{\text{fm}}\right]$ 
      &  $\beta$   & $\gamma$   & $\rho_g$ [fm] & $n$\\\hline
      1.01         &     755    &  1.06  & 0.19 &  0.35   & 2.3 \\
      1.26         &     925    &  1.06  & 0.21 &  0.35   & 2.4 \\
      1.51         &    1110    &  1.06  & 0.23 &  0.34   & 2.4 \\
      1.75         &    1306    &  1.06  & 0.23 &  0.33   & 2.3 \\
      2.00         &    1511    &  1.06  & 0.24 &  0.32   & 2.3 \\\hline\hline

      $g_s$ & $\tau\,\left[\frac{\text{MeV}}{\text{fm}}\right]$ 
      &  $\beta$   & $\gamma$   &  $\rho_g$ [fm] & $n$\\\hline
      0.2          &     100    &  0.94  & 0.20 &  0.21   & 1.9 \\ 
      0.5          &     223    &  0.97  & 0.20 &  0.24   & 2.0 \\
      1.0          &     418    &  1.06  & 0.19 &  0.29   & 2.2 \\
      1.5          &     591    &  1.06  & 0.19 &  0.32   & 2.3 \\
      2.0          &     755    &  1.06  & 0.19 &  0.35   & 2.3 \\
      2.5          &     911    &  1.05  & 0.18 &  0.37   & 2.4 \\
      3.0          &    1062    &  1.05  & 0.18 &  0.39   & 2.4 \\
      3.5          &    1212    &  1.04  & 0.18 &  0.41   & 2.3 \\
      4.0          &    1356    &  1.04  & 0.18 &  0.42   & 2.3 \\
      5.0          &    1637    &  1.04  & 0.17 &  0.43   & 2.2 
    \end{tabular}
    \caption{The string tension $\tau$, the string width $\rho_g$
      and the steepness parameter $n$ for varying model parameters.
      The ratios $\beta$ and $\gamma$ are defined in the text.
      We vary the parameters one by one and keep the other fixed at
      $B^{1/4}=240\,$MeV, $m_g = 1000\,$MeV,
      $\sigma_\text{vac}=1.01\,\text{fm}^{-1}$ and $g_s=2$,
      respectively. Throughout this work we use
      $\kappa_\text{vac}=10^{-4}$.  
    } 
    \label{tab:vary-params}
  \end{ruledtabular}
\end{table}

\paragraph{Bag constant $B$:}
\label{sec:B-variation}

We start with a variation of the scalar potential $U$ (see
eq.~\eqref{eq:uscalar}). 
We vary $B$ from $B=0$ to the maximal value allowed by
eq.~\eqref{eq:inflection} $B^{1/4}=240\,$MeV (first panel in
tab~~\ref{tab:vary-params}). At the lower value
$U(\sigma)$ has two degenerate vacua and at the higher value it has
only an inflection point at $\sigma=0$ (see dashed lines in
fig.~\ref{fig:U_pot}). From the simple bag model,
eq.~\eqref{eq:bag_min}, we would expect the 
string tension to increase quadratically with $B^{1/4}$ and the string
width to decrease with $B^{-1/4}$ for increasing values $B$.  Instead
the string constant $\tau$ grows only weakly over the whole range and
is non-zero at $B=0$. Note that for $B=0$ the profile of the volume
energy $U(\rho)$ is non-zero everywhere and a string of finite width
is formed.
The string radius shows the expected tendency
though the decrease starts only for $B^{1/4} \ge 120\,$MeV. The profile is
steeper than a Gaussian for small $B$ ($n=2.8$) and approaches 
a Gaussian for larger values. 

\paragraph{Glueball mass $m_g$:}
\label{sec:mg-variation}
When we fix the bag constant and vary only
the glueball mass $m_g$ (second panel in tab.~\ref{tab:vary-params})
the string tension increases roughly linearly with $m_g$. The string width
stays constant over the whole tested range $m_g = (1000 \ldots
1800)\,$MeV in tab.~\ref{tab:vary-params}. Simultanously the steepness
$n$ of the profile increases 
strongly up to $n=4$. In the limit $m_g\rightarrow \infty$ the local maximum
$U(\sigma)$ rizes infinetely and the confinement field $\sigma$ is restricted
to be either $\sigma=0$ or $\sigma=\sigma_\text{vac}$. In this case we
expect to reproduce the bag with sharp boundaries discussed in
eq.~(\ref{eq:bag_min}). Of course this is not reached in the CDM
but we observe the tendency in the rise of $n$. 
We see that the string tension is more affected
by the mass $m_g$ than by the bag constant $B$ and that vice versa the
string radius is nearly independent on $m_g$ but decreases with
$B$. Note that the glueball mass has no analog in the bag model, where
the shape of the flux tube is assumed to be rectangular.

\paragraph{Vacuum value $\sigma_\text{vac}$:}
\label{sec:sigvac-variation}

Qualitatively the string quantities should depend in the same way on
the vacuum value $\sigma_\text{vac}$ as on $m_g$. If we increase
$\sigma_\text{vac}$ from its minimal allowed value given by
eq.~\eqref{eq:inflection} to higher values a local maximum in the
scalar potential $U(\sigma)$ develops (see dash-dotted curves in
fig.~\ref{fig:U_pot}). In this case the stiffness $m_g$, i.e.\/ the
curvature of the potential at $\sigma_\text{vac}$ is unchanged. 
The increasing surface energy related to the larger gradient in the
confinement field $\sigma$
reduces the string radius $\rho_g$. In the tested range the radius
decreases from $\rho_g=0.35\,$fm to $\rho_g=0.32\,$fm (see third panel
in tab.~\ref{tab:vary-params}). Though the decrease in the string
radius is not very large, the string tension doubles its value from
755 MeV/fm to 1511 MeV/fm. The steepness of the energy profile is
rather independent from $\sigma_\text{vac}$.

\paragraph{Coupling constant $g_s$:}
\label{sec:param_gs}

\begin{figure}[htbp]
  \centering
  \includegraphics[width=\defaultfig,keepaspectratio,clip]{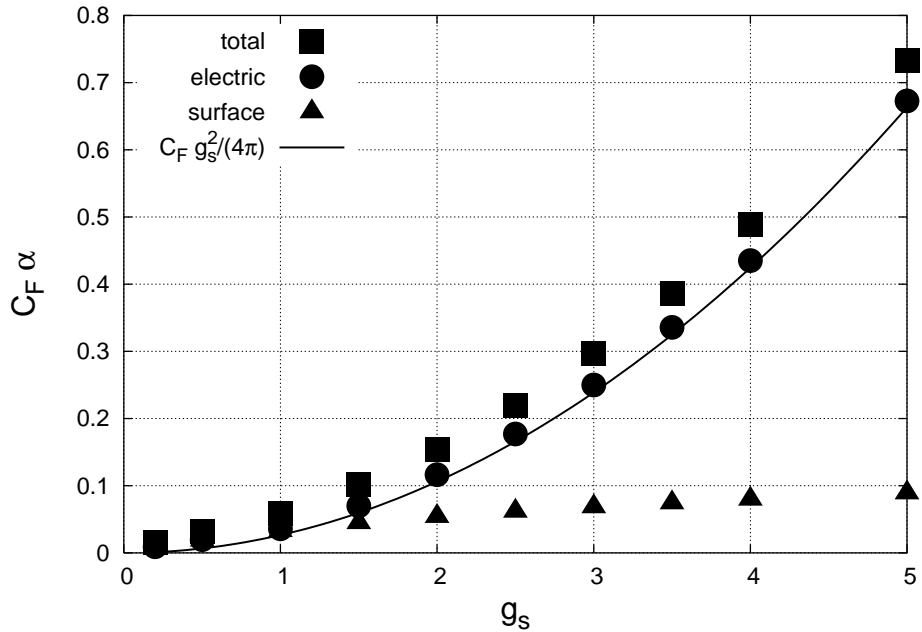}
  \caption{The Coulomb coupling parameter $\alpha$  extracted from the
  Cornell fit for  different $g_s$. The electric part (dots) is in
  perfect agreement with the one-gluon exchange result (line). The
  surface part (triangles) contributes only a small amount to the 
  Coulomb potential and therefore the effective $\alpha$ (squares) rises
  quadratically with $g_s$.} 
\label{fig:alpha}
\end{figure}

The couping $g_s$ determines directly the strength of the electric
flux in the string. From eq.~\eqref{eq:bag_min} we expect the string
tension to rize linearly with $g_s$ and the string radius to increase
proportional to $\sqrt{g_s}$. Indeed we find a nearly perfect linear
behavior of $\tau$ (4th panel in tab.~\ref{tab:vary-params}) and also
the string width has the expected qualitative behavior. The linear
dependence between the string tension and the coupling $g_s\sqrt{C_F}$
is valid in general for all bag-like models
\cite{Lucini:2001nv,Hansson:1986um}. For the coupling $g_s$ we also
look at the $q\bar{q}$ potential at small quark separations $R$,
i.e.\/ we extract the parameter $\alpha$ in the Coulomb term of the
Cornell potential which is dominant for $R<0.2\,$fm. For small $R$ the
bag has almost spherical shape and the $q\bar{q}$ interaction can be
treated perturbatively. In fig.~\ref{fig:alpha} we show the Coulomb
parameter for varying $g_s$. The electric part $\alpha_\text{el}$
(circles) is perfectly described
by the perturbative result $\alpha_\text{el} = \frac{g_s^2}{4\pi}$
(dashed line). The surface contribution $\alpha_\text{sur}$ (triangles) 
does not follow 
the quadratic behavior and seems to grow more linearly and rather
slowly. It thus acts as a non-perturbative correction to the
perturbatively expected Coulomb interaction $\alpha =
\frac{g_s^2}{4\pi} + \alpha_\text{sur}$.  
The volume energy increases linearly with
$R$ and thus the corresponding Coulomb parameter is compatible with
zero and not included in the figure.

\begin{table}[htbp]
  \centering
  \begin{ruledtabular}
    \begin{tabular}{ccccc}
      & $B$ & $m_g$ & $\sigma_\text{vac}$ & $g_s$  \\\hline
      $\tau$ & $\nearrow$ & $\uparrow$ & $\uparrow$ & $\uparrow$
      \\\hline
      $\rho_g$ & $\downarrow$ & --- & $\searrow$ & $\uparrow$ \\\hline
      $n$      & $\downarrow$ & $\uparrow$  & --- & ---
    \end{tabular}
    \caption{The reaction of string tension $\tau$, the string width
      $\rho_g$ and the steepness $n$ on increasing model parameters.}
    \label{tab:reaction}
\end{ruledtabular}
\end{table}

We end this section in summarizing the reaction of the string tension
$\tau$, the string width $\rho_g$ and the steepness of the profile on
the parameter variation in tab.~\ref{tab:reaction}.

\section{$q\bar{q}$ strings}
\label{sec:strings}

\subsection{CDM and lattice results}
\label{sec:cdm-lattice}

After having analyzed the influence of the model parameters on the
shape and the energy of $q\bar{q}$ strings, we compare the CDM
results to lattice calculations. We first describe $q\bar{q}$ strings
to fix the parameters and later on turn to $qqq$ systems in
section~\ref{sec:baryons}. 

In \cite{Bali:1995de} detailed studies of long
flux tubes were made in SU(2) lattice gauge theory. The authors
compared the profile 
of the action and the energy density of strings of varying length. We
will tune the parameters of our model to reproduce the
phenomenological value of the string tension $\tau$ and  to desribe
the energy profile of a 1 fm long $q\bar{q}$ string 
as good as possible within our model. For the profile
a measure for the agreement of our calculations and lattice results is
the quantity $(\Delta\varepsilon)^2 = \sum_i 
[\varepsilon(\rho_i) - \varepsilon_\text{lat}(\rho_i)]^2$, where
$\varepsilon_\text{lat}(\rho_i)$ is the energy density given at
discrete radial distances $\rho_i$ (see fig.24 in
\cite{Bali:1995de}). To find the optimal set of parameters we 
scan a wide range of our parameter space and minimize  
$\Delta\varepsilon$ with the constraint to reproduce the string
tension $\tau=980\,$MeV/fm. 

\begin{figure}[htbp]
  \centering
  \includegraphics[width=\defaultfig,keepaspectratio,clip]{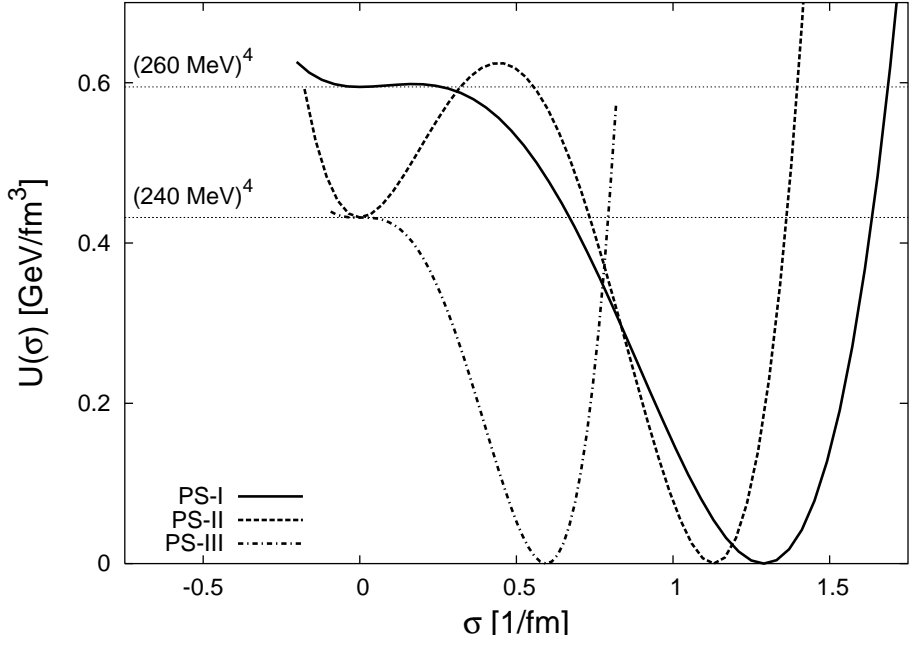}
  \caption{The scalar potential $U(\sigma)$ for the three parameter
  sets given in tab.~\ref{tab:params}.}
  \label{fig:U_ps123}
\end{figure}

Doing this we follow three different strategies.
In the first we treat all model parameters from
tab.~\ref{tab:reaction} as free parameters, determined from the
minimization of $\Delta\varepsilon$ only. In this case we do not lean
on the heuristic interpretation of $B$ as the bag constant and $m_g$
as the glueball mass. This results in
parameter set PS-I in tab.~\ref{tab:params}. In the second we take $B$
and $m_g$ as the bag constant and the glueball mass, respectively. The former is
chosen to be $B^{1/4}=240\,$MeV as found in QCD sum rules
\cite{Shifman:1979bx,Shifman:1979by} and also in another
bag-like analysis \cite{Hasenfratz:1980jv,Hasenfratz:1980ka}.  
The latter is restricted to be at $m_g=(1500-1700)\,$MeV as given by
SU(3)-lattice results for the scalar glueball mass
\cite{Morningstar:1997ff,Michael:1998tr}. This leaves only $g_s$ and 
$\sigma_\text{vac}$ as free parameters and $m_g$ in the small range
mentioned to find the minimum of
$\Delta\varepsilon$ and we get parameter set PS-II in
tab.~\ref{tab:params}. In the last variation we restrict additionally
the coupling constant to $g_s=3.3$ in order to reproduce  not
only the string tension but also the Coulomb parameter $\alpha=0.29$
of the Cornell potential as calculated on the lattice
\cite{Bali:2000gf}. The last free parameter $\sigma_\text{vac}$ is
varied in this case to get the right string tension and we obtain
parameter set PS-III. Of course the
deviation of the CDM energy profile to the lattice result increases
from parameter set PS-I to PS-III. The three different parameter sets
therefore express a quantitative variation of the model predictions.
With the given parameters the scalar potential $U(\sigma)$ changes as
shown in fig.~\ref{fig:U_ps123}. Within PS-I (solid curve) there is
only a negligible relative maximum in the potential, and it vanishes
exactly for PS-III (dash-dotted line). In the second parameter set a
pronounced relative maximum develops (dashed line). 
We note that the values of the bag constant $B$ are much larger than
the value $B^{1/4} = 145\,$MeV chosen in the original MIT bag model
\cite{Chodos:1974je} and also in a previous CDM analysis
\cite{Traxler:1998bk}. Recall that in \cite{Traxler:1998bk} the
string profile is much broader and inconsistent with lattice results.
The vacuum value for the dielectric constant
is $\kappa_\text{vac}=10^{-4}$ for all parameter sets. We show in 
appendix \ref{sec:kappa-dep}, that the physical quantities of the
string do not change anymore for smaller values of $\kappa_\text{vac}$.

\begin{table}[htbp]
  \centering
  \begin{tabular}{c||c|c|c|c|c}
      No. & $B$ & $m_g$ & $\sigma_\text{vac}$ & $g_s$ & $\kappa_\text{vac}$
       \\\hline 
       I   & (260 MeV)$^4$ & 1000 MeV & 1.29 fm$^{-1}$ & 2.0 & $10^{-4}$\\
       II  & (240 MeV)$^4$ & 1500 MeV & 1.13 fm$^{-1}$ & 1.8 & $10^{-4}$\\
       III & (240 MeV)$^4$ & 1700 MeV & 0.59 fm$^{-1}$ & 3.3 & $10^{-4}$
  \end{tabular}
  \caption{CDM parameter sets used in the description of $q\bar{q}$
       strings and $qqq$ baryons.}
  \label{tab:params}
\end{table}

\begin{figure}[htbp]
  \centering
  \includegraphics[width=\defaultfig,keepaspectratio,clip]{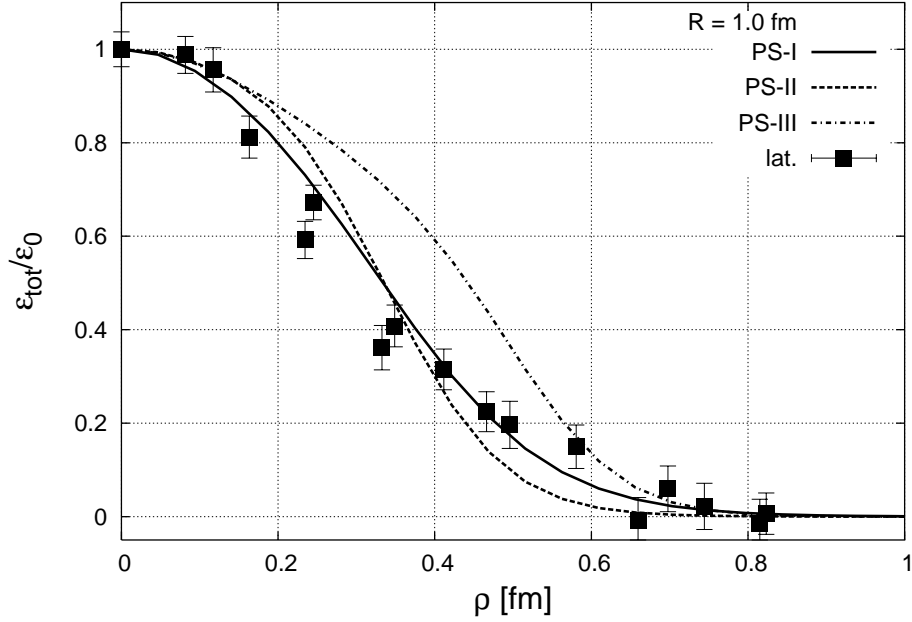}
  \caption{The energy density profile of a 1 fm string. The
    profiles for PS-I and PS-II have a half maximum width  of
    $\rho\approx0.35\,$fm although the slope is steeper for PS-II. 
    PS-III results in a broader string.
    Lattice results taken from \cite{Bali:1995de}.}
  \label{fig:energy_prof}
\end{figure}

In fig.~\ref{fig:energy_prof} we show the result of the three
different fitting procedures for the 1 fm long $q\bar{q}$ string. 
The profile for PS-I runs smoothly through the lattice points showing
the high quality of the fit. The profile of PS-II has roughly
the same half maximum width $\rho_{1/2}=0.32\,$fm but a steeper profile due
to the higher glueball mass. In the last case for PS-III the flux tube
is much broader with a half maximum width $\rho_{1/2}=0.45\,$fm. Due
to the large coupling $g_s$ for PS-III we cannot push the string
radius to smaller sizes while keeping the string tension at the
prescribed value.

\begin{figure*}[htbp]
  \centering
  \includegraphics[width=\widefig,keepaspectratio,clip]{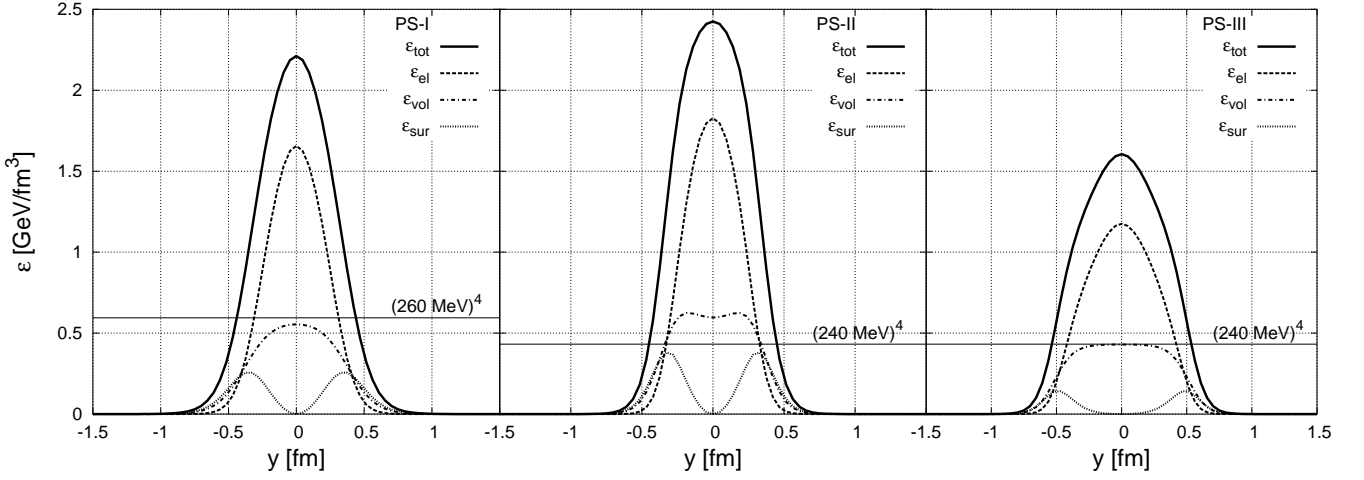}
  \caption{The decomposition of the energy profile into the different
  energy components for $R = 1\,$fm.}
  \label{fig:energy_frac}
\end{figure*}

To analyze further the different parameter sets, we decompose the
total energy according to eq.~\eqref{eq:energy_decomp} into the
different energy parts. The result is
shown in the profiles in fig.~\ref{fig:energy_frac}. In all cases the
electric energy builds up most part of the total energy. Note that
this is not in contradiction to the previous result that the electric
and the volume part of the string tension are of the same magnitude. 
The string tension is by definition $\tau = \frac{dE_\text{tot}}{dR}$
and not the energy of the central slice of the string. This would be the
same only for strings with constant width, which is not the case for
finite quark separations.

For PS-I (left panel) the volume energy (dash-dotted lines) never
exceeds the bag constant, whereas for PS-III (right panel) it has a
wide range inside the bag, where it is constant and equal to $B$. For
PS-II (central panel) we see that on the string axis, the volume
energy comes close to the local minimum at $\sigma=0$, which is seen in the
central dip. The surface energy is strongly pronounced for PS-II and
the two peaks are clearly separated for PS-III. In the last case the
interior of the bag is therefore much more pronounced than in the
other two parameter sets.

\begin{figure*}[htbp]
  \centering
  \includegraphics[width=\widefig,keepaspectratio,clip]{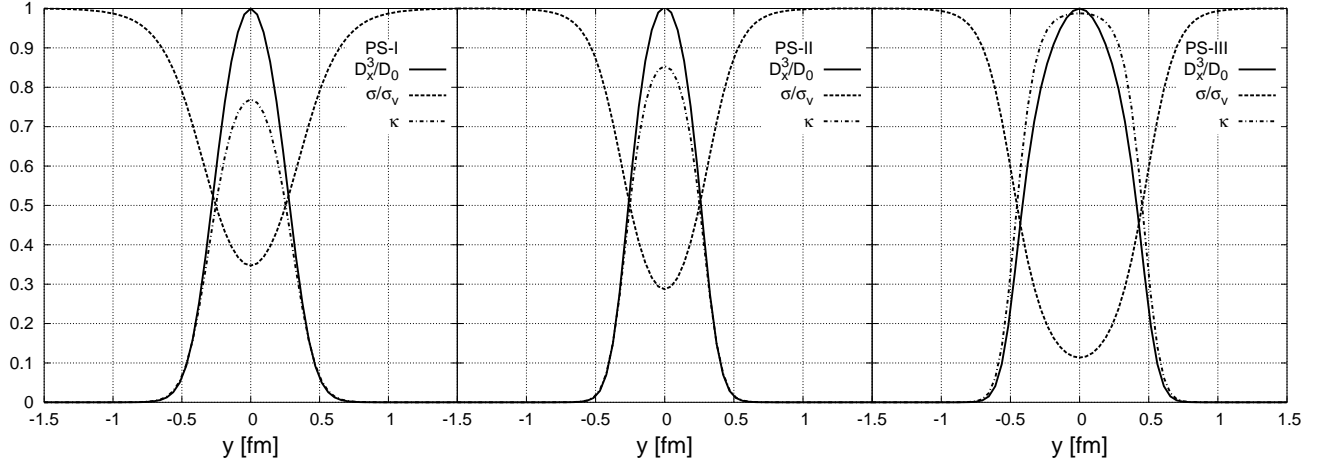}
  \caption{The scaled profiles of $\vec{D}^a$, $\sigma$ and
  $\kappa(\sigma)$ of a 1 fm string. The profiles of 
  $\kappa$ and $\vec{D}^a$ have roughly the same shape as expected
  from   eq.~\eqref{eq:abrikosov} for long flux tubes.
  $\kappa$ reaches 1 only for PS-III and is smaller than 1 for
  PS-I and PS-II.}
  \label{fig:1fm_fields}
\end{figure*}

The underlying fields $D^a(\rho)$, $\sigma(\rho)$ and the dielectric
constant $\kappa(\rho)$ for this 1 fm string are shown in
fig.~\ref{fig:1fm_fields}. For all parameter sets the scalar field
never reaches the perturbative situation $\sigma=0$ within the string.
According to that $\kappa$ is never exactly equal to 1. For PS-I and PS-II
it has only a value of $\kappa=0.75$ and $\kappa = 0.85$
respectively. Only for PS-III the perturbative value $\kappa\apprle 1$
is reached. Of course also the fields are broader as compared
to the other two parameter sets. 

\begin{figure*}[htbp]
  \centering
  \includegraphics[width=\widefig,keepaspectratio,clip]{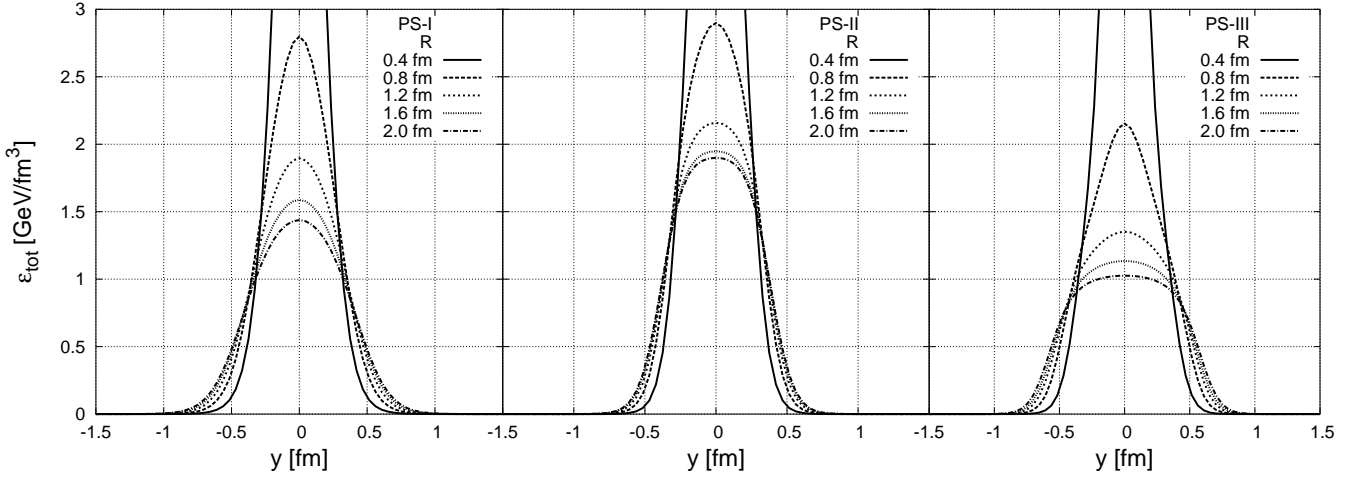}
  \caption{The profile of the energy density for different $q\bar{q}$
  separations $R$ given in the figure. The asymptotic shape 
  is nearly reached for $R\ge 1.2\,$fm. For small $R$ the Coulomb peak
  is seen.}
  \label{fig:string_profile}
\end{figure*}

The dielectric constant $\kappa(\rho)$ has roughly the same shape as
the electric field $D^a(\rho)$. This indicates already that according
to eq.~\eqref{eq:abrikosov} the string picture should be valid for
$R=1\,$fm. 
One can expect that for increasing $q\bar{q}$ distances $R$ the
profiles evolve to some stable shape and that the asymptotic relation
eq.~\eqref{eq:abrikosov} between electric field and dielectric
constant should
become increasingly accurate. To study this issue we show the profile of
the total energy density for different quark separations $R$ in
fig.~\ref{fig:string_profile}. For small $R$ we see the strong Coulomb
peak. But for separations $R>1.2\,$fm the string profile does not
change strongly anymore and the asymptotic profile is nearly reached.

\begin{figure*}[htbp]
  \centering
  \includegraphics[width=\widefig,keepaspectratio,clip]{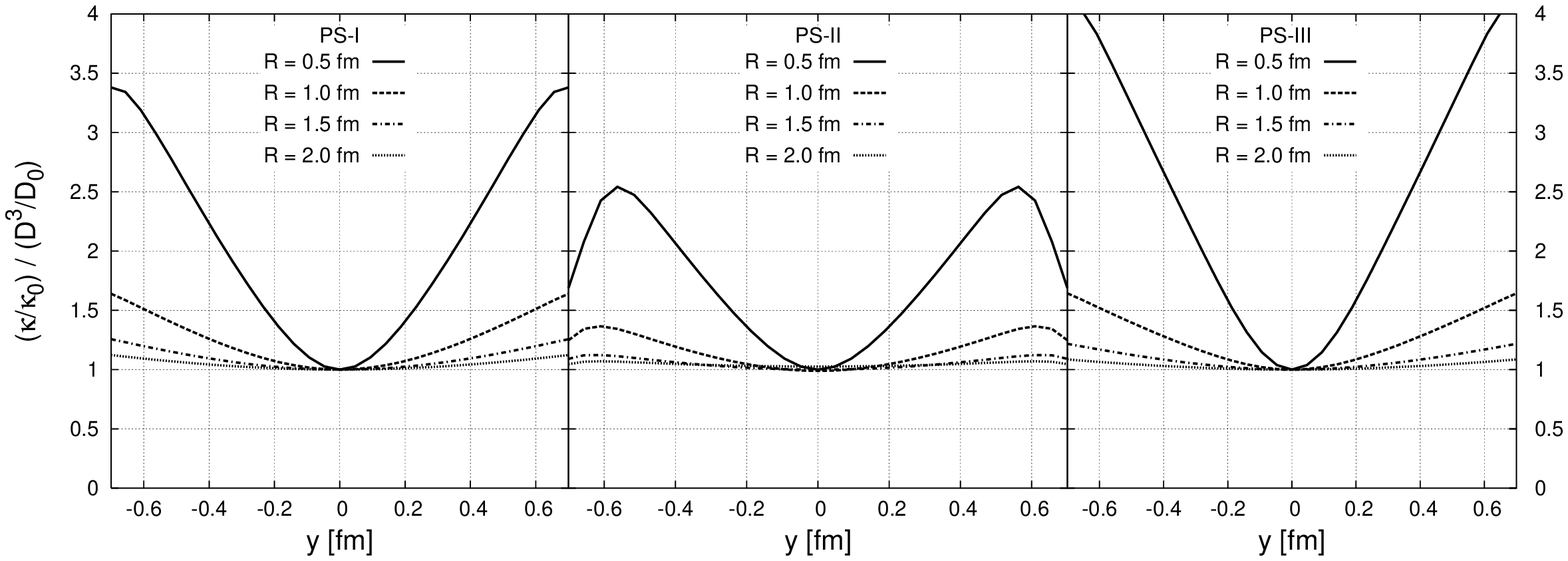}
  \caption{The ratio of the electric field $D^3$ to the dielectric
  constant $\kappa$, each normalized to its central value. The
  asymptotic relation in eq.~\eqref{eq:abrikosov} is nearly reached
  for $q\bar{q}$ separations $R\ge 1.2\,$fm.} 
  \label{fig:string_pic}
\end{figure*}

This can be seen as well in fig.~\ref{fig:string_pic}, where we have
plotted the ratio of the normalized profiles of $D^3$ and
$\kappa$ in the range $|y| \le 0.7\,$fm, where the string is
located. According to eq.~\eqref{eq:abrikosov} this should be constant
equal to unity for large $R$. Indeed 
this ratio becomes increasingly flat and equal to one for increasing
$R$. 

\begin{figure}[htbp]
  \centering
  \includegraphics[width=\defaultfig,keepaspectratio,clip]{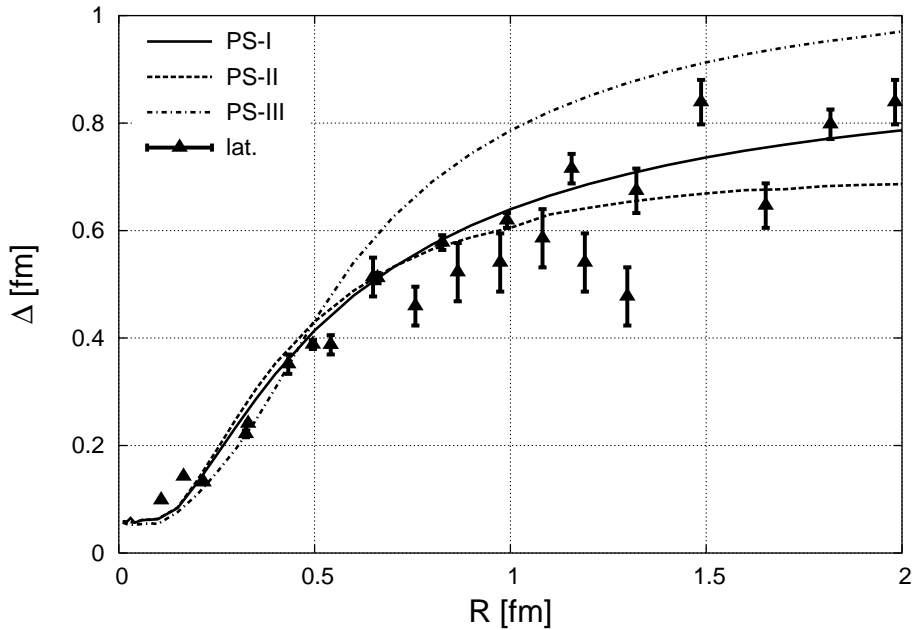}
  \caption{The Gaussian width of the flux tube as a function of quark
  separation. For parameter sets PS-II the width saturates
  at $\Delta \approx 0.7\,$fm. For PS-I and PS-III the width of the
  profile still increases slightly for separations up to $R=2\,$fm.
  Lattice data are taken from \cite{Bali:1995de}.}
  \label{fig:fwhm_string}
\end{figure}

To complete the discussion of the energy density profile we compare
the full half maximum width  $\Delta = 2 \rho_g$ extracted from the
Gaussian fit to the results obtained in lattice SU(2) calculations
\cite{Bali:1995de} for various string lengths $R$. Here we have fixed
the steepness parameter to $n=2$ to be consistent with the analysis
in \cite{Bali:1995de}. In fig.~\ref{fig:fwhm_string} one 
sees that the calculated string widths are compatible with the
lattice data for parameter sets PS-I and PS-II. Again the string width
is overestimated for PS-III due to the large coupling $g_s$. 
First the width increases rapidly for small separations $R
\le 0.8\,$fm. For the largest quark separation $R=2\,$fm the width has
a value $\Delta= 0.78\,$fm and $\Delta = 0.68\,$fm  for parameter sets
PS-I and PS-II, respectively, and the width seems to saturate. The width
for PS-III is $\Delta= 0.97\,$fm and still increases slightly. We note
that qualitatively this behavior is consistent with the lattice string
picture \cite{Kogut:1975ag,Kogut:1981ny}, where a logarithmic increase
of the string width with the quark separation $R$ is predicted.

\subsection{CDM and the Dual Color Superconductor}
\label{sec:dsc}

\begin{figure}[htbp]
  \centering
  \includegraphics[width=\defaultfig,keepaspectratio,clip]{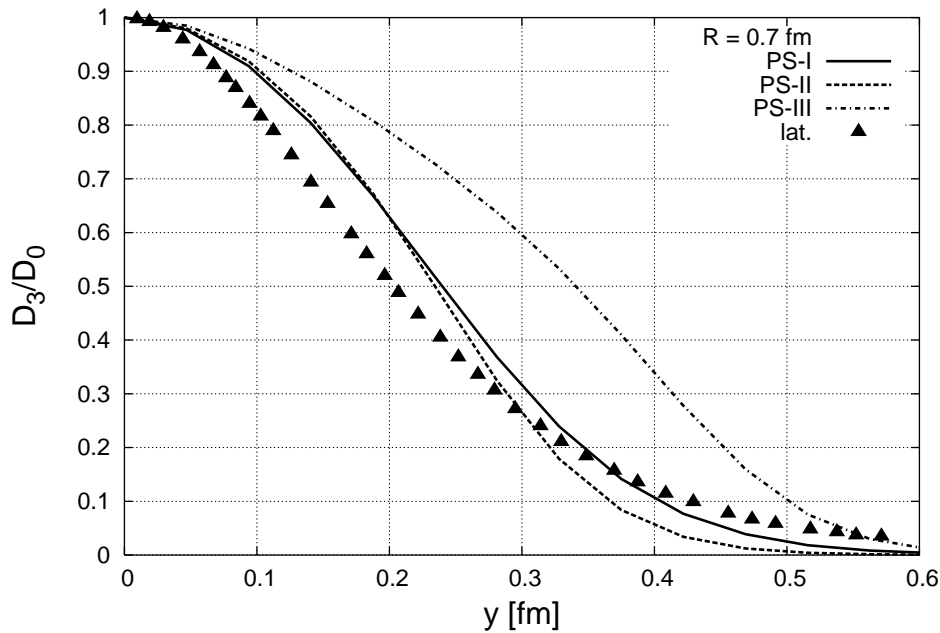}
  \caption{Comparison of the profile of the electric displacement
    $\vec{D}$ within CDM and the profile of the electric field
    $\vec{E}$ found on the lattice.} 
  \label{fig:abri_niels_oles}
\end{figure}

In this part of the discussion of $q\bar{q}$ strings we make
a connection to the \emph{Dual Color Superconductor} model (DCS) known
also as the \emph{dual Ginzburg-Landau} model. In
this model color flux tubes between a quark and an anti-quark are
formed by the interaction to a scalar field carrying a 
magnetic charge. Those Abrikosov-Nielsen-Olesen  vortices
\cite{Abrikosov:1957sx,Nielsen:1973cs,Ripka04:dual_super} are
stabilized by a circular magnetic current 
\begin{equation}
  \label{eq:mag_current}
  \vec{\jmath}_\text{mag} = \vec{\nabla} \times \vec{E}_\text{DCS}
\end{equation}
flowing
around the string axis. It is speculated that such a magnetic current
is produced by the condensation of magnetic monopoles which can be
constructed  in non-Abelian gauge theories
\cite{'tHooft:1974qc,Polyakov:1974ek}. The DCS flux tube has 
a characteristic profile of the electric field $\vec{E}_\text{DCS} =
E_\text{DCS}(\rho) \vec{e}_x$. To compare our results with the DCS
model we first note that by construction of the electromagnetic field
tensor $F^{\mu\nu,a}$ in eq.~\eqref{subeq:gluontensor} the curl of the
electric field $\vec{E}$ vanishes identically for static
configurations. However, by a simple redefinition of the field tensor
$F^{\mu\nu,a} \rightarrow G^{\mu\nu,a} = \kappa(\sigma)
F^{\mu\nu,a}$ one may define a quantity 
\begin{equation}
  \label{eq:cdm_mag_curr}
  j_\text{mag}^{\mu,a} := \partial_\nu
  {\cal G}^{\mu\nu,a} \neq 0  \quad ,
\end{equation}
where ${\cal G}^{\mu\nu,a} = \half
\epsilon^{\alpha\beta\mu\nu} G^{\mu\nu,a}$ is the dual field
tensor to $G^{\mu\nu,a}$. This quantity exactly behaves as a magnetic
current and the spatial part is in our standard notation
$\vec{\jmath}^{\;a}_\text{mag} = \vec{\nabla} \times \vec{D}^a$. In the
absence of magnetic fields there is no magnetic charge density,
i.e. $\rho^a_\text{mag} = \vec{\nabla} \cdot \vec{H}^a = 0$. In our model
the magnetic current is connected to the electric displacement
$\vec{D}^a$ whereas it is connected to the electric field
$\vec{E}_\text{DCS}$. Thus $\vec{D}^a$ and $\vec{E}_\text{DSC}$ are
confined in the two models and we compare $\vec{D}^a$ with
$\vec{E}_\text{DCS}^a$. In
\cite{Bali:1998cp,Gubarev:1999yp,Koma:2003hv} a detailed study 
of the electric profile of a $q\bar{q}$ configuration was made within
lattice gauge theory and it was found that the profile can be
described very well by the DCS model. In addition the authors found a
magnetic current $\vec{\jmath}_\text{mag}$ fulfilling
eq.~\eqref{eq:mag_current}. In fig.~\ref{fig:abri_niels_oles} we show
the lattice results of the 
electric profile (triangles) compared to those obtained within our
calculations. We find only a qualitative agreement, where again PS-I
and PS-II reproduces the electric profile better than PS-III.

\begin{figure}[htbp]
  \centering
  \includegraphics[width=\narrowfig,keepaspectratio,clip]{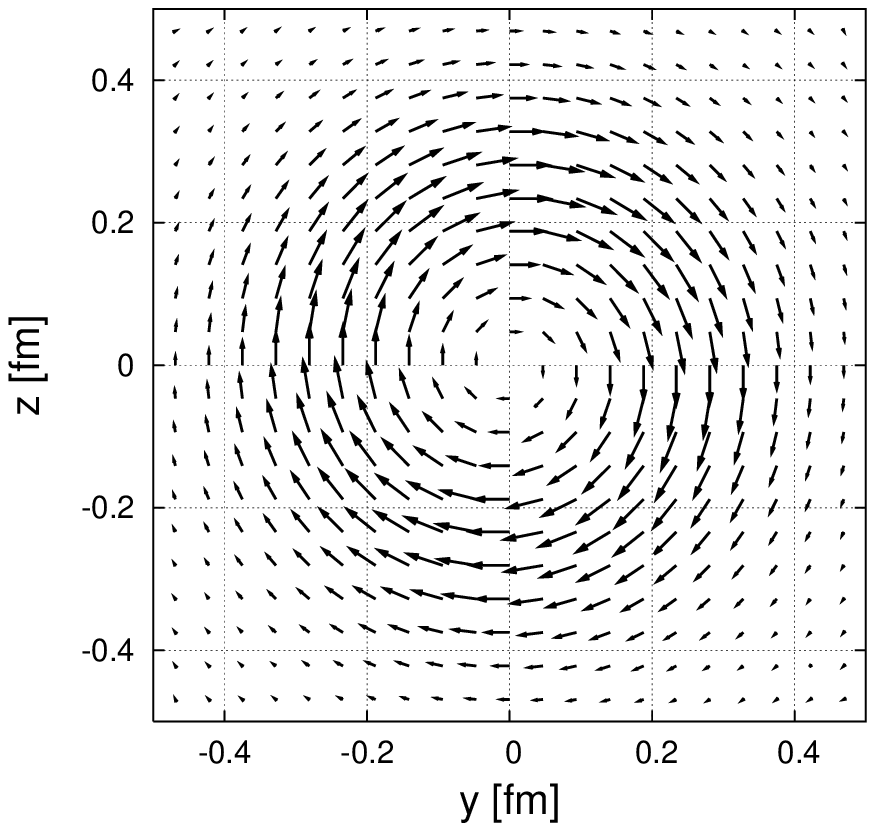}
  \includegraphics[width=\narrowfig,keepaspectratio,clip]{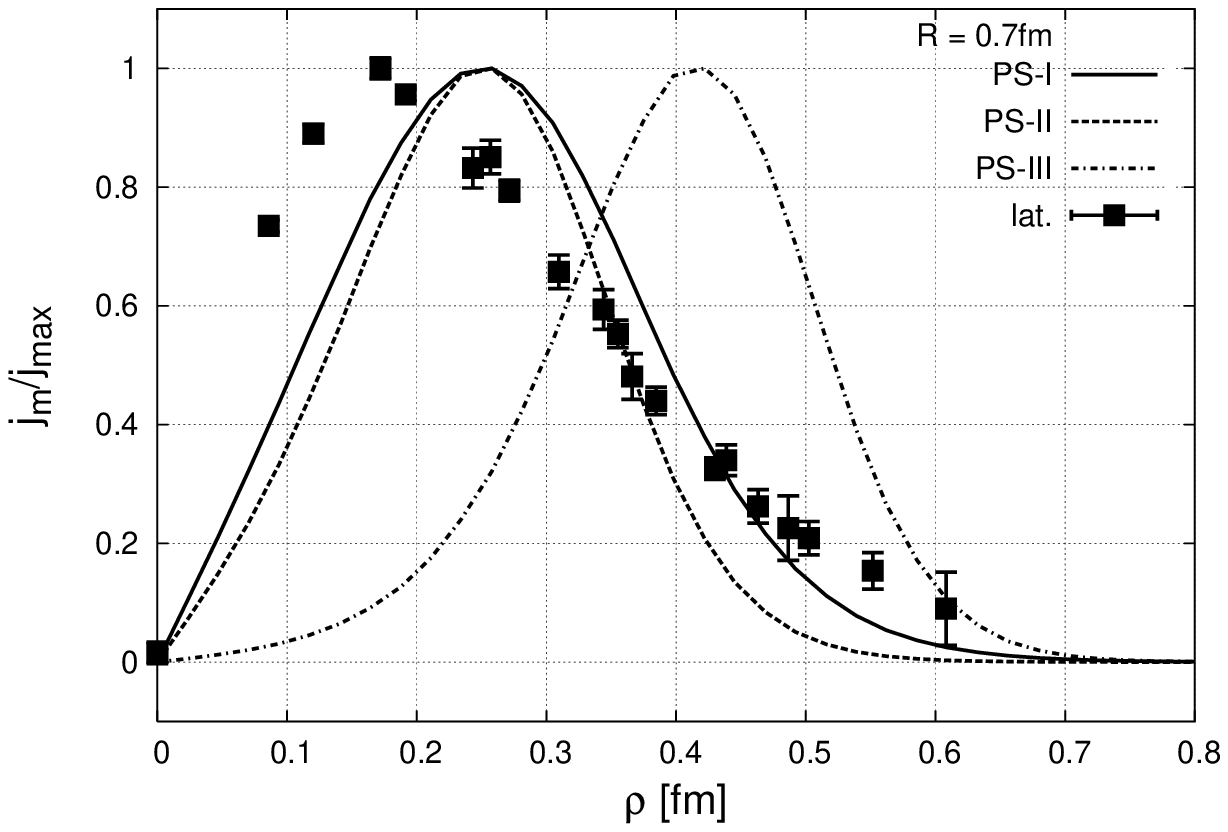}
  \caption{The circular magnetic current $\vec{\jmath}_\text{mag} =
  \nabla\times \vec{D}^a$ (left) and the scaled profile of it
  (right). For PS-III the current is locate at the surface of the bag.
  Lattice results taken from \cite{Bali:1998cp}.}
  \label{fig:mag_current}
\end{figure}

The spatial part of the magnetic current constructed in
eq.~\eqref{eq:cdm_mag_curr} in the central plane between the two
particles at $x = 0$ is displayed in fig.~\ref{fig:mag_current}
(left). In a  cylindrical basis it has only an azimuthal
component. One sees the circulating structure of the current which
gives this flux tube the name \emph{vortex} in the DCS model. 
We show our results for the
profile of the magnetic current with the different parameter sets and
compare it to lattice data \cite{Bali:1998cp} in the right panel of
figure \ref{fig:mag_current}. The maximal values are shifted to larger
values of $\rho$ compared to lattice data. Parameter set PS-III
develops a pronounced current only on the surface of the string.
In identifying the magnetic current in our model we can make a link
between our model and the model of the dual color superconductor.
It should
be noticed that the CDM model is formulated on the basis of the gauge
potentials $A_\mu^a$ which can be directly related to the gluon fields
of QCD, whereas the DCS model is formulated in the \emph{dual} gauge
potentials. In this sense the model gauge fields in the CDM  can be
more easily interpreted as the QCD gluon fields than in the dual
superconductor model.

\subsection{The $q\bar{q}$ potential}
\label{sec:qbarq_potential}

\begin{figure*}[htbp]
  \centering
  \includegraphics[width=\widefig,keepaspectratio,clip]{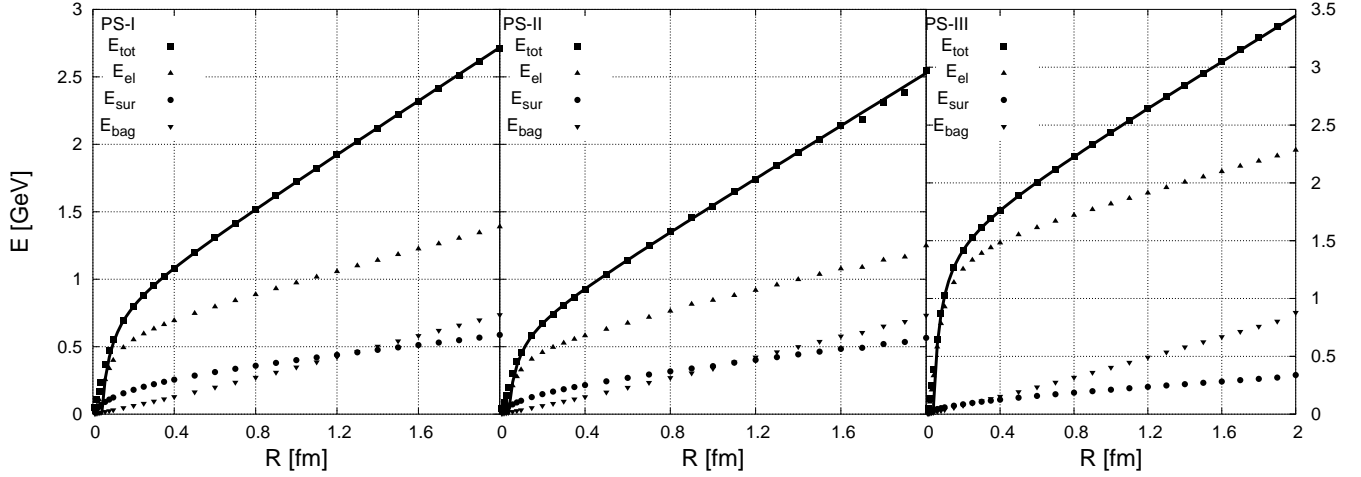}
  \caption{The $q\bar{q}$ potential $E(R)$ for the different parameter
  sets together with the different fractions of the energy and the
  Cornell fit to the total energy.}
  \label{fig:ps_cornell}
\end{figure*}

We can quantify the analysis of a string further by showing the total
energy $E_{\text{tot}}$ of the string as a function of the $q\bar{q}$
distance $R$ in fig.~\ref{fig:ps_cornell} together with the electric,
the volume, and the surface part of the energy. All parts of the
energy show a linear rise with the $q\bar{q}$ separation for $R
\gtrsim 0.2\,$fm. Due to the self energy of the nearly point-like
particles the electric energy is larger than the other two
energy contributions. This is most obvious for PS-III where the
coupling constant $g_s$ is largest. We have 
performed a Cornell fit according to eq.~\eqref{eq:cornell} to the
total energy $E_{\text{tot}}$ and to the single energy parts
separately. The extracted parameters $E_0, \alpha$ and $\tau$ 
and the corresponding values for the electric, volume and surface
energy contributions are listed in the first three columns in
tab.~\ref{tab:cornell_par}.
It should be noted, that the self energy $E_0$ depends on the width of
the particles. The given values correspond to a Gaussian width of $r_0
=0.02\,$fm which can still be resolved on the used computational grid.

The string tension is $\tau \approx 980\,$MeV/fm as the
model parameters were fitted to this value. 
The effective strong coupling $a_F = C_F^\text{ab}\alpha$ isolated
from the Cornell fit to the $q\bar{q}$ potential ranges from 0.12 with
PS-II to 0.30 with PS-III. 
A Cornell fit obtained from meson spectroscopy gives an effective
Coulomb coupling ranging from $a_F = 0.25$ \cite{Eichten:1977jk}
to $a_F = 0.5$ \cite{Quigg:1979vr}. The latter estimate
included both the charm and the bottom quark mesons. In the MIT bag
model \cite{DeGrand:1975cf} and in the CDM hadronization study
\cite{Traxler:1998bk} an effective coupling $a_F = 3.0$,
and $a_F = 2.6$ were used, respectively. Together with the small
value of the bag constant used in these works the string width amounts
to $\rho = 1.9\,$fm and $\rho = 1.5\,$fm, respectively, which is large
as compared to the value of $\rho=0.35\,$fm obtained on the lattice
\cite{Bali:1995de}. In a further bag model analysis
\cite{Hasenfratz:1980jv} the authors extracted a value
$a_F=0.38$ for the coupling constant. The coupling isolated from
a Cornell fit to lattice data prefers a value of $a_F = 0.3$
\cite{Morningstar:1997ff,Bali:2000gf}. The CDM values for the coupling
in tab.~\ref{tab:cornell_par} underestimates this value for
$a_F$ for PS-I and PS-II. Parameter set PS-III was designed in
order to reproduce the Cornell potential, so the agreement is optimal here.
 
\begin{table*}[htbp]
  \begin{ruledtabular}
    \centering
    \begin{tabular}{l|c|c|c||c|c|c||c|c|c}
      & \multicolumn{3}{c||}{fundamental} & \multicolumn{3}{c||}{adjoint}
      & \multicolumn{3}{c}{ad/fund}\\
      & PS-I  & PS-II  & PS-III & PS-I  & PS-II  & PS-III
      & PS-I  & PS-II  & PS-III\\\hline
      $e_D$ [MeV]              & 388  & 294  & 757
                               & 961  & 763 & 2202
                               & 2.5   & 2.6  & 2.9 \\
      $a_D$             & 0.18  & 0.12 & 0.30
                               & 0.41  & 0.31 & 0.87 
                               & 2.3   & 2.7  & 2.9 \\
      $\tau$ [MeV/fm]          & 979   & 982  & 980
                               & 1548  & 1502 & 1586 
                               & 1.6   & 1.5  & 1.6 \\\hline
      $e_D^\text{el}$ [MeV]    & 297   & 235  & 726
                               & 833  & 679 & 2163 
                               & 2.8   & 2.9  & 3.0 \\
      $a_D^\text{el}$   & 0.12  & 0.09 & 0.29
                               & 0.34  & 0.28 & 0.86 
                               & 2.7   & 3.1  & 3.0 \\
      $\tau_\text{el}$ [MeV/fm]& 404   & 390  & 424
                               & 640   & 585  & 695 
                               & 1.6   & 1.5  & 1.6 \\\hline
      $\tau_\text{vol}$ [MeV/fm] & 381   & 380  & 435
                               & 610   & 588  & 717 
                               & 1.6   & 1.5  & 1.7 \\\hline
      $e_D^\text{sur}$ [MeV]   & 108   & 78  & 47
                               & 151   & 108  & 63 
                               & 1.4   & 1.4  & 1.4 \\
      $a_D^\text{sur}$  & 0.07  & 0.05 & 0.03
                               & 0.10  & 0.07 & 0.04 
                               & 1.4   & 1.4  & 1.4 \\
      $\tau_\text{sur}$ [MeV/fm] & 195   & 212  & 122
                               & 299   & 328  & 174
                               & 1.5   & 1.5  & 1.4
    \end{tabular}
    \caption{The parameters $e_D= C_D E_0$, $a_D = C_D\alpha$ and
      $\tau$ extracted from  the Cornell fit to the total energy as well as to the
      different energy contributions separately. The first block of three
      columns belongs to quarks in the fundamental representation
      ($D=F$), the second to quarks in the adjoint representation
      ($D=A$) and the last shows the ratio of both.}
    \label{tab:cornell_par}
  \end{ruledtabular}
\end{table*}

After having shown the CDM results for the $q\bar{q}$ potential, we
would like to comment on the so called \emph{Casimir scaling}
hypothesis \cite{Ambjorn:1984mb, Ambjorn:1984dp, Bali:1999hx,
  Deldar:1999vi, Lucini:2001nv, Koma:2002wm}. According to this
hypothesis the 
$q\bar{q}$ potential should scale with the
quadratic Casimir operator $C_D$ of the static sources $q$ and
$\bar{q}$ in the representation $D$ . For small  
quark separations $R$ the potential is dominated by the perturbative
Coulomb term which already scales with $C_D$. For larger
distances this cannot be deduced from perturbation theory and
must be shown numerically on the lattice. This has been done for
various representations $D$ in \cite{Michael:1998sm,
  Kratochvila:2003zj, Bali:1999hx, Deldar:1999vi}. For the adjoint 
representation for example one should expect the string tension to
scale with $(C_A/C_F)\, \tau_F = 2.25 \tau_F$. This can hold if
at all only for intermediate quark distances, as the adjoint potential
should saturate for separations larger than some critical distance
$R_c$ \cite{Michael:1998sm,Kratochvila:2003zj}. Numerical values
for the string tension $\tau_D$ have been given in \cite{Bali:1999hx}
and in \cite{Deldar:1999vi}. An unambiguous confirmation for the
scaling hypothesis has not been
seen. The deviations from scaling where found
to be (2-5)\% \cite{Bali:1999hx} and (10-15)\% \cite{Deldar:1999vi}
for various representations. Interestingly, all results lie   
consistently under the value predicted by Casimir scaling. For the
ratio of the adjoint string tension to the fundamental one a value
$\tau_A/\tau_F = 1.97\pm 0.01 \pm 0.12$ was calculated in
\cite{Deldar:1999vi}, with the respective statistical and systematic
errors. This corresponds to a deviation from Casimir scaling by
$(12\pm6)\%$. 

In the CDM we can simulate adjoint quarks. We simply assign the
adjoint charge to the three quarks, i.e.\/ the sum of the charges of a
quark and an anti-quark. The numerical values are given in
tab.~\ref{tab:glue_charges}. Note that one can construct only 6
charged adjoint quarks. The other two members of the octet are
uncharged. We have already seen that the Coulomb term in our potential
scales with the square of the coupling constant $g_s$ or,  rephrased in
terms of the Casimir operator, scales with the Casimir Operator. The
string tension, however, was found to scale with $g_s$ and
$\sqrt{C_D^\text{ab}}$, 
respectively. Therefore the adjoint string tension should be $\tau_A =
\sqrt{C_A^\text{ab}/C_F^\text{ab}} \, \tau_F \approx 1.7 \tau_F$,
where we have used the Abelian Casimir values $C_F^\text{ab} = 1/3$ and
$C_A^\text{ab}=1$. 

We have performed the simulation for the adjoint string and isolated
the Cornell parameters $e_A = C_A E_0$ and $a_A = C_A
\alpha$ and $\tau_A$. The results of these values for the different parameter sets
are given in tab.~\ref{tab:cornell_par} (2nd block of three columns)
together with the corresponding ratios (3rd block). The short range
Cornell parameters of the electric energy have approximately the
expected ratio of 3. Also the adjoint string tension as well as the
different parts of the energy scale with the expected value of 
$\sqrt{C_A^\text{ab}/C_F^\text{ab}} = 1.7$ which has to be compared to
the  SU(3) scaling value $C_A/C_F = 2.25$ and the
empirical value $\tau_A/\tau_F = 1.97$ \cite{Deldar:1999vi}. In SU(2)
lattice calculations \cite{Kratochvila:2003zj} a clear deviation from
Casimir scaling was detected. We conclude that Casimir scaling is not seen
in our model which is in accordance with all bag models
\cite{Hansson:1986um,Johnson:1976sg,Lucini:2001nv}. Casimir scaling is
also not seen within the Dual Color Superconductor model, when the
superconductor is on the border between first and second type
\cite{Ball:1988cf,Koma:2002wm}, i.e.\ where lattice results agree with
this model the best. We note that it is also qualitatively 
in line with the tendency of lattice calculations to underestimate
the adjoint string tension systematically. 

We note in this context
that the deviation is an intrinsic feature of the CDM, which is due to
the larger fluxtube cross section in the present of charges in higher
representation. If in the future lattice data will show that Casimir
scaling would hold for the $q\bar{q}$-potential, this would be a
strong argument against all bag-like models in favor of models with
explicit scaling \cite{Shoshi:2002rd}. In the meantime we regard those
type of models as competitive.

\begin{table}[htbp]
  \begin{ruledtabular}
    \begin{center}
      \begin{tabular}{lcr}
        color      & $q^3$ & $q^8$ \\\hline
        $r\bar{g}$ &   $1$   &   $0$          \\
        $r\bar{b}$ &   $1/2$   & $\sqrt{3}/2$ \\
        $g\bar{b}$ &  $-1/2$   & $\sqrt{3}/2$ \\
        $g\bar{r}$ &  $-1$   &   $0$         \\
        $b\bar{r}$ &  $-1/2$   & $-\sqrt{3}/2$\\
        $b\bar{g}$ &   $1/2$   & $-\sqrt{3}/2$\\
      \end{tabular}
      \caption{Color charge of color sources in the adjoint reprentation.}
      \label{tab:glue_charges}
    \end{center}
  \end{ruledtabular}
\end{table}

\section{$qqq$ baryons}
\label{sec:baryons}

\begin{figure}[htbp]
  \centering
  \includegraphics[width=0.5\narrowfig,keepaspectratio,clip]{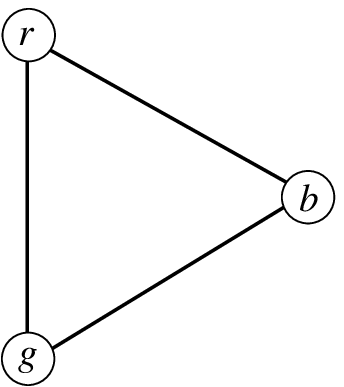}
  \hspace{2em}
  \includegraphics[width=0.5\narrowfig,keepaspectratio,clip]{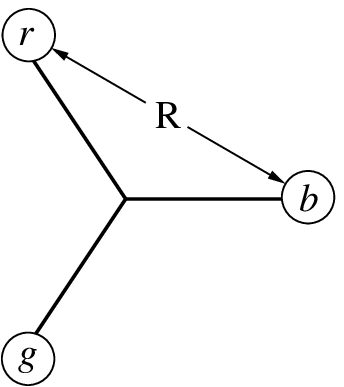}
  \caption{$\mathsf{\Delta}$-shape (left) and \textsf{Y}-shape (right) of the $qqq$
  configuration. In the \textsf{Y}-geometry each of the three quarks is
  connected to the central point with the same string as in the
  $q\bar{q}$ case. In the $\mathsf{\Delta}$-geometry the quarks are connected
  pairwise by a modified flux tube.} 
  \label{fig:delta_y}
\end{figure}

Now that we have fixed the model parameter to reproduce lattice
results for $q\bar{q}$ strings, we go on to describe baryon-like $qqq$
configurations. It has been discussed for a long time
\cite{Cornwall:1977xd,Sommer:1984xq,Cornwall:1996xr,Kuzmenko:2000rt,
  Alexandrou:2001ip, Takahashi:2002bw},  whether 
the flux tubes stretching between the quarks will connect the
particles pairwise with a $qq$ string,  indicating 2-particle
interactions, or whether they are connected via a central point giving
rise to a real 3-particle force. In the former case the geometry of
the system will show a 
triangular or $\mathsf{\Delta}$-like shape, whereas in the latter it will have a
\textsf{Y}-like shape. The situation is depicted in figure
\ref{fig:delta_y}. Although our model is formulated in the Abelian
approximation we must not naively assume that the flux tubes are a
simple superposition of three $qq$ flux tubes. The non-linear
interactions with the dielectric medium might deform the flux tube to
show a \textsf{Y}-like geometry. 
This question has been studied on the level of the potential within
lattice SU(3) \cite{Bali:2000gf,Takahashi:2002bw,Alexandrou:2001qg}
and on the level of the fields
\cite{Ichie:2002dy,Ichie:2002mi,Bornyakov:2002mh}. 
Depending on the shape the potential will scale 
characteristically with the 2-particle distance $R$. To parameterize
it in the spirit of the Cornell potential eq.~\eqref{eq:cornell} we
decompose the potential into a constant term due to the quark self energy, a
Coulomb-like short distance term and a confining linear term. The
constant term scales with the number of particles in the system and
the short range term scales with the sum over the two particle Coulomb
interactions. The Coulomb interaction is accompanied by the same color
factor
$\frac{1}{3!}\epsilon_{\alpha\beta\gamma}\epsilon_{\sigma\tau\gamma}
t^a_{\alpha\sigma}t^a_{\beta\tau} = -C_F^\text{ab}/2$ as in
eq.~\eqref{subeq:bar_energy}.

The confining term scales with the total length of the flux tubes
spanned between the quarks, which is different in the two geometries
(see fig.~\ref{fig:delta_y}). In the case of the \textsf{Y}-geometry
a $q\bar{q}$-like string is connected to each of the three  quarks,
meeting at a central  point. In the case of three quarks sitting on
the corners of an equilateral triangle with length $R$, the length of
each of the flux tubes is equal to $R/\sqrt{3}$. This yields
eq.~\eqref{subeq:pot_mercedes} below. 

In the $\mathsf{\Delta}$-geometry two flux tubes
are connected to each quark, so there is a reduced electric flux in each
$qq$ string compared to that of the $q\bar{q}$ string. Due to the
symmetry of color exchange the total energy per unit length stored in the two
strings must be of equal size. To deduce this 
relative strength of the modified flux tube we apply a limiting
procedure. Think of two of the quarks approaching each other,
e.g. the $r$-quark and the $g$-quark. As they come together we end up
with a quark--diquark system but with two $qq$ strings lying on
top of each other. As the diquark behaves exactly like an anti-quark
the linear part scales with the known string tension $\tau$ and we
conclude that the $qq$ flux tube has a string tension reduced by a
factor of two. This factor is also obtained in
\cite{Cornwall:1996xr}. Note that we have assumed here, that the flux
in each $qq$ string is independent of the third quark position. 
The $qqq$ potential may now be parameterized as:
\begin{subequations}
  \label{eq:qqq-potential}
  \begin{alignat}{1}
    \label{subeq:pot_mercedes}
    V_\mathsf{Y} &= 3 C_F^\text{ab} E_0 - \frac{C_F^\text{ab}}{2} \sum_{i<j}
    \frac{\alpha}{|\vec{r_i} - 
      \vec{r}_j|} + \tau \sum_i |\vec{r_i} - \vec{r}_c| \\
    &= 3 C_F^\text{ab} E_0 - 3 \frac{C_F^\text{ab}}{2}
    \frac{\alpha}{R} + \sqrt{3} \tau R \nonumber\\
    \label{subeq:pot_delta}
    V_\mathsf{\Delta} &= 3 C_F^\text{ab} E_0 
    - \frac{C_F^\text{ab}}{2} \sum_{i<j} \frac{\alpha}{|\vec{r_i} - \vec{r}_j|} 
    + \frac{1}{2}\tau \sum_{i<j} |\vec{r_i} - \vec{r}_j| \displaybreak[2]\\
    &= 3 C_F^\text{ab} E_0 - 3 \frac{C_F^\text{ab}}{2}
    \frac{\alpha}{R} + \frac{3}{2} \tau R \nonumber
  \end{alignat}
\end{subequations}
Here $\vec{r}_i$ are the positions of the particles and $\vec{r}_c$ is the
position of the flux tube junction in the \textsf{Y}-case. The parameters
$E_0$, $\alpha$ and $\tau$ are the ones extracted from the fit to the
$q\bar{q}$ string in the previous section and listed in
tab.~\ref{tab:cornell_par}. The respective last equality in
eq.~\eqref{eq:qqq-potential} holds for 
the quarks sitting on the corners of an equilateral triangle.
The effective string tension $\tau_\mathsf{\Delta} = (3/2)\, \tau$ is smaller than
$\tau_\mathsf{Y} = \sqrt{3} \tau$ making it the preferable configuration
although it is only an effect of 14\%.

This picture is surely oversimplified, since the flux tubes are not
strings with zero transverse extent. The single flux tubes will
overlap due to their finite width $\rho$ and the true fields
will lead to a smearing of the two extreme configurations at least for
quark distances only slightly greater than the flux tube width. Both
the finite size of the flux tubes and the tininess of the overall
effect makes it hard to decide on the lattice, which might be the
more appropriate description. In references
\cite{Bali:2000gf,Alexandrou:2001ip} 
a $\mathsf{\Delta}$-like scaling of the potential was found
suggesting an effective 2-particle interaction, whereas in
\cite{Takahashi:2002bw,Bornyakov:2004uv} the baryonic string tension
was better described by \textsf{Y}. In another calculation
\cite{Alexandrou:2002sn} neither ansatz give a proper description of the
potential for all quark separation. Instead the authors stated that
the potential is of the $\mathsf{\Delta}$-type for small separations $R$ and of
$Y$-type for large $R$. 

In the following we first show the fields for the three-quark system
with the symmetry shown in figure \ref{fig:delta_y} as calculated in
CDM. Then we will 
compare the $qqq$ potential obtained within CDM to those constructed
in eq.~\eqref{eq:qqq-potential} using the Cornell parameters $E_0,
\alpha$ and $\tau$ obtained in section 
\ref{sec:strings}. Finally we will perform a new fit of our Cornell
parameters to the $qqq$ potential. We would like to stress here, that
in a renormalization group derivation of the dielectric model
\cite{Mack:1983yi, Chanfray:1989vz, Mathiot:1989tu} an additional
colorless vector field appears. This vector field is relevant in the
calculations 
of systems with non-vanishing baryon-density. However in the spirit
of the present work, we are interested in the pure glue $qqq$-type
flux tubes with fixed external charges, where the neglect of this
additional term is justified.  

\begin{figure}[htbp]
  \begin{center}
    \includegraphics[width=\narrowfig,keepaspectratio,clip]{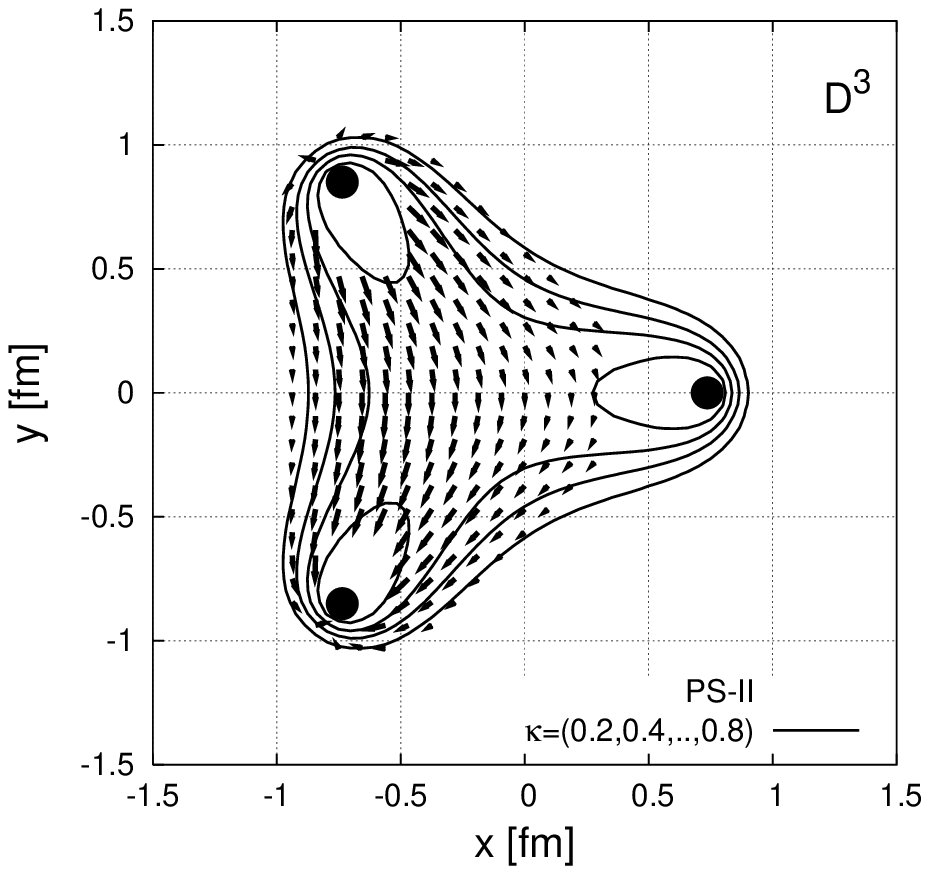}
    \includegraphics[width=\narrowfig,keepaspectratio,clip]{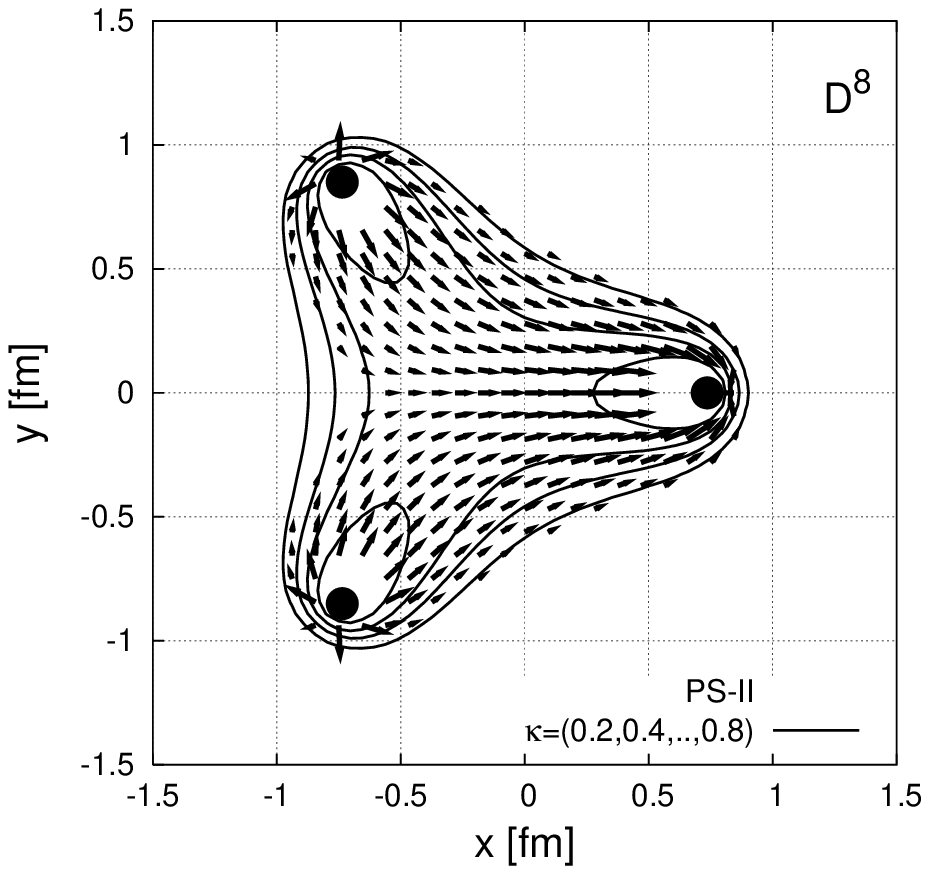}
    \caption{The color fields of a $qqq$ baryon. The red/green/blue
      quark is the upper/lower/rightmost one. The $\vec{D}^3$ flux
      (left) connects only the red and the green quark, whereas the
      $\vec{D}^8$ flux (right) connects all three quarks. The flux is
      bent into the center of the configuration. Solid lines are the
      equipotential values of $\kappa$.}
    \label{fig:qqq-color-fields}
  \end{center}
\end{figure}

In figure \ref{fig:qqq-color-fields} we show the color electric field
of a configuration, where the rightmost, uppermost and lowermost
quarks have color blue ($b$), red ($r$) and green ($g$),
respectively. The pairwise quark distance is equal to $R=1.7\,$fm.
In the left (right) part of the figure the electric
3-field (8-field) $\vec{D}^{3/8}$ is shown. As the $b$ quark has no
3-charge component the 3-field connects only the $r$ and the
$g$ quark. But as can be seen clearly, due to the existence of the
$b$ quark, the confinement field $\sigma$ is reduced in the
whole region between the three quarks and the electric 3-flux is bent
towards the $b$ quark. The electric flux of the 8-field connects all
three quarks. The $r$ and the $g$ quark are equal sources of this
field and the flux ends on the $b$ quark. Again the flux is deformed
compared to the $q\bar{q}$ flux tube and is pushed towards the center
of the $qqq$ configuration.
We have included in the same picture the
contour lines for the dielectric function $\kappa(\sigma)$. One can
see that the maximal value of $\kappa$ (neglecting the peak positions
at the quarks) shows a \textsf{Y}-type shape, and falls off to zero
towards the sides of the triangle within a range of 0.5$\,$fm.

As we have argued in section \ref{sec:cdm} the electric field
shown in figure \ref{fig:qqq-color-fields} is not an invariant
quantity under the global color symmetry. The strength and the
direction of the fields depend on the specific arrangement of the
colors in the $qqq$ state. We have included the picture of the
electric field only to give an insight to the underlying mechanism
of flux tube formation. The relevant quantity again is the energy
density of the system. As in the $q\bar{q}$ case we split the total
energy into electric, volume and surface terms
\eqref{eq:energy_decomp}.
\begin{figure*}[htbp]
  \centering
  \includegraphics[width=\narrowfig,keepaspectratio,clip]{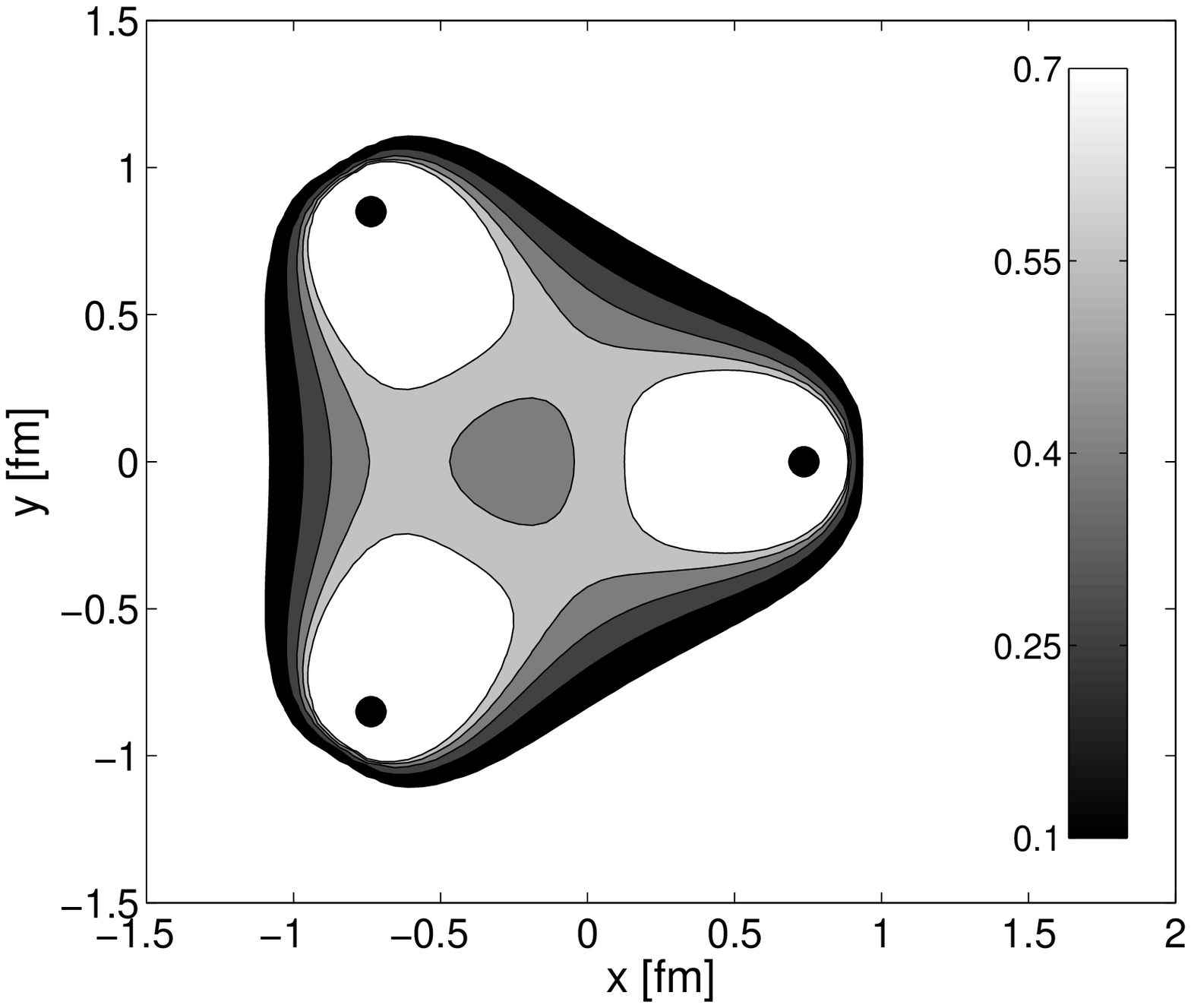}
  \includegraphics[width=\narrowfig,keepaspectratio,clip]{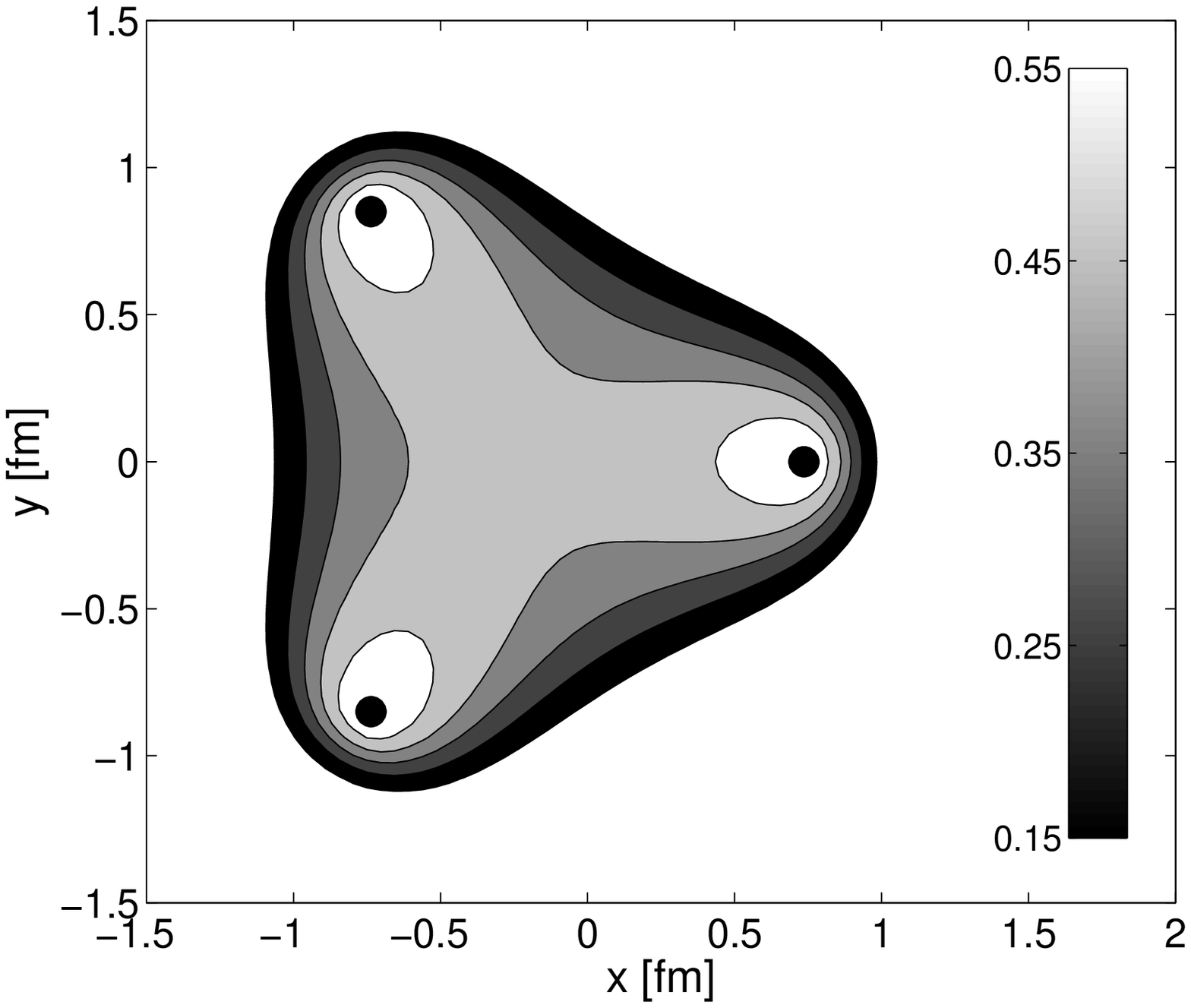}\\
  \includegraphics[width=\narrowfig,keepaspectratio,clip]{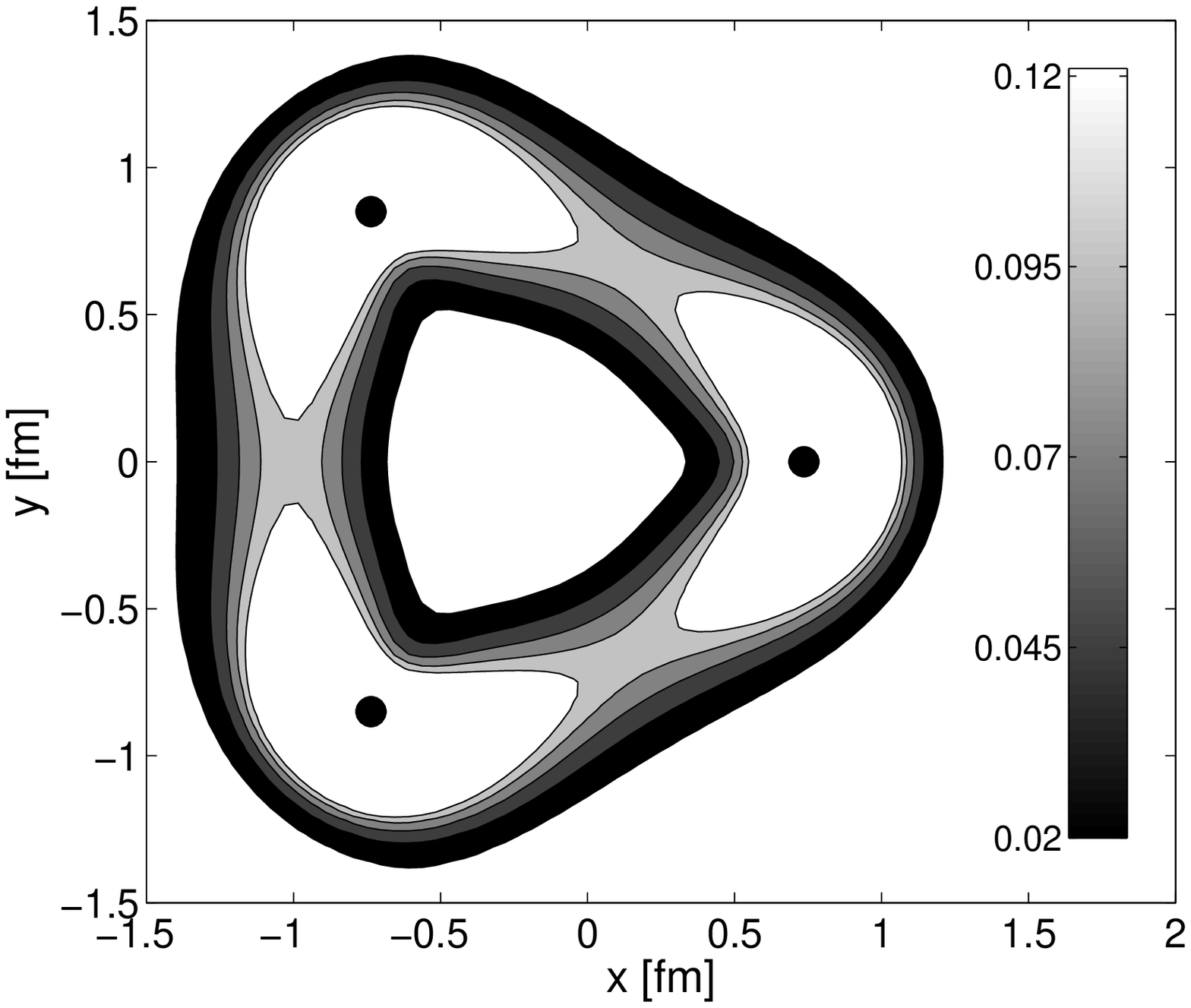}
  \includegraphics[width=\narrowfig,keepaspectratio,clip]{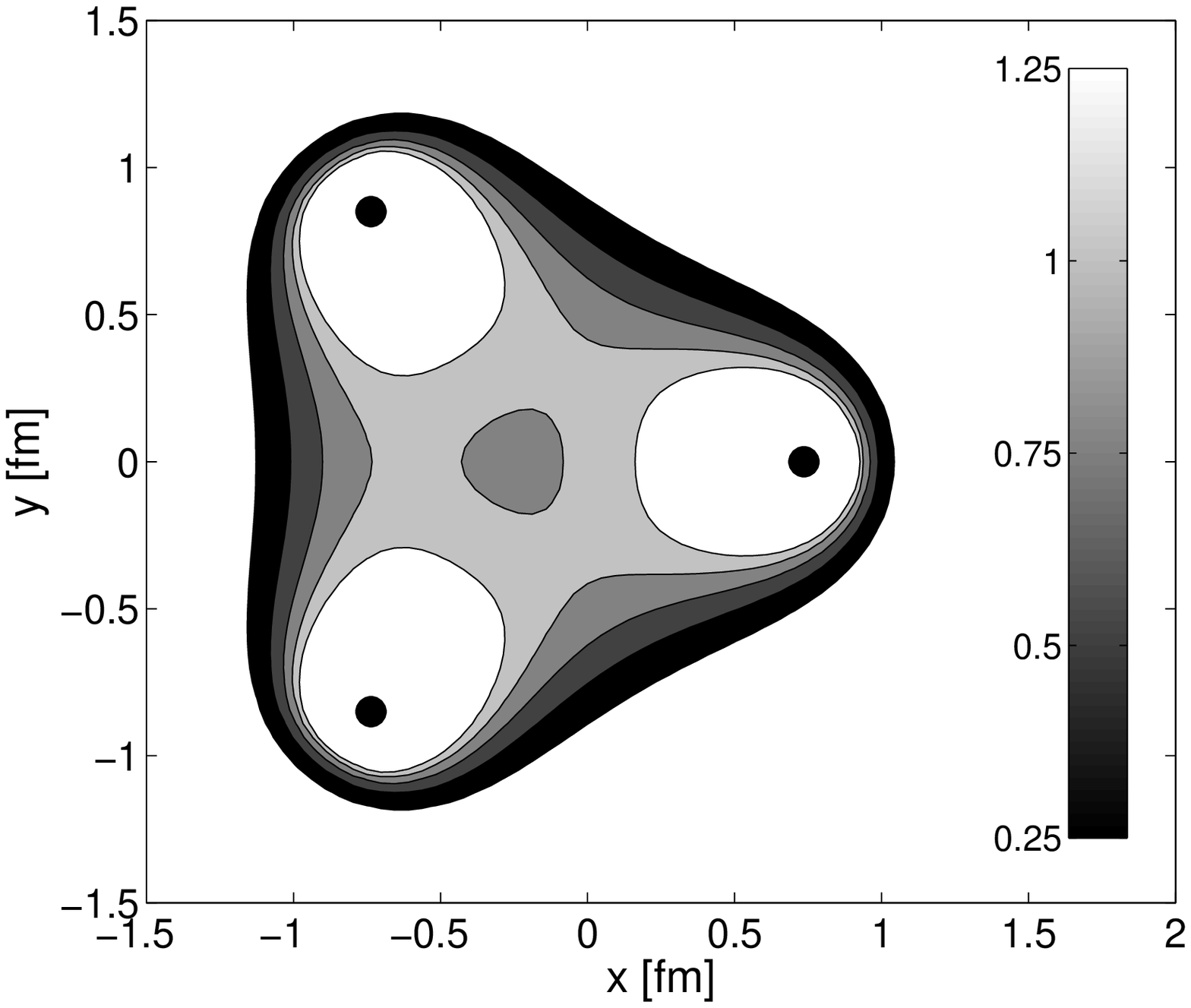}
  \caption{Energy density of the $qqq$ configuration. Black dots show
    particle positions. White voids at particles are spikes in the
    energy density due to the particles Coulomb peaks which are cut
    out. Upper left: $\epsilon_\text{el}$, upper right:
    $\epsilon_\text{vol}$, lower left: $\epsilon_\text{sur}$, lower
    right: $\epsilon_\text{tot}$. Lines are equidistant in energy
    density with the values given in the figures.}
  \label{fig:qqq_en_dens}
\end{figure*}
In figure \ref{fig:qqq_en_dens} we show the energy density of the
different types together with the total energy density obtained with
parameter set PS-I for a $qqq$ state with quark separation
$R=1.7\!$~fm. A clear difference in the geometries is seen for the
different energy fraction. The electric energy distribution 
(upper left panel) shows pairwise electric flux tubes between the
three quarks, bent into the center of the baryon. The geometry is
nearly $\mathsf{\Delta}$-shaped due to the dip at the center. Going along the
$x$-axis from the center into the negative direction, there is a well
in the energy density of about $95\,$MeV/fm$^3$, corresponding to 20\%
of the central value. This is qualitatively the same for all parameter
sets, but the energy barrier is larger in PS-II ($120\,$MeV/fm$^3$) and
smaller in PS-III ($90\,$MeV/fm$^3$). 

In contrast to the electric part of the energy the volume part has a
clear \textsf{Y}-shaped structure (upper right panel) which is the same as
for the dielectric function $\kappa(\sigma)$ shown in figure
\ref{fig:qqq-color-fields}.  The surface part (lower left panel) is
only relevant where $\sigma$ varies spatially. This is true on the
edges of the triangle and one sees therefore a pure $\mathsf{\Delta}$-like
distribution. The absolute value of the surface energy density in the flux
tubes is small compared to the values in the electric and the volume
fluxes. 

The sum of all three energies results in the total energy
density shown in the lower right panel. The picture shows
qualitatively the same structure as the one for the electric
energy. But the barrier of the total energy is only 10\% of the
central value and the competing structures of the electric energy and
the volume energy are smeared out. 

\begin{figure*}[htbp]
  \centering
  \includegraphics[width=\widefig,keepaspectratio,clip]{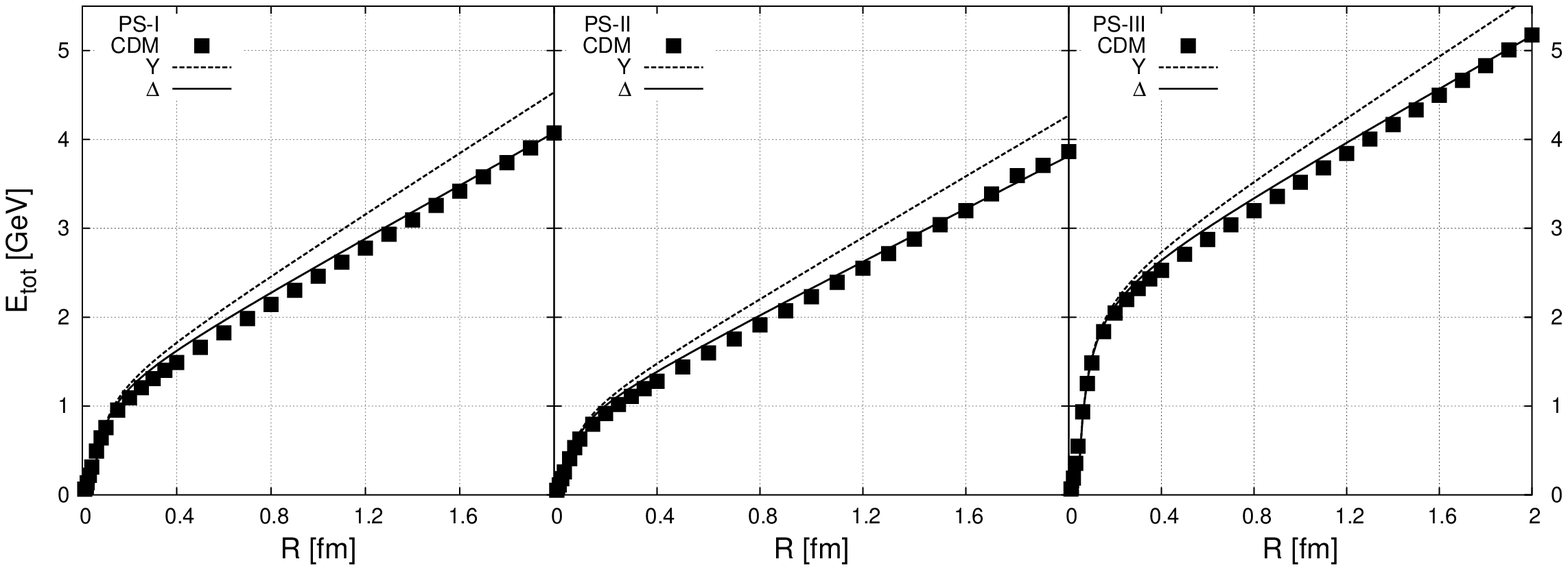}
  \caption{The total energy of the $qqq$ state compared to the \textsf{Y} and
    the $\mathsf{\Delta}$-forms of the Cornell parameterization.} 
  \label{fig:qqq_pot}
\end{figure*}

We will study now whether this picture of
superimposed structures can be found in the baryonic potential when
the size of the $qqq$ state is varied. The result is shown in figure
\ref{fig:qqq_pot} together with the \textsf{Y} and the
$\mathsf{\Delta}$-parameterization of the Cornell potential
\eqref{eq:qqq-potential}. 
The absolute values of the total energy agree very nicely with the
$\mathsf{\Delta}$-parameterization which is in line with the results in
\cite{Bali:2000gf,Alexandrou:2001ip}. However, it is the string
tension $\tau_{qqq}$ that distinguishes 
the two parameterizations and not the absolute value
$E_\text{tot}$. If one takes a 
closer look at the potential, one sees a slightly larger slope than
expected from the $\mathsf{\Delta}$-picture. To analyze the baryon potential
with respect to $\tau_{qqq}$ we make a fit of the form
\begin{equation}
  \label{eq:qqq-fit}
  V_{qqq} = 3 C_F^\text{ab} E_0 - \frac{3}{2} C_F^\text{ab}
  \frac{\alpha}{R} + \tau_{qqq} R 
\end{equation}
to the CDM results and extract the string tension $\tau_{qqq}$ from
the fit. Here $E_0$ and $\alpha$ are free fit parameters as well. As in
the $q\bar{q}$ case we make this fit to the total energy
\eqref{eq:energy_1} as well as to the electric, the volume and the
surface parts of the energy
\eqref{eq:energy_el}--\eqref{eq:energy_sur} to see if the different 
contributions show the same behavior as the energy distributions
in figure \ref{fig:qqq_en_dens}.

\begin{table}[htbp]
  \begin{ruledtabular}    
    \begin{center}
      \begin{tabular}{l||l|c|c}
        & PS-I & PS-II & PS-III \\\hline
        $\tau_{qqq}$[MeV/fm] & 1544 & 1568 & 1517 \\
        $\delta \tau$ & 0.33 & 0.41 & 0.21 \\\hline
        $\tau_{qqq}^\text{el}$[MeV/fm] & 616 & 588 & 607 \\
        $\delta \tau_\text{el}$ & 0.11 & 0.05 & -0.29 \\\hline
        $\tau_{qqq}^\text{vol}$[MeV/fm] & 650 & 670 & 725 \\
        $\delta \tau_\text{vol}$ & 0.88 & 1.11 & 0.72 \\\hline
        $\tau_{qqq}^\text{sur}$[MeV/fm] & 278 & 311 & 184 \\
        $\delta \tau_\text{sur}$ & -0.3 & -0.016 & 0.01
      \end{tabular}
      \caption{The $qqq$ string tension of the baryon Cornell
        potential and the deviation $\delta\tau$ from the expected \textsf{Y}
        ($\delta\tau = 1$) and $\mathsf{\Delta}$-picture ($\delta\tau = 0$).}
      \label{tab:qqq_pot}
    \end{center}
  \end{ruledtabular}
\end{table}
The result of the fits is summarized in table \ref{tab:qqq_pot}. Here
we show only the values for the $qqq$ string tension as both the
constant term and the Coulomb term scale in the same way in
the $\mathsf{\Delta}$ and the \textsf{Y}-geometry. Together with $\tau_{qqq}$ we
show the deviation 
$\delta\tau = (\tau_{qqq}-\tau_\mathsf{\Delta})/(\tau_\mathsf{Y} - \tau_\mathsf{\Delta})$ from
the expected values in both pictures. A value $\delta\tau = 0\;(1)$
indicates perfect agreement with the $\mathsf{\Delta}\;(\mathsf{Y})$ picture whereas
$\delta\tau \approx 0.5$ means that neither the \textsf{Y} nor the $\mathsf{\Delta}$
is realized but a transition between both geometries. 

First we confirm the visual impression that the electric and the
surface part of the string tension $\tau_{qqq}^\text{el}$ and
$\tau_{qqq}^\text{sur}$ are well described by the
$\mathsf{\Delta}$-picture. All deviations from the $\mathsf{\Delta}$
parameterization 
$\delta\tau$ are close to or even undershoot zero. In the surface part
of PS-I this can be explained by the respective energy density
(lower left panel in fig.~\ref{fig:qqq_en_dens}). There most part of
the energy is located outside the region defined by the triangle of
the quarks.
The opposite is true for the volume part $\tau_{qqq}^\text{vol}$. Here the
deviations are close to one, indicating a better description by the
\textsf{Y}-geometry. In the $qqq$ case the equality between the
electric and the surface part of the string tension is not true
anymore. For PS-III $\tau_{qqq}^\text{vol}$ is larger than
$\tau_{qqq}^\text{el}$ by 20\% with the same tendency in the other two
parameter sets.

The total string tension $\tau_{qqq}$ as a superposition of the three parts
takes on a value in between the $\mathsf{\Delta}$ and the
\textsf{Y}-value. All parameter sets have a slight preference for the  
$\mathsf{\Delta}$-picture with parameters $\delta\tau$ between 0.21 (PS-III)
and 0.41 (PS-II).

Another useful check for the scaling is to look at the $qqq$-potential
for arbitrary geometry of the $qqq$-triangle. With the
parameterizations for the \textsf{Y} and the $\mathsf{\Delta}$-picture
one would expect a universal behavior of the long range part of the
potential with the total string length $L$. For the
\textsf{Y}-geometry we have $L_\mathsf{Y}= \sum_i
|\vec{r}_i-\vec{r}_c|$ and for the $\mathsf{\Delta}$-geometry
$L_\mathsf{\Delta}=\sum_{i<j}|\vec{r}_i - \vec{r}_j|$. For triangles
with no angle exceeding $120^\circ$ the minimal string length
$L_\mathsf{Y}$ can be calculated \cite{Bornyakov:2004uv} with
\begin{equation}
  \label{eq:Y-length}
  L_\mathsf{Y} = \left(
    \frac{1}{2} \sum_{i<j} (\vec{r}_i-\vec{r}_j)^2 + 2\sqrt{3} S_\mathsf{\Delta}
  \right)^{1/2} \quad ,
\end{equation}
where $S_\mathsf{\Delta}$ is the area of the triangle spanned by the
three quarks.
\begin{figure}[htbp]
  \centering
    \includegraphics[width=\widefig,keepaspectratio,clip]{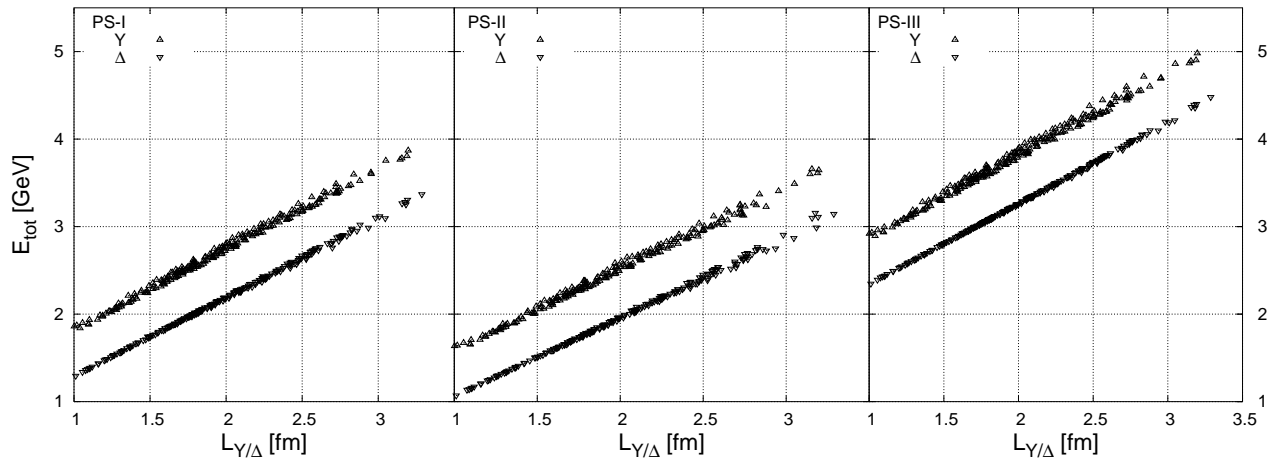}
  \caption{The $qqq$-potential for arbitrary placed quarks as a
    function of the total string length $L_\mathsf{Y}$ and
    $L_\mathsf{\Delta}$ respectively. $L_\mathsf{\Delta}$ is scaled by
    a factor $1/\sqrt{3}$ and the corresponding potential is shifted by
    -0.5 GeV for better visibility.}
  \label{fig:qqq_Vpot}
\end{figure}
We show in fig.~\ref{fig:qqq_Vpot} the $qqq$ potential for arbitrary
quark positions. We have selected only those triangles, where all
angles are smaller than $120^\circ$ and where the 2-particle distance
$|\vec{r}_i-\vec{r}_j|>0.4\,$fm. The visual impression agrees with the
previous result, that the $qqq$-potential is better described by the
$\mathsf{\Delta}$-geometry. The potential plotted as a function of
$L_\mathsf{Y}$ scatters more strongly in the energy than in the
$\mathsf{\Delta}$-case where it follows a rather straight line. Note
that this result is different to that obtained in \cite{Bornyakov:2004uv}. 

In summary the $qqq$ configuration shows two competing geometries visible in
the energy distribution and in the potential. The scalar field  has a
\textsf{Y}-like structure while the electric energy follows a
$\mathsf{\Delta}$-like 
distribution. In the former case the bag energy is minimized by
minimizing the volume of the three-quark bag. However, the flux
tubes have a finite transverse extent with no sharp boundaries.
Therefore the electric flux is not restricted to the \textsf{Y}-shaped
flux tube but may evolve in a region outside the scalar \textsf{Y}-volume.
The superposition of both the electric and the scalar energy
distributions leads to a potential right 
in-between the two ans\"atze, which is qualitatively
the same result as obtained in
a lattice calculation \cite{Alexandrou:2002sn}. 

\section{Conclusions and Discussion}
\label{sec:conclusion}

In this work we have shown that the Chromo Dielectric Model is able to
describe confinement of overall color neutral systems by the formation
of color flux tubes in a perfect dielectric vacuum. We studied the
dependence of the string tension and the string profile on the model
parameters. The parameters of
the model were chosen carefully to reproduce the profile of $q\bar{q}$
strings as well as the $q\bar{q}$ potential obtained in lattice
calculations and from heavy-meson spectroscopy. Three different
parameter sets were found that describe the profile of the $q\bar{q}$
flux tube obtained on the lattice optimally under given constraints.
All parameter sets reproduce the phenomenological value of the string
tension $\tau = 980\,$MeV/fm. Parameter set PS-I additionally describes
both the width as well as the shape of the flux tube rather well. In
PS-II we fixed the bag constant and the glueball mass parameters to
values from the literature and found that only the width of the profile
can be reproduced. The shape develops a steeper profile than the
lattice profile. In a final set we have fixed as well the coupling
constant to reproduce also the Coulomb parameter of the Cornell
potential. With this parameter set PS-III the resulting string width
is larger by 20\% than the value obtained on the lattice.
The profiles of the strings reach their asymptotic shapes for quark
separations larger than $R=1.2\,$fm. However, the width does not
saturate at a constant value but increases slowly, which is in line
with the lattice expectation, where the string width increases
logarithmically. It would be desirable if there were SU(3) lattice data
available of the same accuracy as obtained within SU(2) lattice
calculations. Then one could compare the CDM results to a somewhat
more realistic theory. 

We have also made a comparison of the $q\bar{q}$ flux tubes obtained
in CDM to the calculations within the Dual Color
Superconductor. We were able to 
extract a magnetic current in the CDM showing the same vortex-like
behavior as in the DCS. Both the profile of the electric field and
the magnetic current were in qualitative agreement with the profiles
calculated within the Dual Color Superconductor model and also
calculated on the lattice. 

Finally we have shown that the $q\bar{q}$ string tension does not obey
the Casimir scaling hypothesis. The scaling of the adjoint string
tension with $\sqrt{C_A/C_F}$ and not with $C_A/C_F$ is a general
feature of all bag-like models. Casimir scaling has not been proven
unambiguously on the lattice and therefore does not exclude our model. 

For $qqq$ configurations we have calculated the color 
electric fields, the color invariant energy density and the
$qqq$ potential to discuss the geometric form of those baryonic
states. We have found two competing pictures in the electric and the scalar
part of the system. The scalar energy distribution is confined into a
clear \textsf{Y}-like form whereas the electric energy distribution is of the
$\mathsf{\Delta}$-type form. In the total energy the overall values are in good
agreement with the $\mathsf{\Delta}$-type of the $qqq$ Cornell parameterization
but the string tension $\tau_{qqq}$ shows the same competing behavior
as the energy distributions: $\mathsf{\Delta}$-like in the electric sector and
\textsf{Y}-like in the volume sector of the energy. The total string tension
thus has neither the \textsf{Y} nor the $\mathsf{\Delta}$-like value but lies rather
in-between the two pictures.

Out of the three parameter sets PS-I to PS-III the first one describes
best all lattice data of $q\bar{q}$ flux tubes shown in the present
work, although the values of the model parameters differ 
from that found in the literature. It should be noted, that the
identification of $m_g$ with the glueball mass and of $B$ with the
gluon condensate obtained in QCD sum rules is motivated from heuristic
arguments only. Thus the results obtained with PS-I give a good
agreement to lattice data and the other two serve for a comparison if
the space of model parameters is restricted to the values given
above. The Chromo Dielectric Model gives therefore an adequate
mechanism of confinement, which can be understood dynamically. 

It would be interesting to see the influence of the color electric
fields within a self-consistent treatment including also the quarks
dynamically. Those calculations were done in \cite{wilets:1989}
including a direct interaction between the confinement field and the
quark field only. In our model the interaction with the quark field
is only indirect via the quark-gluon interaction $-g_s
\bar{\psi}\gamma_\mu t^a A_a^\mu\psi$ and the interaction of the gluons
with the confinement field. Here $\psi$ is the quark field obeying the
Dirac equation. 

A natural extension of our calculations is to include also
time-dependent fields according to eq.~\eqref{eq:eom}. Those studies
were done in \cite{Traxler:1998bk} where only the confinement field
was treated dynamically. Within those time-dependent calculations one
could study the hadronization out of a gas of colored quarks and gluon
fields which might be produced in a relativistic heavy-ion collision. 
Hadronization is up to now treated as an instantaneous process at the
freeze-out temperature. Within our model one could test this assumption as
a time-resolved process. Of course this is numerical challenging
due to the complexity of the equations but might be feasible within
nowadays computing power.

\appendix

\section{Numerics}
\label{sec:numerics}

\subsection{The FAS-Algorithm}
\label{sec:fas}

To solve the set of equations \eqref{eq:static_eoms} we define a
rectangular box with fixed volume $V = L_x \times L_y \times L_z = L^3$ and
discretize the equations on a Cartesian grid  $\Omega_h$ with $N =
n^3$  nodes and $n = 2^m = 128$. The grid spacing is therefore $h =
L/n$. The particle positions are not restricted to this grid, but can
be placed arbitrary. To avoid spurious oscillations in observable
quantities when changing the quark position, we assign a spatial width
to the quarks given by $r_0$. For the grids used in this work with
$L=3\,$fm we have a grid spacing $h\approx0.02\,$fm and this
determines the width $r_0=0.02\,$fm. 

We use the \emph{Full
  Approximation  Storage} (FAS) multigrid algorithm described
in \cite{brandt82,briggs87,joppich91,nrc96}. The FAS algorithm is
especially suited for our purposes 
because both the needed memory resources and the computing time scale
only with the number of grid points $N$ which is mandatory for
those large systems. It is like any other multigrid algorithm an
iterative solution technique that improves on an initial guess for the
solution step by step. Conventional relaxation methods cease to
converge for grids with increasing $n$ (decreasing $h$). This
means that the convergence get worse if one wants to decrease the
discretization error. To be more precise it is the low frequency part
of the system which does not converge anymore. The grid introduces an
infrared cutoff due to the finite dimensions $L$ of the box and an
ultraviolet cutoff due to the finite number of grid points $n$ (known
as \emph{Brillouin zone} in solid state physics). The Fourier spectrum
of the solution is built up of modes with wave vector $k_i = (\pi/L
\ldots i \pi/L \ldots k_\text{max} = n \pi/L)$. Modes of the error made
in approximating the 
true solution with wave vector $k_i$ and $i>n/2$ (oscillating
modes) are damped out fast, but modes with $i<n/2$ (smooth
modes) do not die out due to the locality of the difference operator. 

To cure this problem the algorithm introduces temporary coarser grids
$\Omega_\ell$ with $n_\ell = (n/2, n/4, \ldots, 2^{-\ell}n, \ldots,
2)$ nodes per dimension. On those grids the spectrum is reduced to
modes with maximal wave vector $k_\text{max}^\ell = 2^{-\ell}
k_\text{max}$ and smooth modes on the fine grid become oscillating
modes on the coarse grid. Further relaxation of the solution on the
coarse grid allows to solve for those modes and in turn to get an
estimate for the error made on the fine grid.

Thus the FAS algorithm consist of three building blocks: (i) the
relaxation method on each grid, (ii) the transport of the approximation
from a fine grid to the next coarser grid and (iii) the transport back
from the coarse to the fine grid. The recursion starting from the
finest grid $\Omega_0=\Omega_h$ down to the coarsest $\Omega_{m-1}$ and
back is called a V-cycle. In contrast to linear multigrid algorithms
the FAS algorithm allows also for non-linear differential equations.
 
Both the electric potentials $\phi^a$ and the confinement field
$\sigma$ are solved  within the same algorithm at once. As a
smoothing method we use $\mu = 4$ Gauss-Seidel relaxations with
red-black ordering for the smoothing update on each grid both on the
downward and the upward stroke of the V-cycle.
For the discretized version of
eq.~\eqref{eq:static_sigma} we have to include a Newton--Raphson
approximation in order to cope with the non-linearity. 
The transfer of all discretized
field quantities from coarse to fine grids and back is performed with 
linear interpolation and full weighting reduction operators
respectively. To improve the convergence of the algorithm we have to find a
good initial guess to start the V-cycle on the finest grid
$\Omega_0=\Omega_h$. To this end we first find a solution on the next coarser
grid $\Omega_1$ which is much less expensive in the computational costs. In
general we find a solution on every grid $\Omega_\ell$ with an initial guess
found on the grid $\Omega_{\ell+1}$.

The computational costs of the multigrid V-cycle in $d$ space dimensions can
be estimated as 
follows \cite{briggs87}. The $\gamma = 3$ independent fields ($\sigma, \phi^a)$ and
the corresponding residuals defined below in eqs.~\eqref{eq:residuals}
are stored on a grid 
with $n^d$ nodes. Thus the minimal amount of memory needed is given by
$M_1 = m_1 \gamma n^d$ float value units with $m_1=2$. On each of the
$n^d$ nodes the discretized 
equations for the $\gamma$ fields are solved locally $2 \mu$ times and
the $\gamma$ residuals are calculated once. If we estimate the amount of
work for the transfer between the fine and the coarse grid with
another $2 n^d$ operations, the minimal number of computational
operations on each grid is given by $M_2 = (2\mu+3)\gamma n^d 
= m_2 \gamma n^d$ with $m_2=11$. We may call the minimal amount of memory
$M_1$  and the minimal number of numerical operations $M_2$ one memory unit
and one working 
unit respectively. The equivalent numbers on a coarse grid $\Omega_\ell$ are
$M_{1/2}^\ell= m_{1/2} \gamma (n/2^\ell)^d$. An upper limit for the
numerical costs summed over all coarse grids can be estimated by a
geometrical series $\sum_{\ell=0}^\infty M_{1/2}^\ell = m_{1/2} \gamma n^d
(1-2^d)^{-1} = \frac{8}{7} m_{1/2} \gamma n^d$ for $d=3$. The algorithm is
surprisingly more efficient in higher dimensions $d$. In $d=3$ space
dimensions the amount of memory needed is therefore only
$\frac{8}{7}m_1$ times larger than that for traditional relaxation
methods and the computational work for a whole V-cycle sums up to
$\frac{8}{7} m_2$ times that for one relaxation sweep. One sees that
the total computational costs scale only linearly with the number
$\gamma n^d$ unknowns.

We impose Dirichlet boundary conditions on the
confinement field $\sigma = \sigma_{\text{vac}}$ and von-Neumann
boundary conditions on the electric potentials, i.e. $\frac{\partial
  \phi^a}{\partial n} = 0$ where $\vec{n}$ is a normal vector to the
boundary. This forbids any electric flux out of the computational
domain and is in line with the observed fact, that the electric flux
is compelled into the flux tube.

\begin{figure}[htbp]
  \centering
  \includegraphics[width=\narrowfig,keepaspectratio,clip]{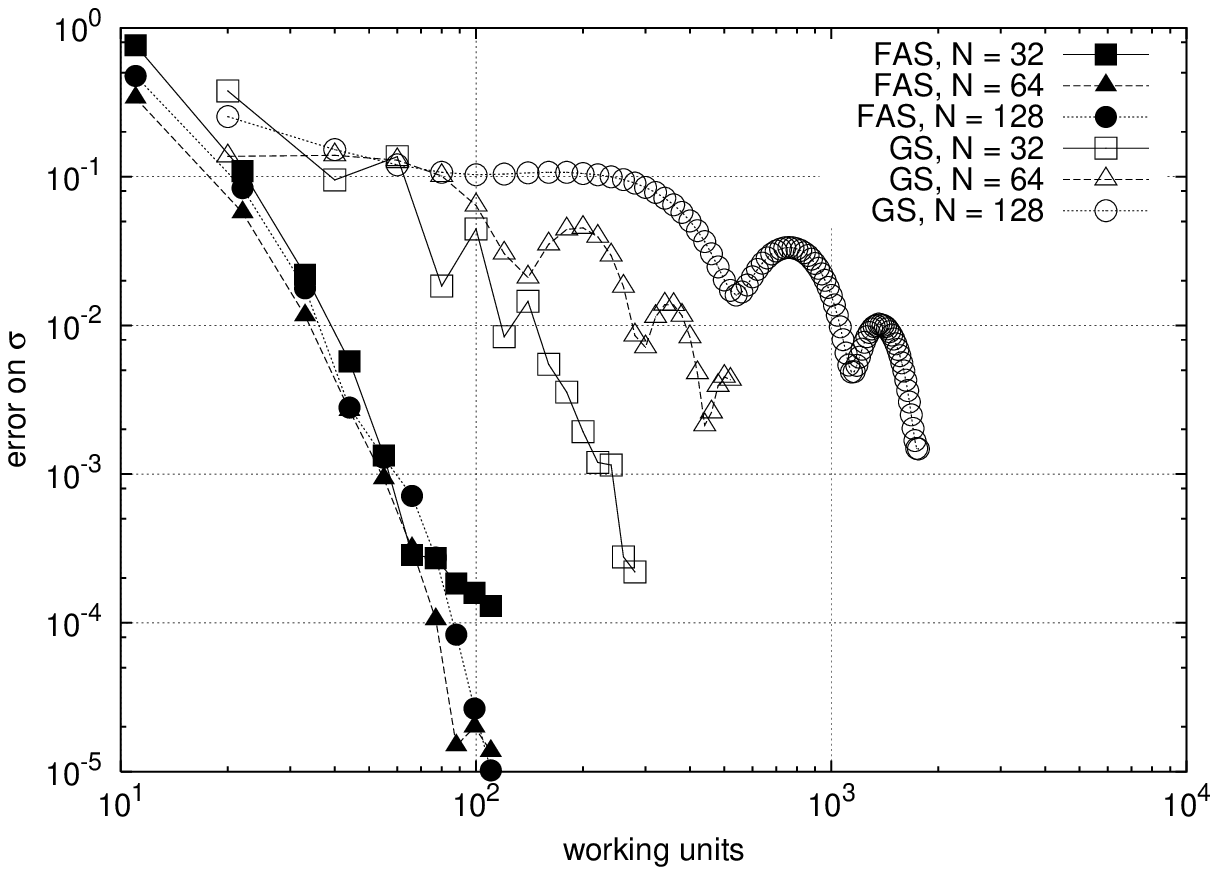}
  \includegraphics[width=\narrowfig,keepaspectratio,clip]{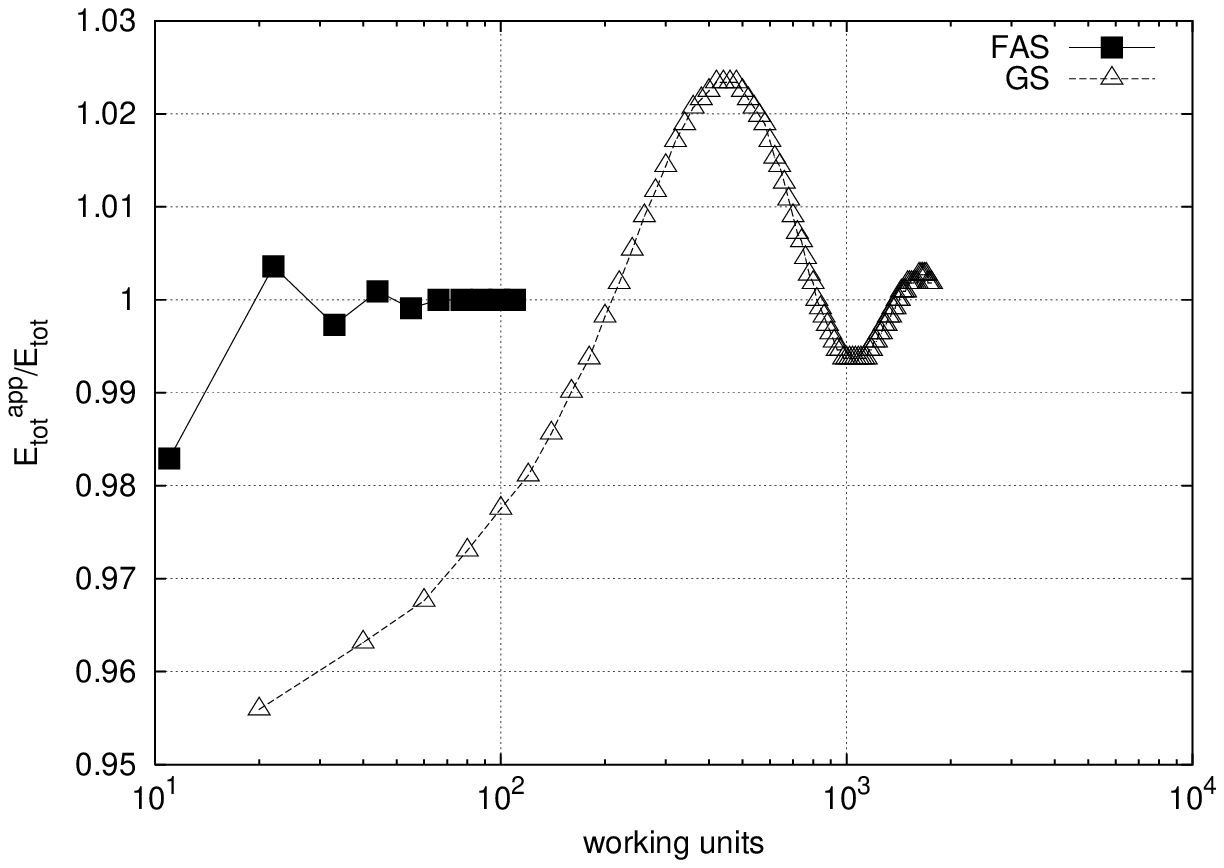}
  \caption{The residual $r_\sigma$ (left) and the total energy (right)
    as a function of the number of V-cycles in the FAS
    algorithm. Gauss--Seidel relaxation are the open symbols, the FAS
    results are given with solid symbols. Note the logarithmic scale
    for the computational work.} 
  \label{fig:convergence} 
\end{figure}

We compute after each V-cycle the residuals
\begin{subequations}
  \label{eq:residuals}
  \begin{eqnarray}
    \label{subeq:phi_res}
    r_\phi^a &=&
    \nabla\cdot(\kappa(\wt{\sigma})\vec{\nabla}\wt{\phi}^a) 
    + g_s \rho^a \\
    \label{subeq:sigma_res}
    r_\sigma &=& \nabla^2 \wt{\sigma} - U'(\wt{\sigma})
    + \kappa'(\wt{\sigma}) 
    \vec{\wt{E}}^a\cdot\vec{\wt{E}}^a  \quad . 
  \end{eqnarray}
\end{subequations}
with the approximations $\wt{\phi}^a$ and $\wt{\sigma}$
and the corresponding energies given in eq.~\eqref{eq:energy_decomp}.
We observe a rapid decrease of $|r_\phi|$ and $|r_\sigma|$ and
simultaneously a fast convergence to the asymptotic values of the
energy. We show as an example $r_\sigma$ and $E_\text{tot}$ as a
function of the computational work in figure \ref{fig:convergence} for
the multigrid algorithm (solid symbols) and for pure Gauss-Seidel
relaxation (open symbols). For both methods we have started with the
same initial guess obtained on the next coarser grid.
On the left panel one sees, that the rate
of convergence is slowing down for Gauss-Seidel relaxation when going
from small grids ($n = 32$) to larger grids ($n=128$) whereas it is
independent of the discretization within the FAS algorithm. On the
right panel we see that the total energy converges very rapidly to
the asymptotic one in the FAS algorithm. In practice we stop the
algorithm if the total energy stays constant within a given
error of $10^{-3}$. This is reached usually after three to five
V-cycles on the largest grid. The time needed for the solution is
reduced by nearly 2 orders of magnitude as compared to ordinary
relaxation. On a 2.2 GHz Pentium 4 machine a typical configuration is
solved on a ($N=128^3$) grid within 5-6 minutes.

\subsection{The vacuum value of the dielectric constant}
\label{sec:kappa-dep}

\begin{figure}[htbp]
  \centering
  \includegraphics[width=\narrowfig,keepaspectratio,clip]{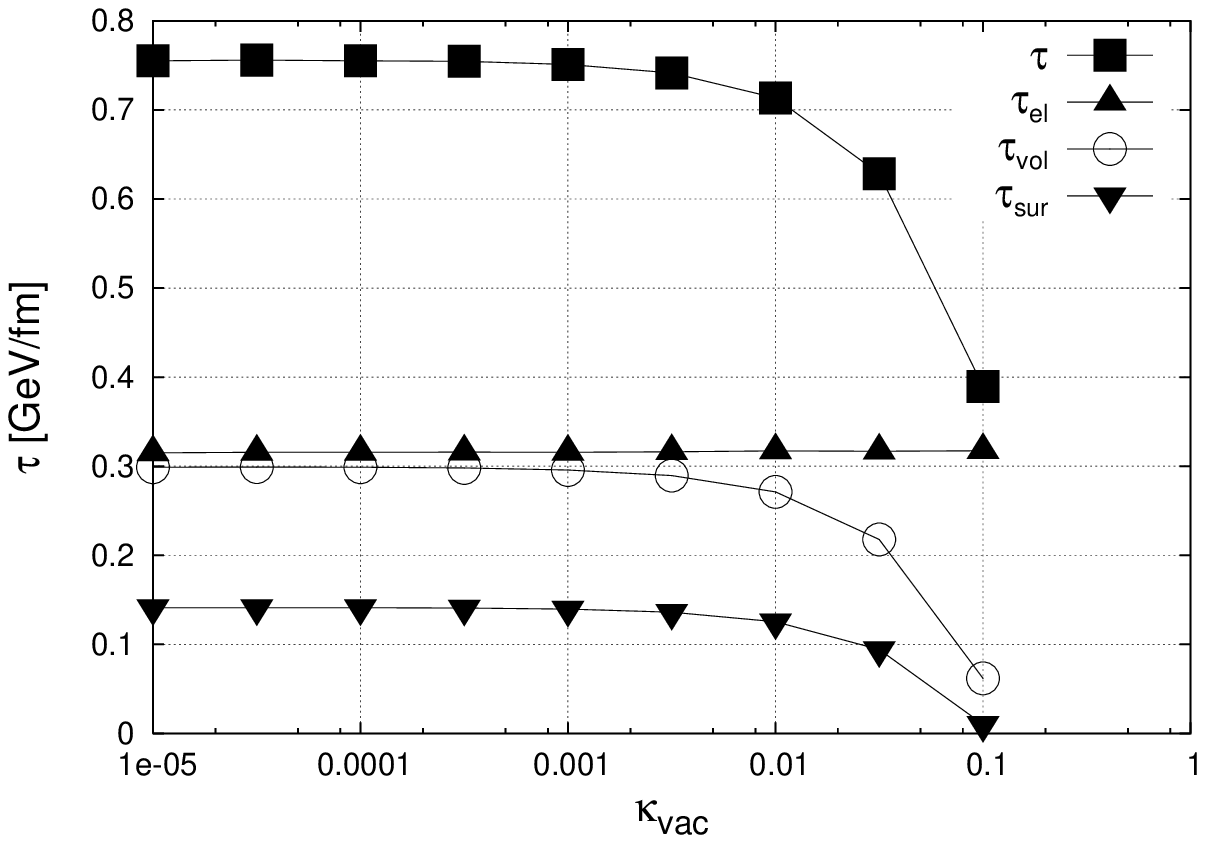}
  \includegraphics[width=\narrowfig,keepaspectratio,clip]{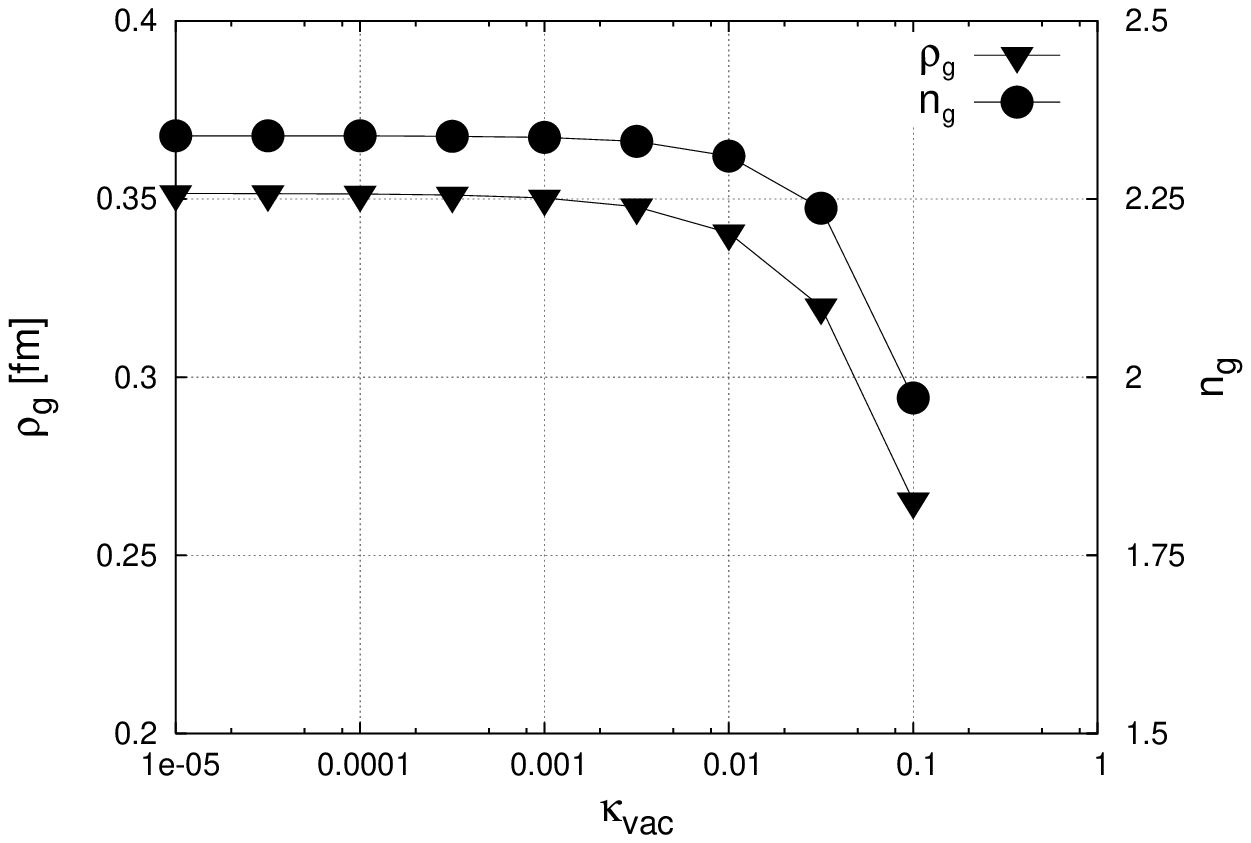}
  \caption{All parts of the string tension (left) and parameters
  defining the shape of the profile (right) saturate very fast for
  decreasing values of $\kappa_\text{vac}$. In all results shown in
  this work, we have used a value $\kappa_\text{vac}= 10^{-4}$.}
  \label{fig:kappa-dep}
\end{figure}

In the spirit of the model one should set
$\kappa_\text{vac}=0$. However the Poisson eq.~\eqref{eq:poisson} is
ill-defined in the perturbative phase in this limit.  In
\cite{Bickeboeller:1985xa} it was shown that in order to get a
consistent solution for the flux tube, one also must have
$\kappa'(\sigma_\text{vac})=0$. Therefore we 
have parameterized the dielectric constant according to
eqs.~\eqref{eq:kappa} and \eqref{eq:kappa_koeff} with a finite but
small value for $\kappa_\text{vac}$. With this form also
$\kappa'(\sigma_\text{vac})\rightarrow 0$ for
$\kappa_\text{vac}\rightarrow 0$. As we are dealing with a boundary
value problem we could not start right away with
$\kappa'(\sigma_\text{vac})= 0$ as in this limit
$\sigma=\sigma_\text{vac}$ everywhere is a solution of
\eqref{eq:static_sigma} which will be found by the
algorithm. Therefore we have chosen a rather large value
$\kappa_\text{vac}=10^{-2}$ when finding an initial guess on
coarser grids with $n\le32$. We then decrease the value of
$\kappa_\text{vac}$ step by step on the finer grids until a prescribed
minimal value is reached. We have analyzed the dependence of the
string quantities on $\kappa_\text{vac}$. In fig.~\ref{fig:kappa-dep}
we show that the string tension as well as the profile parameters do
not depend on $\kappa_\text{vac}$, once it is smaller than say
$\kappa_\text{vac}=10^{-3}$. The numerics were still stable and fast
for $\kappa_\text{vac}=10^{-4}$ which we have used throughout this
work. Of course it is not possible to calculate the energy of a
single-quark configuration, as the radius of the bag diverges and most
of the electric energy is stored outside the bag. But we can estimate
this energy in the same spirit as for the cylindrical flux tube with
sharp boundaries in eq.~\eqref{eq:bag_min} and obtain $E_q =
\frac{1}{3}[(2g_s^2 C_F B)/(\pi^2 \kappa_\text{vac}^3)]^{1/4}$. With
the smallest values used in our model ($B^{1/4}=240\,$MeV and $g_s=2$)
we obtain $E_q = 66\,$GeV which is large on a hadronic scale.

\begin{acknowledgments}
  G.~M. wants to thank G.~Ripka for valuable discussions about the
  magnetic current in the CDM during a visit at the ECT$^\star$, Trento,
  Italy. 
\end{acknowledgments}

\bibliography{bag,cdm,dual_CSC_GL,gluon_conf,hadron_potential,numeric}

\end{document}